\documentclass[amsmath,amssymb,aps,12pt,superscriptaddress]{revtex4-2}
\usepackage[breaklinks=true]{hyperref}
\usepackage{graphicx,fancyhdr}
\usepackage{geometry,color,bm}
\usepackage{natbib}
\usepackage{psfrag,overpic,tikz,breakcites}
\usepackage[export]{adjustbox}
\usepackage[mathlines,pagewise]{lineno}
\usepackage{float}
\usepackage{multirow}

\hypersetup{
    colorlinks = true,
    urlcolor   = blue,
    citecolor  = blue,
}

\begin{document}

\title{Momentum and kinetic energy transport in supersonic particle-laden turbulent boundary layers}

\author{Ming Yu}
\affiliation{State Key Laboratory of Aerodynamics, Mianyang, 621000, China}
\author{Yibin Du}
\affiliation{School of Aeronautic Science and Engineering, Beihang University, Beijing, 100191, China}
\affiliation{State Key Laboratory of Aerodynamics, Mianyang, 621000, China}
\author{Qian Wang}
\affiliation{Tianjin University, Tianjin 300072, China}
\affiliation{State Key Laboratory of Aerodynamics, Mianyang, 621000, China}
\author{Siwei Dong}
\affiliation{State Key Laboratory of Aerodynamics, Mianyang,  621000, China}
\author{Xianxu Yuan}
\email{Corresponding author: yuanxianxu2023@163.com}
\affiliation{State Key Laboratory of Aerodynamics, Mianyang,  621000, China}

\date{\today}

\begin{abstract}
In the present study, we conduct direct numerical simulations of two-way force-coupled 
particle-laden compressible turbulent boundary layers at the free-stream Mach number of $2.0$
for the purpose of examining the effects of particles on the transport of momentum and 
kinetic energy. 
By analyzing turbulent databases with various particle Stokes numbers and mass loadings, 
we observe that the presence of particles suppresses turbulent fluctuations and can even laminarize 
flow under high mass loading conditions. 
This is reflected by the wider and more coherent near-wall velocity streaks, 
reduced Reynolds stresses, and diminished contributions to skin friction and 
turbulent kinetic energy production. Additionally, the particle feedback force becomes 
more dominant in turbulent production near the wall 
and at small scales as mass loadings increase, which is found to be caused by 
the residual velocity fluctuations from particles swept down from the outer region. 
Furthermore, we identify that particle dissipation, resulting from the relative velocity 
between the fluid and particles, accounts for less than 1\% of mean kinetic energy 
viscous dissipation and less than 10\% of turbulent kinetic energy dissipation 
in the case with the highest mass loading. This suggests a modest impact on 
the internal energy variation of the fluid if two-way heat coupling is introduced. 
The elevated mean temperature is found in the near-wall region and is ascribed to the influence of 
the particle feedback force and reduced turbulent diffusion in high mass loading cases.
\end{abstract}

\maketitle

\section{Introduction} \label{sec:intro}

Particle-laden wall-bounded turbulent flows are commonly encountered in nature and 
engineering applications~\citep{rudinger2012fundamentals,cheng2012stochastic,liu2021large}.
The wall-bounded turbulent flows, characterized by a range of scales and 
containing energetic eddies whose sizes are influenced by their distances from the walls, 
carry the dispersed phase particles to acceleration and deceleration, 
to deposition and resuspension, to clustering and segregation
~\citep{saw2008inertial,soldati2009physics,m2016point,balachandar2010turbulent}.
In return, despite their small size, these particles impact turbulent motions by 
exerting feedback forces that extend to scales significantly larger than their own diameters
~\citep{brandt2022particle}.
The complexity of these processes is further increased in the presence of flow compressibility, 
particularly in scenarios such as aerospace engineering where high-speed vehicles travel through 
extreme weather conditions~\citep{cao2014effects,loth2021supersonic}.

When the particles are significantly smaller than the smallest turbulence scale and 
the total volume fractions are low, particle-laden flows can be simulated 
using the Eulerian-Lagrangian point-particle method.
In this method, turbulent fluid flows are treated as continuous phases simulated 
in the Eulerian coordinate system, while the motions of dilute phase particles 
are tracked in the Lagrangian frame~\citep{m2016point,brandt2022particle},
which has been proven to be capable of replicating the experimental results
~\citep{vreman2007turbulence,li2022high}.
Despite such simplification, there are still multiple parameters affecting the statistics and
dynamics of both the particles and turbulent fluid flows,
including the Reynolds and Mach numbers from the perspective of the flows,
and the particle inertia and mass loading from the perspective of particles.
The particle inertia can be evaluated by the particle Stokes number, 
which is represented by the ratio between the particle response time $\tau_p$ and 
the characteristic time scale of the flow $St = \tau_p/\tau_f$.
In wall-bounded turbulence, 
the characteristic timescale of the flow is commonly determined by the wall flow quantities,
such as the wall shear stress $\tau_w$, the wall viscosity $\mu_w$ and 
wall fluid density $\rho_w$, hereinafter referred to as $St^+$.
Mass loading, on the other hand, refers to the ratio between the mass of the particles 
$M_p$ and the fluid $M_f$ in the flow system $\varphi_m = M_p/M_f$.
In the limit of low mass fraction, the impact of feedback forces from particles to 
the fluid can be neglected, resulting in one-way coupling where particles are solely 
transported by the fluid. 
It has been observed that in both incompressible and compressible wall-bounded turbulence
~\citep{xiao2020eulerian,wang2024turbophoresis},
light particles act as tracers, almost following the motions of the fluids,
distributed uniformly in the turbulent channels, pipes and within the boundary layers.
The heavy particles with finite $St^+$, on the other hand, are less responsive to
high-frequency small-scale turbulent motions, but effectively respond to eddies with similar 
timescales to $\tau_p$, leading to the near-wall preferential accumulation and
the clustering beneath the low-speed streaks of the fluid~\citep{rouson2001preferential,
sardina2012wall,soldati2005particles,narayanan2003mechanisms}.
This behavior is most pronounced for particles with $St^+ = 10 \sim 100$ 
at low to moderate Reynolds numbers but are gradually alleviated when $St^+$ further increases
~\citep{picciotto2005characterization,bernardini2014reynolds,milici2016statistics}.
At high Reynolds numbers, the emergence of large-scale and very-large-scale motions,
along with the hierarchical structures in the logarithmic region,
further enhances particle streaks of varying spanwise widths,
depending on the consistency between $\tau_p$ and eddy turn-over time
~\citep{bernardini2013effect,saw2008inertial,jie2022existence,motoori2022role}.

As the mass loading $\varphi_m$ surpasses a threshold higher than $O(0.1)$,
the feedback force from the particle to turbulence commences to play an important role
~\citep{eaton2009two}.
This modulation of turbulence by particles is evident flow characteristics such as mean velocity,
Reynolds stress, and the morphology and intensity of the turbulent structures
\citep{balachandar2010turbulent,brandt2022particle}.
In terms of turbulent kinetic energy transfer, 
\citet{zhao2013interphasial} demonstrated that the particle force acting on the fluid enhances
the near-wall turbulence while attenuates it in the core region of the channel.
They also highlighted the imbalance in work exchange between particles and the fluid 
due to slip velocity, leading to additional dissipation of kinetic energy known as
the `particle dissipation'.
However, this dissipation is insufficient to impede the enhancement of streamwise velocity 
fluctuations.
\citet{pan2020kinetic} investigated the impact of particles on mean and 
turbulent kinetic energy, concluding that particles extract energy from the mean flow 
and redistribute it to turbulent kinetic energy throughout the channel.

The influence of particles on turbulence depends on both the particle Stokes number
$St^+$ and the mass loading $\varphi_m$, with the latter being more pivotal in comparison
~\citep{vreman2007turbulence,richter2013momentum,muramulla2020disruption}.
\citet{zhao2010turbulence} performed direct numerical simulation (DNS) for a turbulent channel flow
laden with spherical particles with $St^+=30$ and mass loading of $1.0$, observing drag reduction and 
turbulence attenuation compared to regular turbulent flows without particles.
In the experimental study of \citet{kulick1994particle}, it is found that 
particles generally attenuate turbulence, particularly affecting specific wavenumbers or scales 
of motion.
Turbulent suppression is more pronounced in channels than in isotropic turbulence.
\citet{li2001numerical} demonstrated that turbulent Reynolds stress exhibits increased anisotropy 
as mass loading rises, attributed to modifications in the velocity-pressure gradient term, 
specifically the pressure-strain term in the turbulent kinetic energy (TKE) budget. 
As a matter of fact, all the transport terms of turbulent kinetic energy and enstrophy transport
are reduced in the presence of particles~\citep{dritselis2016direct}.
Moreover, the turbulent modification is dependent on both the wall-normal location and scales.
Particles tend to diminish production terms across all scales and throughout the channel, 
with the particle feedback effect being comparatively minor~\citep{richter2015turbulence}.
Particle clustering, dependent on the Stokes number, acts as either an energy source or sink,
affecting turbulent motions at certain scales~\citep{richter2013momentum,richter2015turbulence}.
By conditional averaging the flow field surrounding the streamwise vortices,
\citet{dritselis2008numerical} observed that particle feedback forces generate a torque opposing 
the rotation of streamwise vortices, thereby reducing vorticity magnitude.
In terms of flow topology, \citet{mortimer2020density} noted that heavy particles lower
the possibility of finding particles in unstable focus/compressing regions and
stable focus/stretching regions across the channel,
corresponding to the streamwise vortices.

The modulation of turbulence by the Stokes number $St^+$ is primarily influenced by changes 
in preferential concentration, which impacts the spatial distribution and magnitude of 
the feedback force. 
\citet{wu2022effect} investigated the effects of $St^+$ at a mass loading of 0.2 and 
observed that turbulent production and viscous dissipation exhibit non-monotonic variations, 
with the most significant suppression occurring for particles with $St^+=40$. 
This suppression is achieved by accelerating the low-speed and decelerating 
the high-speed fluid~\citep{lee2015modification}.
Additionally, the size of the streamwise vortices increases with longer particle response times,
while the intensities of velocity and vorticity fluctuations are diminished,
yet show an increasing trend with the Stokes number~\citep{dritselis2011numerical}.
Conversely, particles with $St^+=0.5$ enhance the instability of streaks and more effectively 
generate quasi-streamwise vortices~\citep{lee2015modification}.
\citet{richter2014modification} found that in turbulent Couette flows, 
the maximum feedback force is realized at a Stokes number of $O(1)$, under the Kolmogorov scale. 
\citet{wang2019modulation} demonstrated that high-inertia particles reduce the stretching 
and lift-up effects in the regeneration cycle of near-wall turbulence,
thereby suppressing the formation of streaks and vortices.
Furthermore, the two-way coupling mechanism decreases particle velocity fluctuations 
and near-wall particle concentration, with the extent of these reductions increasing 
with mass loading~\citep{nasr2009dns}.

There exists a critical mass loading threshold beyond which turbulence is entirely disrupted. 
At low mass fractions, particles suppress vortical motions, thereby diminishing sweeping 
and ejection events and reducing Reynolds stresses, which are nearly offset by 
the momentum imparted by the particles~\citep{richter2013momentum}.
This effect is stronger with the increasing mass fractions $\varphi_m$~\citep{zhou2020non}.
However, the intensity of streamwise velocity fluctuations varies non-monotonically 
with mass loading, first decreasing and then increasing, with the critical variation caused by
the direction in which the turbulent kinetic energy is transferred~\citep{zhou2020non}.
\citet{vreman2007turbulence} extensively discussed the impacts of particle mass loading,
noting that at low mass loadings, the dissipation due to particle forces is minor compared 
to other turbulent kinetic energy transfer mechanisms. 
Conversely, at moderate to high mass loadings, where turbulent intensities are lower, 
turbulent motions are predominantly induced by particle dynamics rather than conventional 
turbulent mechanisms. 
This leads to a gradual weakening of coherent structures, accompanied by more uniform 
particle distributions in turbulent pipe flows, both radially and along wall-parallel planes.
\citet{muramulla2020disruption} further showed that the critical values for laminarization 
depend on the Stokes number but are less influenced by the choice of force model.
Beyond this critical value, the abrupt decreasing production term, rather than
the particle drag dissipation, is primarily responsible for flow relaminarization.

The aforementioned conclusions pertain solely to incompressible turbulent flows. 
In the realm of compressible flows, such research is relatively sparse. 
In compressible homogeneous isotropic turbulence, it has been observed that 
particles not only accumulate in regions of high shear but also behind shocklets, 
thereby diminishing both rotational and dilatational motions
~\citep{yang2014interactions,zhang2016preferential}.
Fluctuations are mitigated at low wavenumbers while being enhanced at high wavenumbers
\citep{zhang2016preferential}.
Thermodynamic flow properties are reduced, yet the mean temperature and pressure are elevated, 
effects that intensify with the Stokes number, attributed to increased internal energy production
~\citep{dai2017direct}.
In compressible homogeneous shear turbulence, larger particles diminish cross-stream 
velocity fluctuations, curtail the growth rate of the mixing layer, 
and weaken both vortices and shocklets. 
~\citep{dai2018direct,dai2019direct1,dai2019direct}.
\citet{chen2022two} examined particle-laden compressible turbulent boundary layers 
involving combustion, finding that large and heavy particles suppress streamwise vortices, 
leading to a homogeneous particle distribution, yet these particles still predominantly 
concentrate in strain-dominant regions. 
The fluid temperature and wall heat flux are substantially decreased due to 
particle heat transfer.
Additionally, studies focusing solely on the one-way coupling in particle-laden 
compressible wall-bounded turbulence have revealed no significant differences from 
incompressible wall-bounded turbulence in terms of particle statistics and 
preferential accumulation~\citep{xiao2020eulerian,yu2024transport,wang2024turbophoresis}.

The preliminary literature review indicates that while particle-laden flows have been 
extensively studied in the context of incompressible wall-bounded turbulence, 
there is a notable gap in understanding when it comes to compressible flows. 
This gap serves as the motivation for the current study. 
In this paper, we conduct direct numerical simulations of particle-laden compressible 
turbulent boundary layer flows at a free-stream Mach number of $2.0$, 
utilizing the Eulerian-Lagrangian point-particle method. Our analysis is confined to 
considering the feedback force exerted by the particles on the fluid, 
deliberately omitting the heat transfer between them. 
The primary focus is on the modulation of momentum and kinetic energy transfer by the particles, 
the variations in turbulent statistics influenced by the particle Stokes number $St^+$ and 
mass loading, and the impact of particle feedback forces on turbulent motions across 
different scales. 
The findings presented here are intended to enhance the accuracy of modeling and 
predicting such flows.

The remainder of this paper is organized as follows.
In section~\ref{sec:num} we introduce the numerical methods adopted in conducting the two-way
coupling of particle-laden compressible turbulent boundary layers and the flow parameters.
In section~\ref{sec:results} we discuss in detail the modulations of particles on
the instantaneous flow fields, the mean velocity and Reynolds stress, momentum balance and 
kinetic energy transfer and the mean temperature of the fluid.
The conclusions are given in section~\ref{sec:con}.

\section{Numerical Method} \label{sec:num}

We consider the compressible turbulent boundary layers carrying dilute phase small spherical 
particles with negligible volume fraction but finite mass loading.
The Eulerian-Lagrangian point-particle approach is still valid for simulating such flows.
Both the forces acting on the particles and their feedback to the fluid should be considered,
constituting a two-way coupling between them.
The compressible turbulent boundary layer flows are governed by the three-dimensional
Navier-Stokes equation for perfect Newtonian gases,
\begin{equation}
\frac{\partial \rho}{\partial t} +\frac{\partial (\rho u_i)}{\partial x_i} = 0,
\end{equation}
\begin{equation}
\frac{\partial (\rho u_i)}{\partial t} +\frac{\partial (\rho u_i u_j)}{\partial x_j} = 
-\frac{\partial p}{\partial x_i} + \frac{\partial \tau_{ij}}{\partial x_j} + F_{p,i},
\end{equation}
\begin{equation}
\frac{\partial (\rho E)}{\partial t} +\frac{\partial (\rho E u_j)}{\partial x_j} = 
-\frac{\partial (p u_j)}{\partial x_j} + \frac{\partial (\tau_{ij} u_i)}{\partial x_j}
-\frac{\partial q_j}{\partial x_j} + F_{p,i} u_i.
\end{equation}
Here, $\rho$ is the fluid density,
and $u_i$ ($i=1$,2,3) is the velocity component of the fluid in the $x_i$ direction,
(also $x$, $y$ and $z$, corresponding to the streamwise, wall-normal and spanwise directions).
The $p$ is pressure and $E$ is total energy, following the state equations of the perfect gases
\begin{equation}
p=\rho R T,~~ E=C_v T + \frac{1}{2} u_i u_i,
\end{equation}
with $T$ being the temperature, $R$ being the gas constant and $C_v$ being the constant 
volume specific heat.
The $\tau_{ij}$ is the viscous stress tensor and $q_j$ is the molecular heat conductivity,
associated with the velocity and temperature gradients as follows
\begin{equation}
\tau_{ij}= \mu \left( \frac{\partial u_i}{\partial x_j}+\frac{\partial u_j}{\partial x_i} \right)
-\frac{2}{3} \mu \frac{\partial u_k}{\partial x_k} \delta_{ij},~~
q_j = - \kappa \frac{\partial T}{\partial x_j}
\end{equation}
in which $\mu$ is the fluid viscosity, determined by Sutherland's law,
and $\kappa= C_p \mu /Pr$ is the molecular heat conductivity,
with $C_p$ the constant pressure specific heat and $Pr=0.71$ the molecular Prandtl number.
$F_{p,i}$ is the feedback force from the particles to the fluid, which will be introduced 
subsequently.

The boundary conditions are specified as follows.
At the lower wall, no-slip and no-penetration conditions are given for velocity and 
the isothermal condition is given for temperature.
The outflow and no-reflecting conditions are adopted at the upper and streamwise outlet boundaries.
The periodic conditions are used in the spanwise direction.
The turbulent inflow is given at the inlet boundary, composed of the mean velocity profiles
given according to the empirical formula by~\citet{musker1979explicit}, 
the mean temperature profiles obtained by the generalized Reynolds analogy
~\citep{zhang2014generalized}, and the velocity and temperature fluctuations 
by digital filtering method~\citep{klein2003digital}.

The governing equations are solved directly by the finite difference method using the open-source
code developed by \citet{bernardini2021streams}.
The convective terms are approximated by the sixth-order kinetic energy preserving scheme
~\citep{pirozzoli2010generalized} and the viscous terms are expanded into Laplacian forms and 
approximated by the sixth-order central scheme~\citep{pirozzoli2011numerical}.
Time advancement is achieved by the third-order low-storage Runge-Kutta scheme
~\citep{wray1990minimal}.

The motions of the particles are solved by the following equations
\begin{equation}
\frac{{\rm d} r_{p,i}}{{\rm d}t} = v_i, ~~
\frac{{\rm d} v_{i}}{{\rm d}t} = \frac{f_{D}}{\tau_p}(u_i-v_i).
\end{equation}
Here, $r_{p,i}$ and $v_i$ are the position and velocity of the particles, and $\tau_p$ is the
particle relaxation time, defined as $\tau_p = \rho_p d^2_p/(18 \mu)$, with
$\rho_p$ the particle density and $d_p$ the particle diameter.
The coefficient $f_D$ incorporates variation of the drag force with the particle Reynolds number
$Re_p = \rho |u_i - v_i| d_p/\mu$ and the particle Mach number 
$M_p = |u_i - v_i|/\sqrt{\gamma R T}$ following the suggestion of~\citet{loth2021supersonic} 
writes as
\begin{equation}
f_D = (1+0.15 Re^{0.687}_p) H_M,
\label{eqn:cd}
\end{equation}
\begin{equation}
\begin{aligned}
H_M =
\begin{cases}
      0.0239 M^3_p + 0.212 M^2_p - 0.074 M_p + 1,~~~ & M_p \le 1 \\
      0.93 + \frac{1}{3.5 + M^5_p}, ~~~ & M_p >1   
\end{cases}
\end{aligned}
\label{eqn:hm}
\end{equation}
In our previous study regarding the one-way coupling between the fluid and particles
~\citep{yu2024transport},
we have demonstrated that both the particle Knudsen numbers and particle Reynolds numbers are low,
so the terms associated with the rarefaction effects and high Reynolds number effects 
that incorporate fully turbulent wakes can be neglected.
The time advancement in numerically solving the position and velocity of the particles is
achieved using the same third-order Runge-Kutta scheme as the fluid flow.
The information of the fluid at the positions of the particles is obtained using trilinear
interpolation, which has been proven to be sufficiently accurate in direct numerical simulations
~\citep{bernardini2014reynolds,xiao2020eulerian}.
The spanwise periodic condition is also adopted for particles.
When the particles go through the streamwise outlet and upper boundaries,
they are recycled to the turbulent inlet at a random position within the boundary layer thickness.

The feedback force is obtained using the particle-in-cell (PIC) approximation
~\citep{dritselis2008numerical,zhao2013interphasial}, 
summing the forces of the particles within a mesh cell 
\begin{equation}
F_{p,i} = -\frac{m_p}{V_{cell}} \sum^{n_p}_{l=1} \frac{f_{D}}{\tau_p}(u_i-v_i)
\end{equation}
with $m_p$ the mass of the particle, then interpolating onto the grid points using the trilinear
interpolation.
The heat transfer between the particles and fluid is neglected here.
In other words, the particles are considered to be in a thermal equilibrium state, changing their
temperature instantaneously with the ambient fluid.
The effects of the particles absorbing and releasing heat from and to the fluid will be considered
in our future work.
The DNS solver is validated by comparing the statistics of a two-way coupling turbulent channel
flow simulations at low Mach numbers with those of incompressible flows, 
which is given in Appendix~\ref{sec:val}.

Other notations are listed as follows.
The subscript $0$ refers to the flow quantities in the free-stream and $w$ to those at the wall.
The ensemble average of a flow quantity $\phi$ is represented by $\bar \phi$ and the corresponding
fluctuations as $\phi'$.
The density-weighted average is denoted by $\tilde \phi$ and the corresponding fluctuations
as $\phi''$.
The superscript $+$ represents the normalization by the viscous scales, defined by the
wall shear stress $\tau_w = (\mu_w \partial \bar u/\partial y)|_w$, wall mean density
$\bar \rho_w$ and viscosity $\mu_w$, which further gives the friction velocity $u_\tau = \sqrt{
\tau_w /\bar \rho_w}$, the friction length scale $\delta_\nu = \mu_w/(\bar \rho_w u_\tau)$ and
the friction Reynolds number $Re_\tau = \bar \rho_w u_\tau \delta/\mu_w$, with $\delta$
the boundary layer thickness.

We consider the compressible turbulent boundary layers at the free-stream Mach number
$M_0 = U_0 / a_0 = 2.0$ and the free-stream Reynolds number
$Re_0 = \rho_0 U_0 \delta_0/\mu_0 = 9919$, with $a_0$ the free-stream sound speed
and $\delta_0$ the nominal boundary layer thickness at the turbulent inlet.
The wall temperature $T_w$ is set as the recovery temperature $T_r=1.72 T_0$, defined as
$T_r = T_0(1+(\gamma-1) Pr^{1/2} M^2_0/2)$.
Correspondingly, the friction Reynolds number $Re_{\tau,0}$ at the turbulent inlet is 200.
The sizes of the computational domain in the streamwise, wall-normal and spanwise directions
are $L_x = 80 \delta_0$, $L_y=9 \delta_0$ and $L_z = 8 \delta_0$,
discretized by 2400, 280 and 280 grids, respectively,
with the mesh sizes of $\Delta x^+ \approx 6.67$, $\Delta y^+_w=0.5$ and $\Delta z^+ \approx 5.71$
under viscous scales at the turbulent inlet.
According to a preliminary examination of the spatial development features, we found that
the turbulence is statistically equilibrium downstream of $x=50\delta_0$ after the simulations have
been run for $1000 \delta_0/U_0$.
Henceforth, the turbulent statistics are given within the streamwise range $x=(60\sim 70)\delta_0$
and a time period of $200 \delta_0/U_0$.

\begin{table}[tbp!]
\centering
\caption{Computational parameters. Here, $\eta_w$ is the Kolmogorov length scale at the wall,
$N_p$ is the particle number and $\varphi_m$ is the particle mass loading.}
\begin{tabular*}{\textwidth}{@{\extracolsep{\fill}}ccccccccc}
\hline
Case & $Re_\tau$ & $\rho_p/\rho_0$ & $d_p/\delta_0$ & $d_p/\eta_w$ & $St^+$ & $N_p$ & $\varphi_m$ &
Line type \\
\hline
P0-F00 & 353 & 8000 & 0.0025 & 0.318 & 162 & 1,000,000 & - & 
\tikz[baseline]{\draw[black,solid,thick](0,0.5ex) --++ (1.5,0);} \\
P1-F02 & 338 & 8000 & 0.002 & 0.208 & 103 & 6,000,000 & 0.17 & 
\tikz[baseline]{\draw[cyan!80!black,dash pattern={on 7pt off 2pt},thick](0,0.5ex) --++ (1.5,0);}\\
P1-F06 & 296 & 8000 & 0.002 & 0.214 & 79  & 20,000,000 & 0.57 & 
\tikz[baseline]{\draw[cyan!80!black,dash pattern={on 7pt off 2pt},thick](0,0.5ex) --++ (1.5,0);
\protect\draw[-,draw=cyan!80!black, fill=cyan!80!black,line width=1.0pt](7.0mm,0.75mm) -- (8.25mm,2mm) -- (9.5mm,0.75mm) -- (8.25mm, -0.5mm) -- (7.0mm, 0.75mm);}\\
P2-F02 & 358 & 8000 & 0.003 & 0.351 & 230 & 1,800,000 & 0.17 &
\tikz[baseline]{\draw[green!80!black,dash pattern={on 7pt off 2pt on 1pt off 2pt},thick](0,0.5ex) --++ (1.5,0);}\\
P2-F06 & 337 & 8000 & 0.003 & 0.280 & 225 & 6,000,000 & 0.57 &
\tikz[baseline]{\protect\draw[green!80!black,dash pattern={on 7pt off 2pt on 1pt off 2pt},thick](0,0.5ex) --++ (1.5,0); 
\protect\draw[-,draw=green!80!black, fill=green!80!black,line width=1.0pt](7.0mm,-0.5mm) -- (7.0mm,2mm) -- (9.5mm,2mm) -- (9.5mm, -0.5mm) -- (7.0mm, -0.5mm);}\\
P3-F02 & 356 & 8000 & 0.004 & 0.476 & 410 & 750,000   & 0.17 &
\tikz[baseline]{\draw[red,dash pattern={on 7pt off 2pt on 1pt off 2pt on 1pt off 2pt},thick](0,0.5ex) --++ (1.5,0);}\\
P3-F14 & 301 & 8000 & 0.004 & 0.276 & 349 & 6,000,000 & 1.36 &
\tikz[baseline]{\draw[red,dash pattern={on 7pt off 2pt on 1pt off 2pt on 1pt off 2pt},thick](0,0.5ex) --++ (1.5,0); 
\protect\draw[-,draw=red, fill=red,line width=1.0pt](7.0mm,-0.5mm) -- (7.0mm,2mm) -- (9.5mm,1mm) -- (7.0mm, -0.5mm);}\\
 \hline
\end{tabular*}
\label{tab:param}
\end{table}

The particle parameters are listed in Table~\ref{tab:param}.
We consider a one-way coupling case P0-F00 without feedback force from the particle to the fluid, 
and six two-way coupling cases with different particle populations and mass loadings.
The particle density is set to be identical as $\rho_p = 8000 \rho_0$ and the particle diameters
as $d_p = 0.002 \delta_0$, $0.003 \delta_0$ and $0.004 \delta_0$, 
hereinafter referred to as particle populations P1, P2 and P3, respectively.
The particle Stokes numbers under viscous scales $St^+ = \tau_p u_\tau/\delta_\nu$
are different, due to the disparity in their diameters.
For particle P1, the number of particles $N_p$ in case P1-F02 is set to be $6,000,000$, 
resulting in the mass loading, 
hereinafter defined as $\varphi_m = m_p N_p / (\rho_0 L_x L_z \delta)$ ($\delta$ the boundary layer
thickness within $x=(60\sim 80)\delta_0$), being approximately $0.17$.
To retain the mass loading $\varphi_m$, we adjust the particle number for populations P2 and P3
in cases P2-F02 and P3-F02.
We also design two simulations with the same particle numbers $N_p$ as in case P1-F02,
thus giving higher mass loading of $\varphi_m=0.57$ for case P2-F06 and 
$\varphi_m=1.36$ for case P3-F14, and another case P1-F06 to compare with case P2-F06 for the Stokes
number effects.
As reported in Table~\ref{tab:param}, the friction Reynolds numbers $Re_\tau$ are only 
slightly modified, except for case P3-F14 where it is reduced by approximately $15\%$.

\section{Results and discussion} \label{sec:results}

\subsection{Instantaneous distributions} \label{subsec:inst}

\begin{figure}[tb!]
\centering
\begin{overpic}[width=0.8\textwidth,trim={0.1cm 0.1cm 0.1cm 0.1cm},clip]{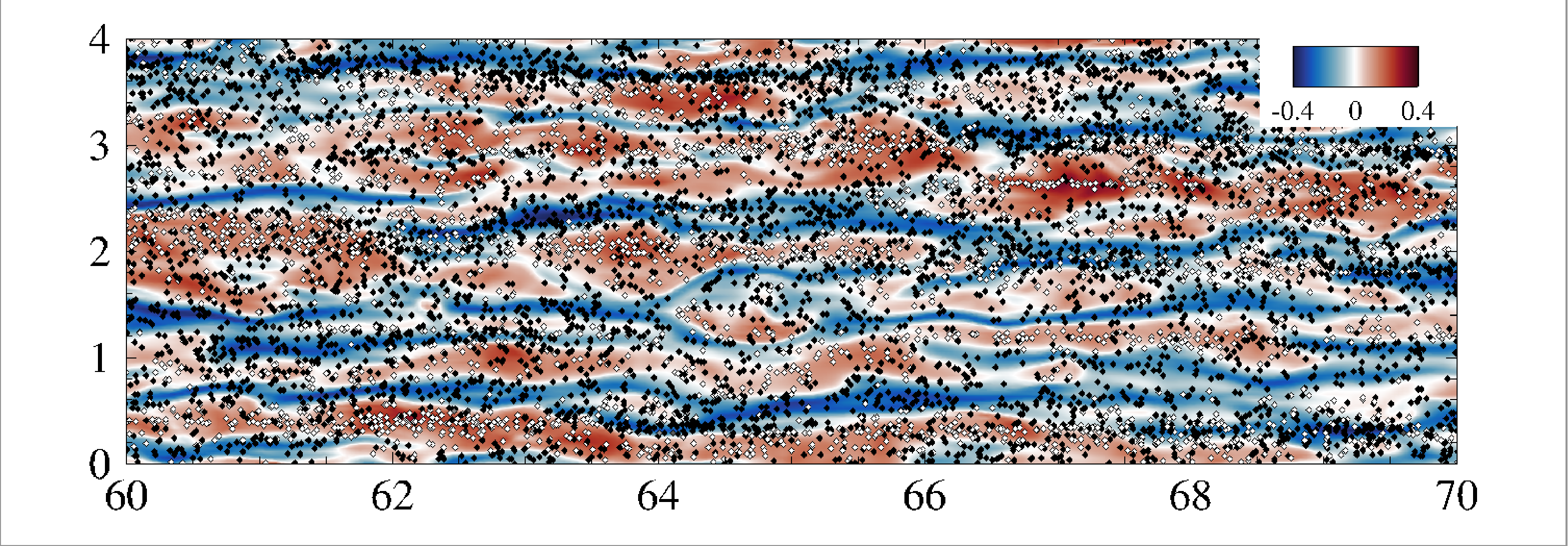}
\put(0,32){(a)}
\put(50,0){$x/\delta_0$}
\put(0,17){\rotatebox{90}{$z/\delta_0$}}
\end{overpic}\\[2.0ex]
\begin{overpic}[width=0.8\textwidth,trim={0.1cm 0.1cm 0.1cm 0.1cm},clip]{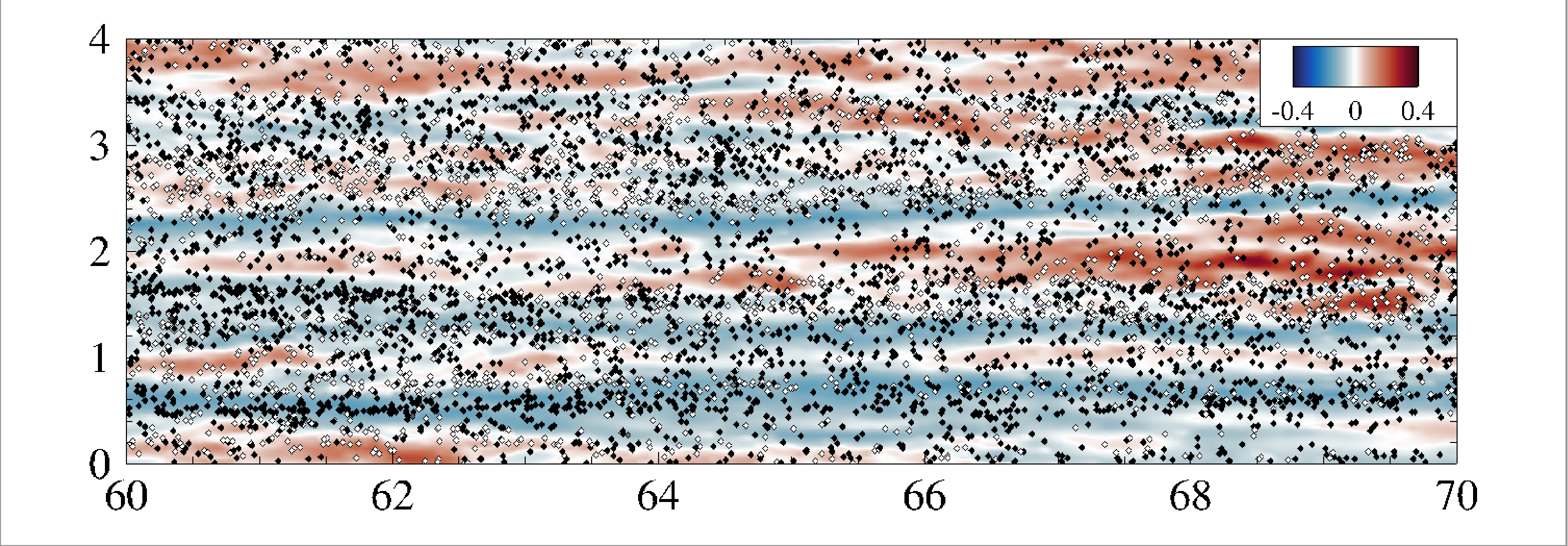}
\put(0,32){(b)}
\put(50,0){$x/\delta_0$}
\put(0,17){\rotatebox{90}{$z/\delta_0$}}
\end{overpic}\\[2.0ex]
\begin{overpic}[width=0.8\textwidth,trim={0.1cm 0.1cm 0.1cm 0.1cm},clip]{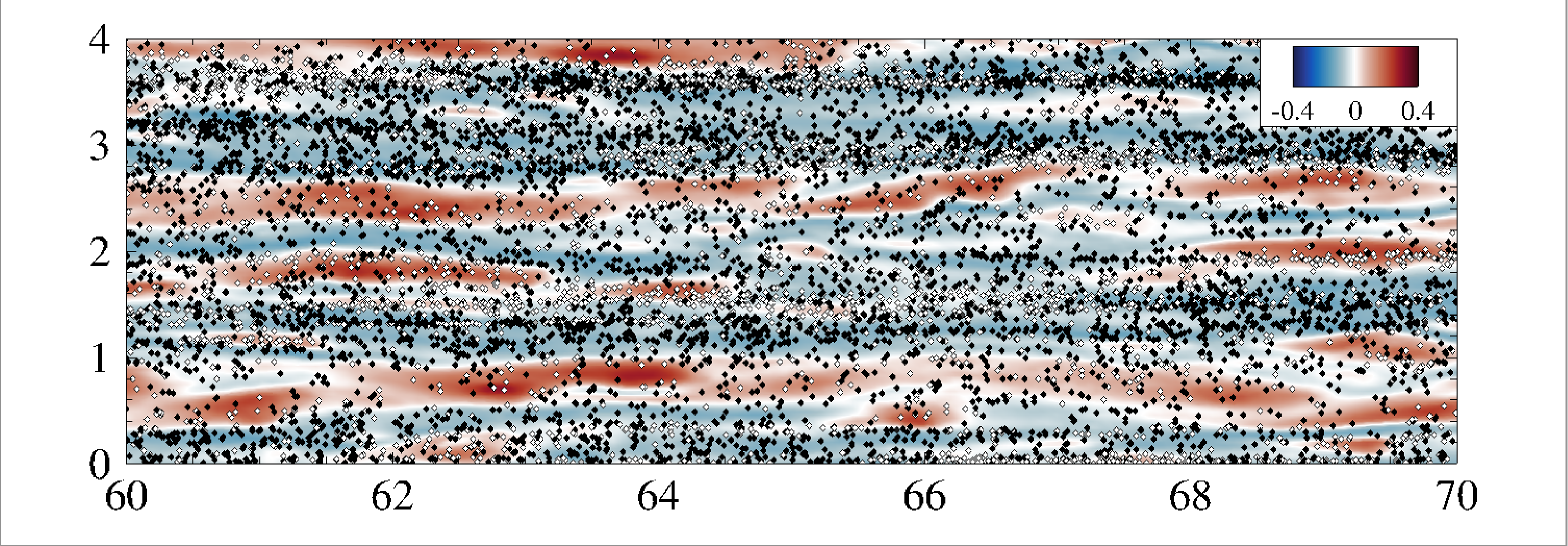}
\put(0,32){(c)}
\put(50,0){$x/\delta_0$}
\put(0,17){\rotatebox{90}{$z/\delta_0$}}
\end{overpic}\\[2.0ex]
\begin{overpic}[width=0.8\textwidth,trim={0.1cm 0.1cm 0.1cm 0.1cm},clip]{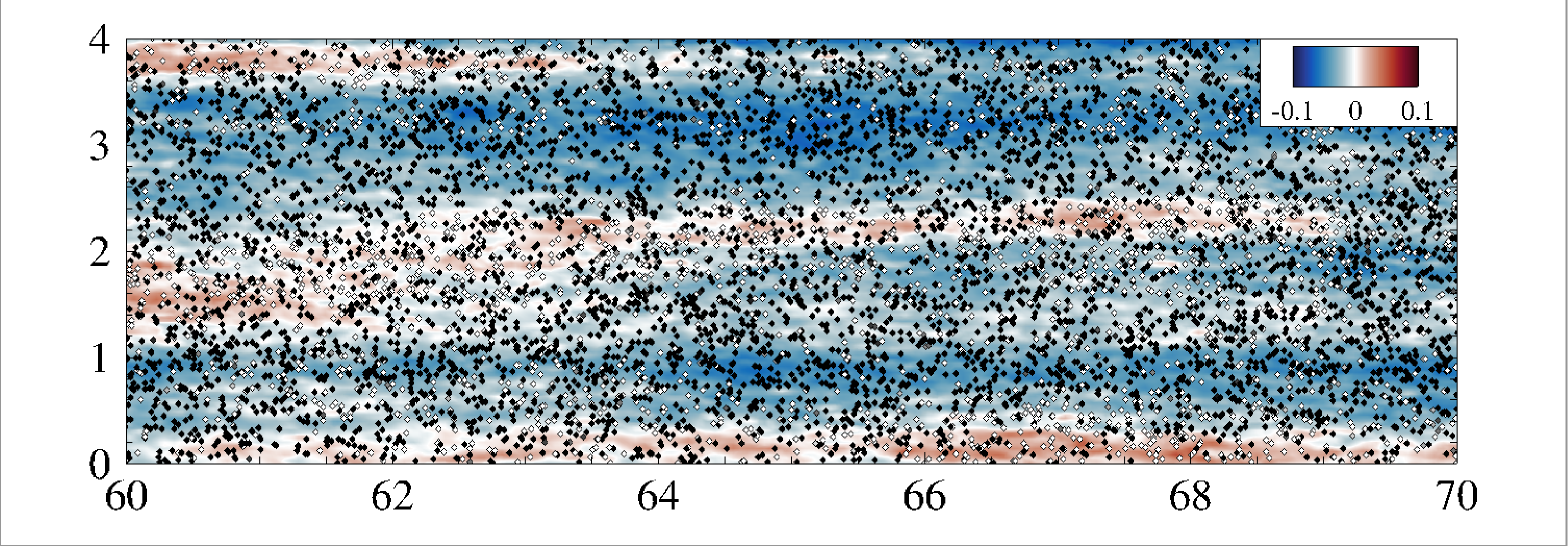}
\put(0,32){(d)}
\put(50,0){$x/\delta_0$}
\put(0,17){\rotatebox{90}{$z/\delta_0$}}
\end{overpic}\\
\caption{Instantaneous distributions of velocity fluctuations at $y^+=8$ 
and particles within $y^+=3 \sim 15$, (a) case P0-F00, (b) case P1-F06, (c) case P2-F06, 
(d) case P3-F14. White particles: $a_1>0$, black particles: $a_1<0$.}
\label{fig:instxz}
\end{figure}
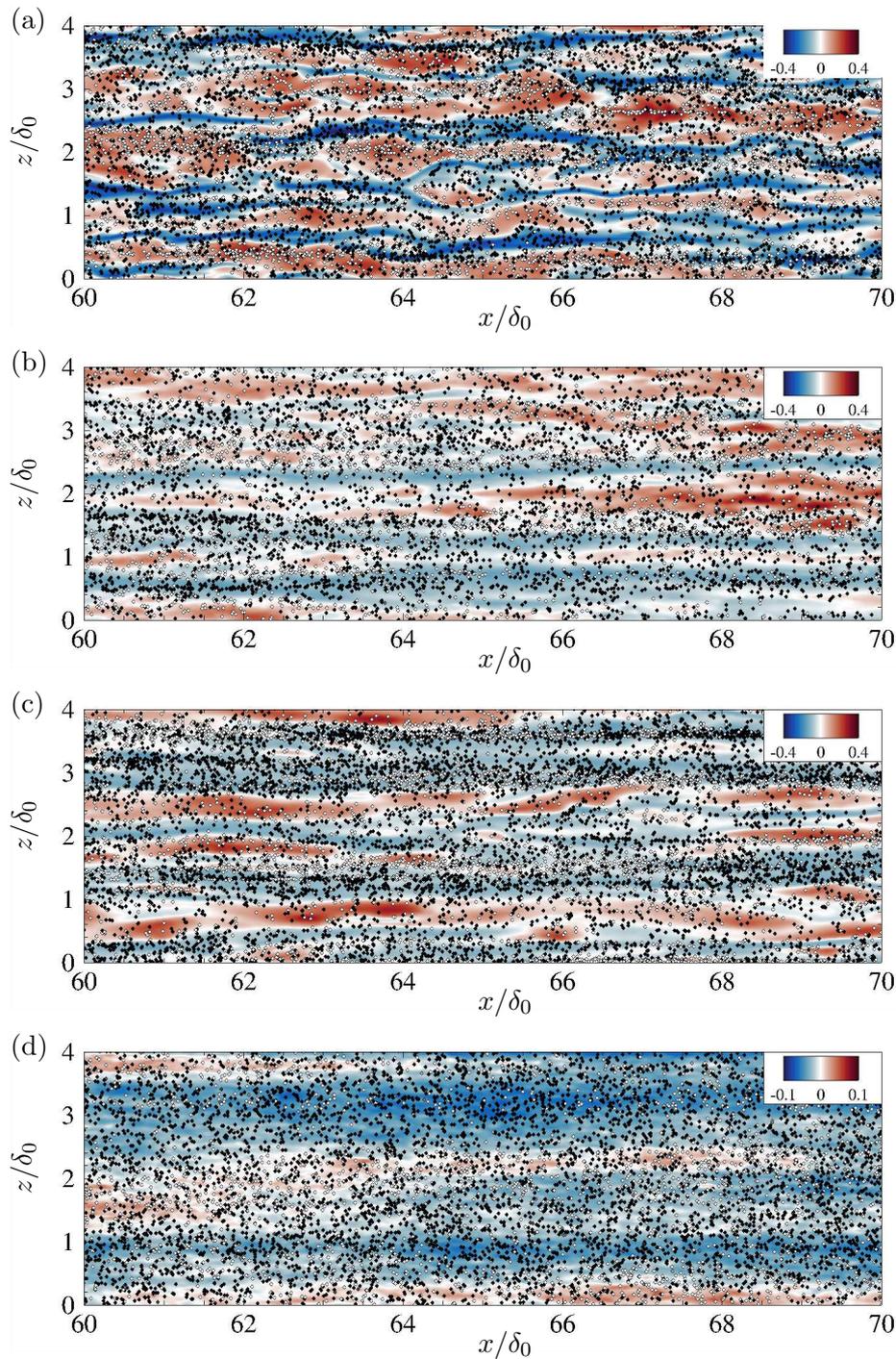

In figure~\ref{fig:instxz} we present the instantaneous distributions of 
the streamwise velocity fluctuations in the wall-parallel plane in the buffer region at $y^+=8$ 
and particles within the wall-normal range of $y^+=3 \sim 15$ for cases P0-F00,
P1-F06, P2-F06 and P3-F14.
In the one-way coupling case P0-F00, the negative regions of the velocity fluctuations are
organized as the streamwise elongated streaky structures, namely the low-speed streaks, whereas
the high-speed regions are comparatively shorter in the streamwise direction and wider in the
spanwise direction. The particles clustered within the low-speed regions manifest similar
streaky structures as the velocity, forming the streamwise-elongated structures
~\citep{sardina2012wall}.
In such regions, the particles are mostly being decelerated.
In the high-speed regions, on the other hand, the particles are more frequently accelerated.
Similar phenomena can be observed in cases P1-F02, P2-F02 and P3-F02 (not shown here for brevity),
except for the different particle numbers near the wall due to the disparity in
the particle Stokes number.
In cases P1-F06 and P2-F06 with a moderate mass loading, 
the velocity fluctuations are slightly weakened.
In these cases, the accelerated particles and decelerated particles tend to be mixing,
showing less tendency to congregate within the low- and high-speed regions respectively
~\citep{vreman2007turbulence,nasr2009dns}.
This is consistent with the findings of the previous studies that the near-wall accumulation
is weakened for higher $St^+$ particles~\citep{lee2015modification}.
In case P3-F14 with the highest mass loading, the flow is almost completely laminarized,
displaying only very weak velocity fluctuations. The corresponding flow structures are wider and
less meandering. It seems that the near-wall velocity fluctuations are induced by
the particles swept towards and reflected from the wall, instead of retaining their near-wall
self-sustaining cycles in canonical wall-bounded turbulence, as pointed out by
~\citet{vreman2007turbulence}.
This will be further demonstrated in the subsequent discussions.

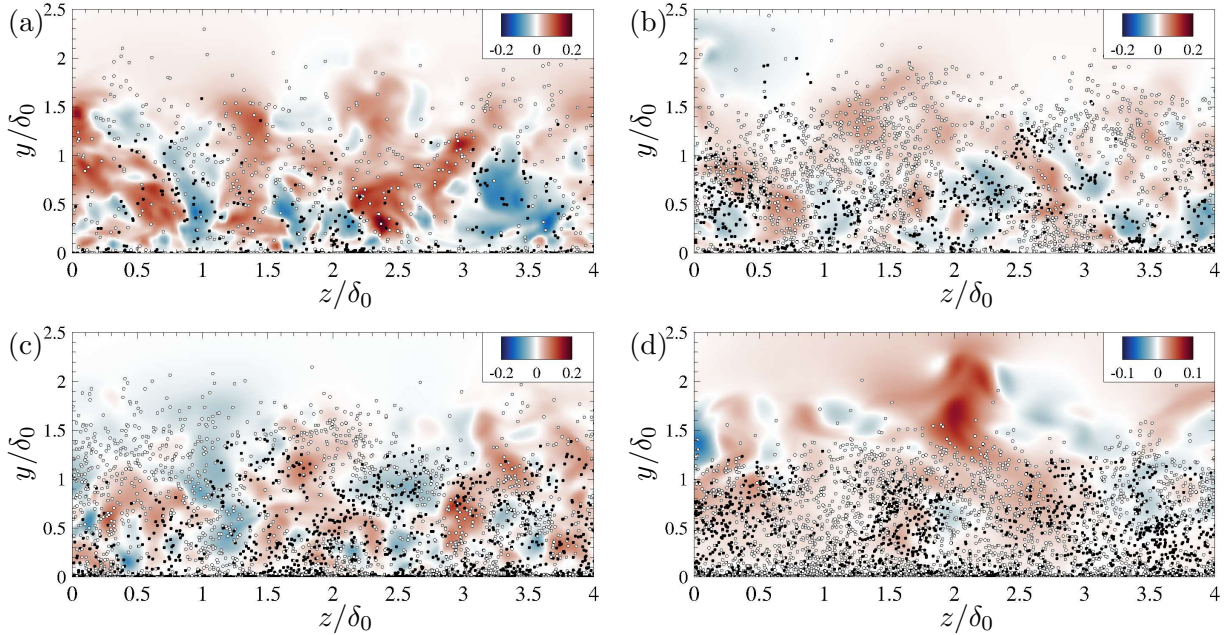
\begin{figure}[tb!]
\centering
\begin{overpic}[width=0.5\textwidth,trim={0.2cm 0.2cm 0.2cm 0.2cm},clip]{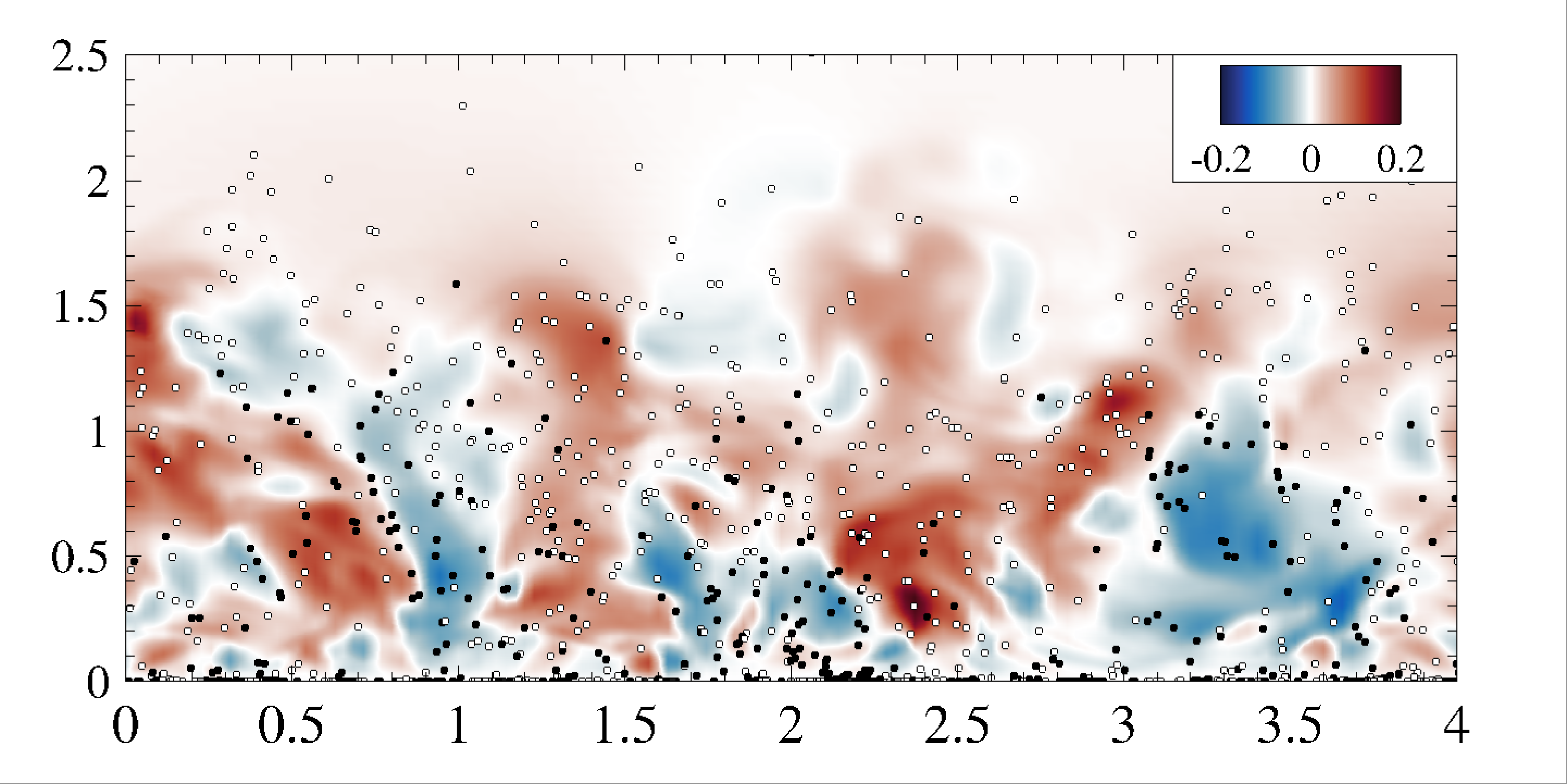}
\put(-3,43){(a)}
\put(48,-2){$z/\delta_0$}
\put(-3,22){\rotatebox{90}{$y/\delta_0$}}
\end{overpic}~
\begin{overpic}[width=0.5\textwidth,trim={0.2cm 0.2cm 0.2cm 0.2cm},clip]{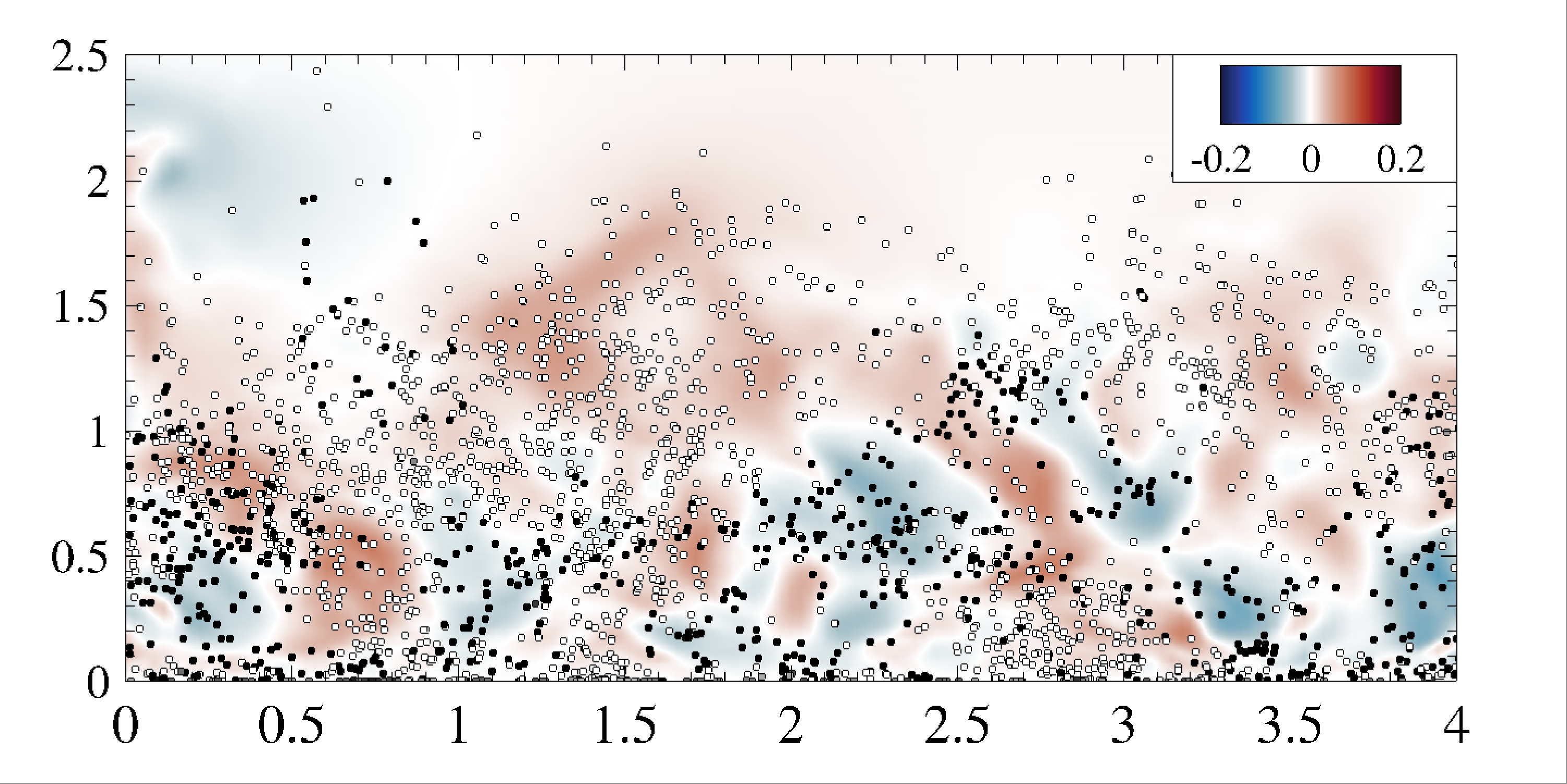}
\put(-3,43){(b)}
\put(48,-2){$z/\delta_0$}
\put(-3,22){\rotatebox{90}{$y/\delta_0$}}
\end{overpic}\\[2.0ex]
\begin{overpic}[width=0.5\textwidth,trim={0.2cm 0.2cm 0.2cm 0.2cm},clip]{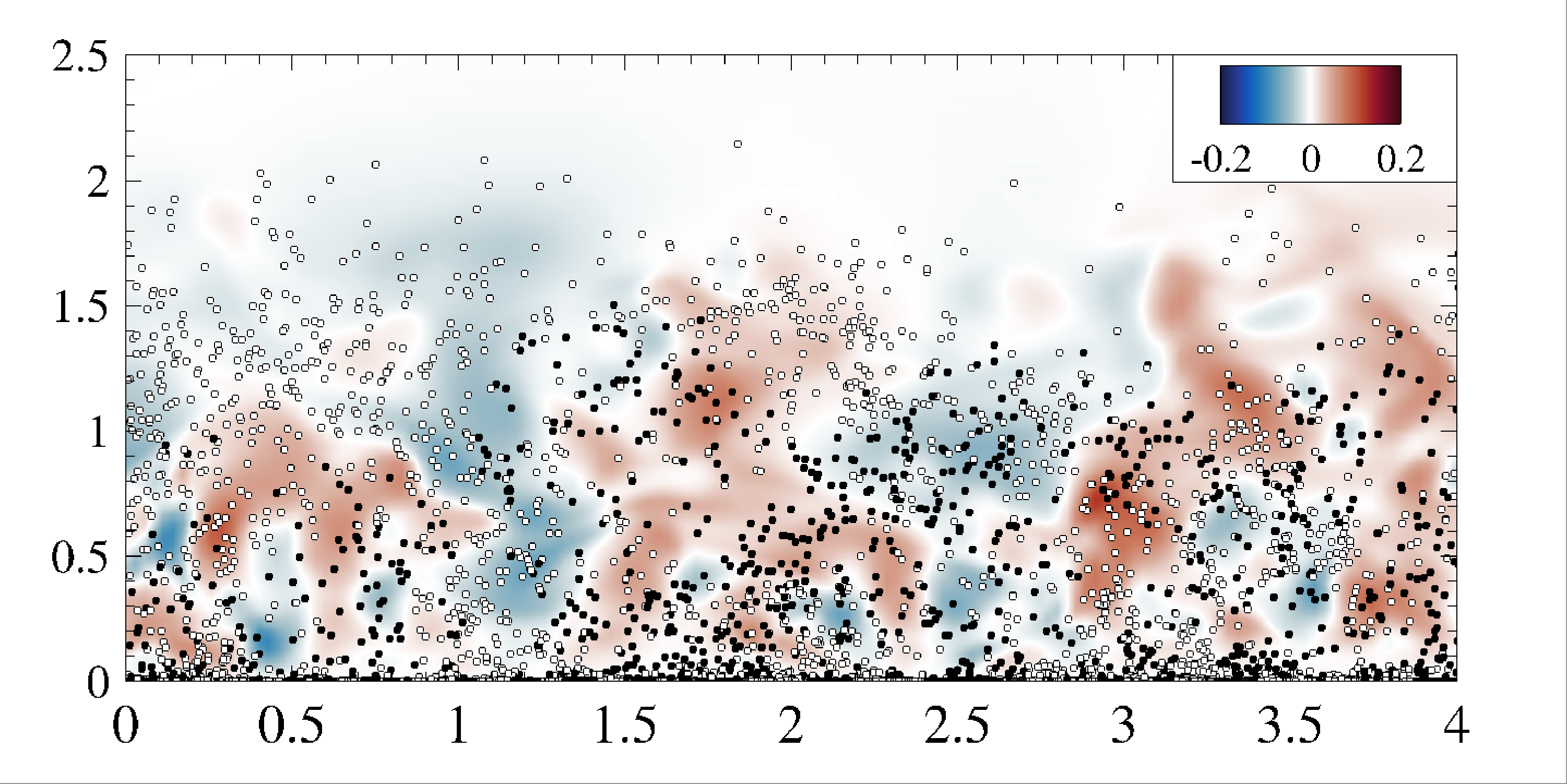}
\put(-3,43){(c)}
\put(48,-2){$z/\delta_0$}
\put(-3,22){\rotatebox{90}{$y/\delta_0$}}
\end{overpic}~
\begin{overpic}[width=0.5\textwidth,trim={0.2cm 0.2cm 0.2cm 0.2cm},clip]{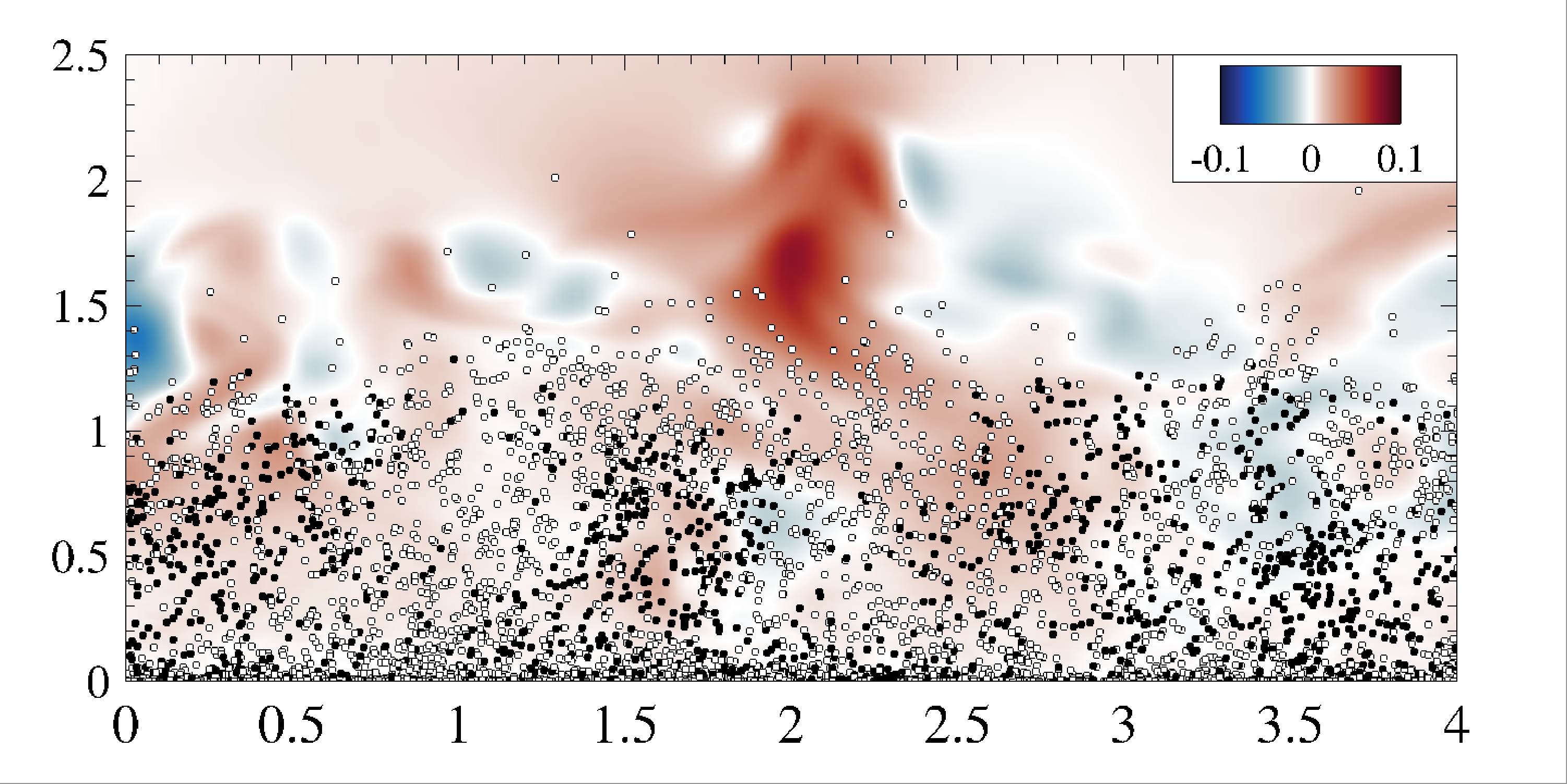}
\put(-3,43){(d)}
\put(48,-2){$z/\delta_0$}
\put(-3,22){\rotatebox{90}{$y/\delta_0$}}
\end{overpic}\\
\caption{Instantaneous distributions of velocity fluctuations $u''_1/U_\infty$ in the cross-stream plane 
$x=60\delta_0$ and particles within $x=(60\sim 60.02)\delta_0$, 
(a) case P0-F00, (b) case P1-F06, (c) case P2-F06, (d) case P3-F14.
White particles: $v_2>0$, black particles: $v_2<0$.}
\label{fig:instyz}
\end{figure}

The cross-stream distributions of the wall-normal velocity of fluid and particles are shown 
in figure~\ref{fig:instyz}.
The vertical motions of the particles are in general agreement with the trend of their
streamwise accelerations, in that they are accelerated when moving away from the wall ($v_2>0$)
and decelerated when moving towards the wall ($v_2<0$).
In the one-way coupling case P0-F00, the particles are accumulating close to the wall,
leaving only a small fraction in the outer region,
where the ascending and descending particles are located almost within the ejections and sweeping 
regions, respectively.
In cases P1-F06 and P2-F06, the degrees of the near-wall accumulation are mitigated, 
and the particles show a higher probability of moving in the opposite direction of
the vertical motions from the fluid.
The near-wall velocity fluctuation intensities are weaker, but not completely interrupted
by the presence of particles.
In case P3-F14, however, the particles are almost evenly distributed within the boundary layer.
The wall-normal velocity fluctuations almost disappear below $y=  0.3 \delta_0$, only showing
comparatively intense fluctuations near the outer edge of the boundary layer,
resembling the features of the turbulent mixing layers.

The observations of the instantaneous fields seem to indicate that the turbulent dynamics 
are strongly modulated by the particles, especially for the high mass loading case P3-F14.
In the following discussions, we will present in detail the variation of flow statistics related
to momentum and kinetic energy transport.

\subsection{Velocity statistics and mean momentum balance} \label{subsec:vel}

\begin{figure}[tb!]
\centering
\begin{overpic}[width=0.5\textwidth]{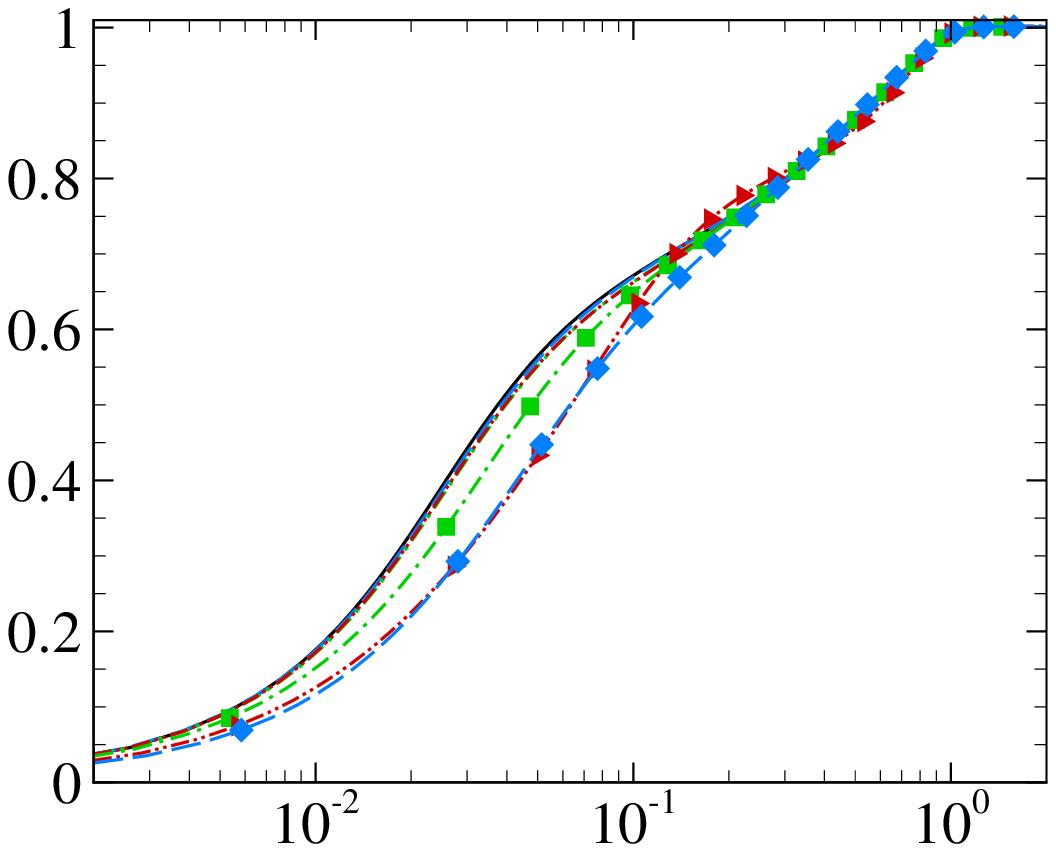}
\put(0,70){(a)}
\put(50,0){$y/\delta$}
\put(0,35){\rotatebox{90}{$\bar u_1/U_0$}}
\end{overpic}~
\begin{overpic}[width=0.5\textwidth]{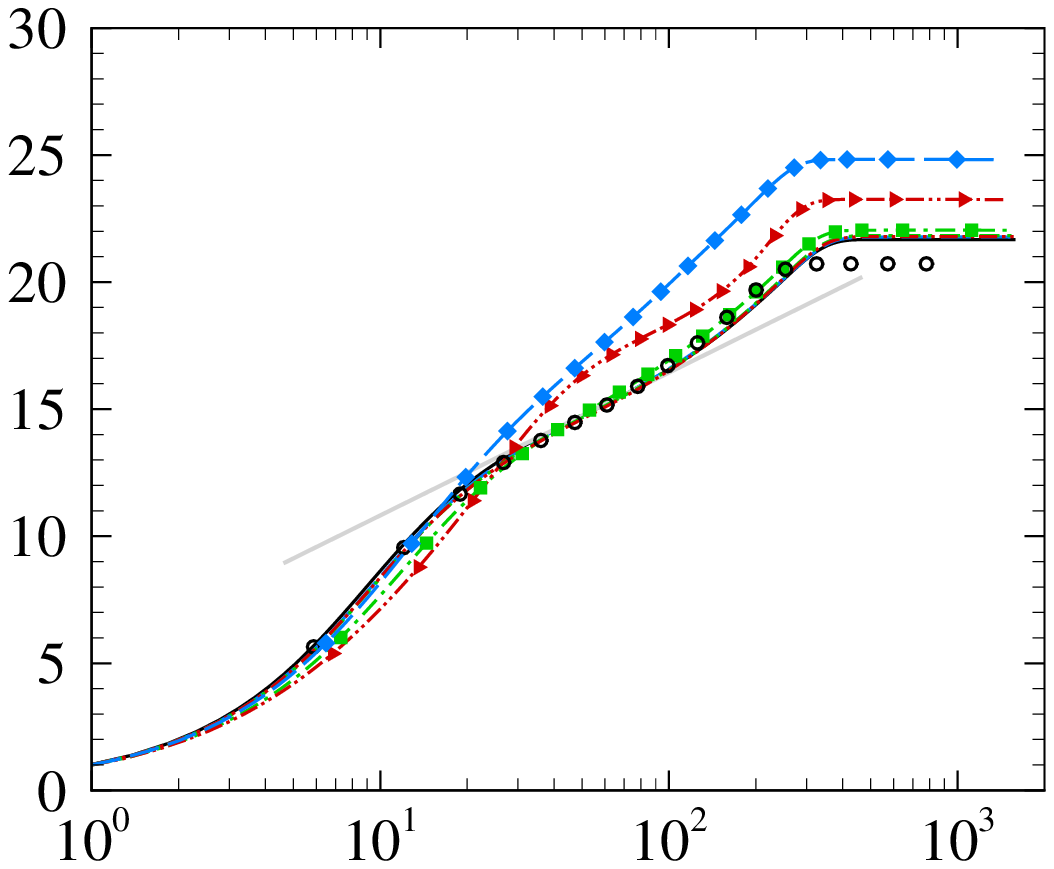}
\put(0,70){(b)}
\put(48,0){$y^+$}
\put(0,35){\rotatebox{90}{$u^+_{VD}$}}
\end{overpic}\\[2.0ex]
\begin{overpic}[width=0.5\textwidth]{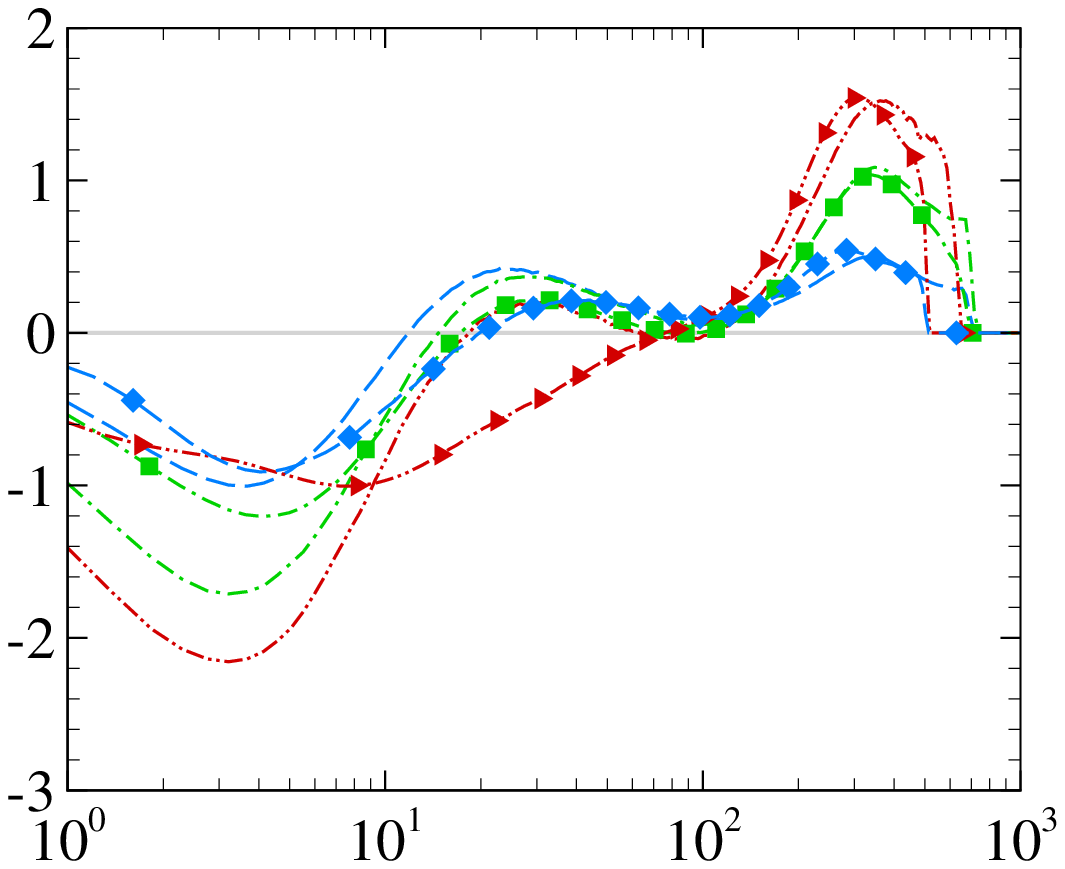}
\put(0,70){(c)}
\put(50,0){$y^+$}
\put(0,35){\rotatebox{90}{$\bar v^+_{1,s}$}}
\end{overpic}~
\begin{overpic}[width=0.5\textwidth]{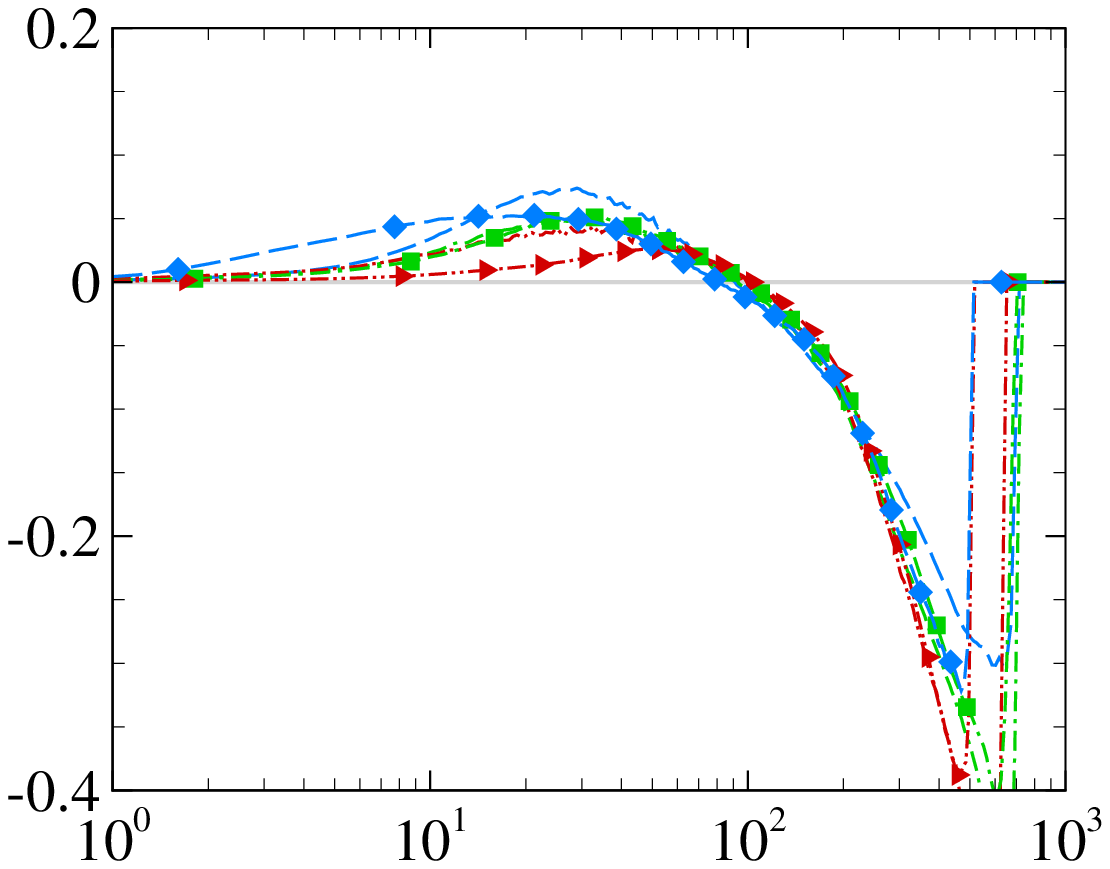}
\put(0,70){(d)}
\put(50,0){$y^+$}
\put(-1,35){\rotatebox{90}{$\bar v^+_{2,s}$}}
\end{overpic}
\caption{Wall-normal distributions of (a) mean velocity $\bar u_1/U_0$ against $y/\delta$,
(b) van Driest transformed mean velocity $u^+_{VD}$,
(c) streamwise mean slip velocity $\bar v_{1,s}/U_0$ and 
(d) wall-normal mean slip velocity $\bar v_{2,s}/U_0$.
Line legends refer to Table~\ref{tab:param}.
Black circles: data reported by \citet{pirozzoli2011turbulence} at $M_0=2$ and $Re_\tau=250$.}
\label{fig:uf}
\end{figure}

In figure~\ref{fig:uf}(a,b) we present the wall-normal distributions of the mean velocity
in outer scales and under van Driest transformation
\begin{equation}
u^+_{VD} = \int^{\bar u^+}_0 \sqrt{\frac{\bar \rho}{\bar \rho_w}} {\rm d} \bar u,
\end{equation}
along with the data reported by~\citet{pirozzoli2011turbulence} at $M_0=2$ and $Re_\tau \approx 250$.
Compared with the one-way coupling case P0-F00, the mean fluid velocity in cases P1-F02,
P2-F02 and P3-F01 with low mass loadings are not much affected, suggesting the weak influences of 
the particles in these cases.
For higher mass loadings cases P1-F06, P2-F06 and P3-F14, on the other hand, the mean velocities 
are significantly lower in the near-wall region but higher in the outer region,
corresponding to the reduction of the skin friction.
The particle population P1 with $St^+ \approx 100$ with the highest degree of
near-wall preferential accumulation seems to leave the strongest impact,
even though the mass loading in case P1-F06 is less a half of that in case P3-F14.
Similar phenomena have been found in the previous studies of turbulent flows
~\citep{mortimer2020density,richter2014modification,lee2015modification,muramulla2020disruption},
in which it is pointed out that both the Stokes number and the mass loadings are
responsible for turbulent modulation.
Despite the similarity of the near-wall mean velocity in cases P1-F06 and P3-F14,
the origin of turbulent motions, as we will demonstrate later, are completely different.

Due to the finite inertia of the particles, the mean particle velocities are different from
those of the fluid.
This can be evaluated by the mean slip velocity in the average sense, 
as shown in figures~\ref{fig:uf}(c,d).
It is defined as $\bar v_{i,s} = \bar u_{i,p} - \bar v_i$, where $u_{i,p}$ is the fluid velocity 
at the particle positions.
The streamwise component $\bar v_{1,s}$ is negative below $y^+ \approx 10$
and positive above that location for cases with low and moderate mass loadings, 
while the negative region extends to $y^+ \approx 80$ for case P3-F14 with the highest mass loading.
This suggests that the particles are moving faster than the fluid in the near-wall region but slower 
in the outer region in the average sense.
The wider extent of the negative $\bar v_{1,s}$ in case P3-F14 can be attributed to
the less effective deceleration effects in the comparatively quiescent flows,
considering their laminarization in the near-wall region.
From another perspective, the negative slip velocity $\bar v_{1,s}$ also implies that 
the fluid is being accelerated in the near-wall region, introducing extra shear rates.
This explains why the skin frictions are only reduced slightly when the near-wall turbulence
is much weakened.
The wall-normal slip velocities $\bar v_{2,s}$ are positive below $y^+ \approx 100$.
It can be inferred that the particles are more inclined to congregate within the ejection regions,
the tendency of which is alleviated by the higher Stokes number $St^+$ and mass loading $\varphi_m$.

\begin{figure}[tb!]
\centering
\begin{overpic}[width=0.5\textwidth]{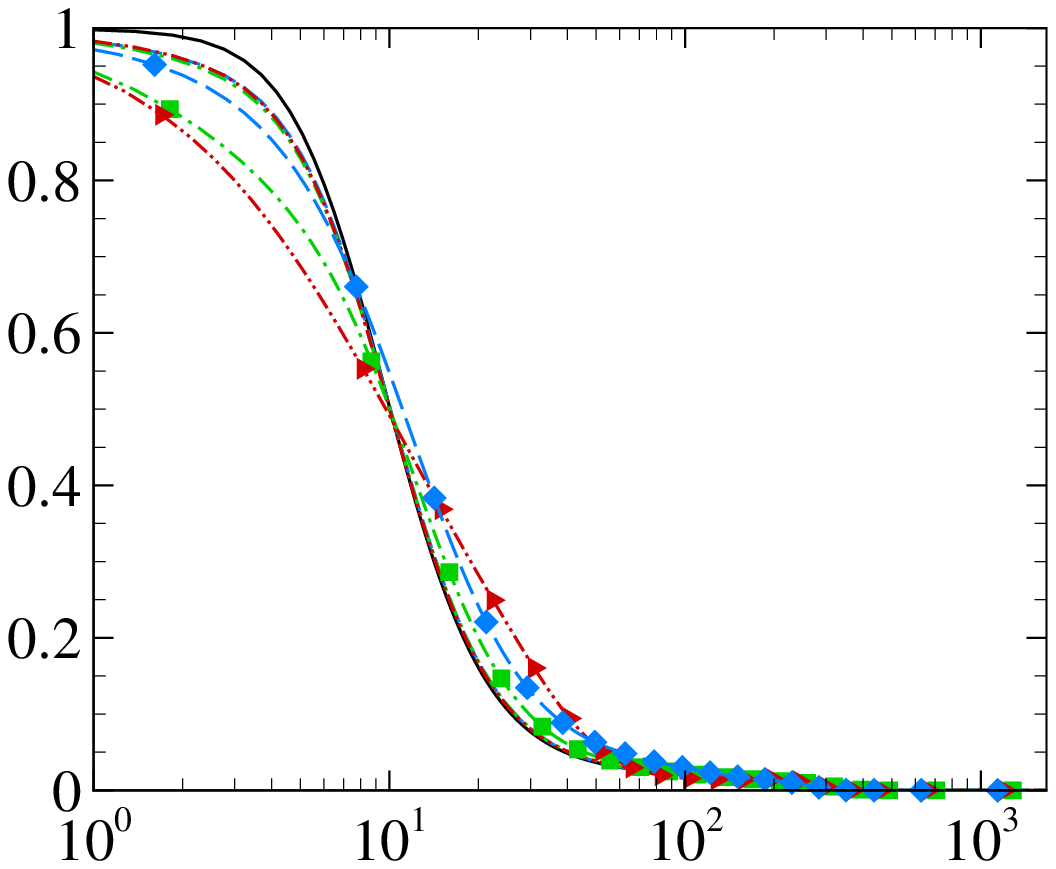}
\put(0,70){(a)}
\put(50,0){$y^+$}
\put(0,38){\rotatebox{90}{$\tau^+_v$}}
\end{overpic}~
\begin{overpic}[width=0.5\textwidth]{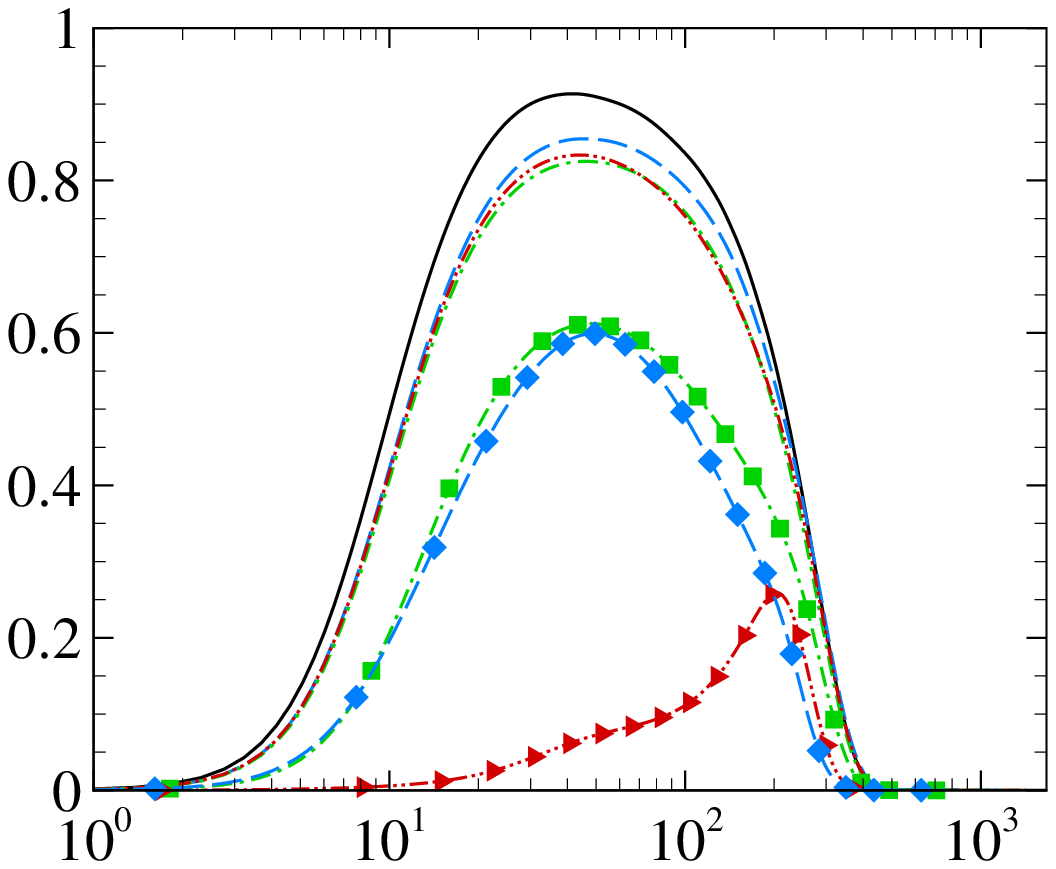}
\put(0,70){(b)}
\put(48,0){$y/\delta$}
\put(0,38){\rotatebox{90}{$\tau^+_r$}}
\end{overpic}\\[2.0ex]
\begin{overpic}[width=0.5\textwidth]{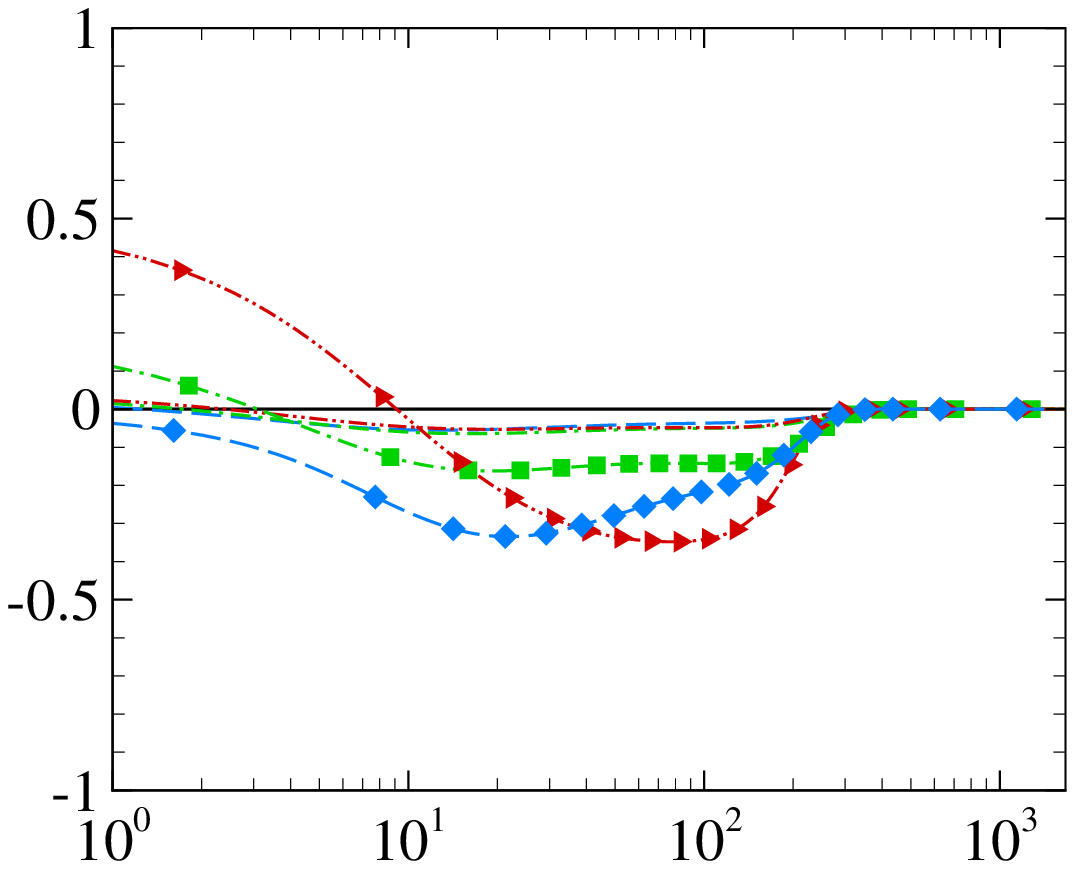}
\put(0,70){(c)}
\put(50,0){$y^+$}
\put(0,38){\rotatebox{90}{$\tau^+_p$}}
\end{overpic}~
\begin{overpic}[width=0.5\textwidth]{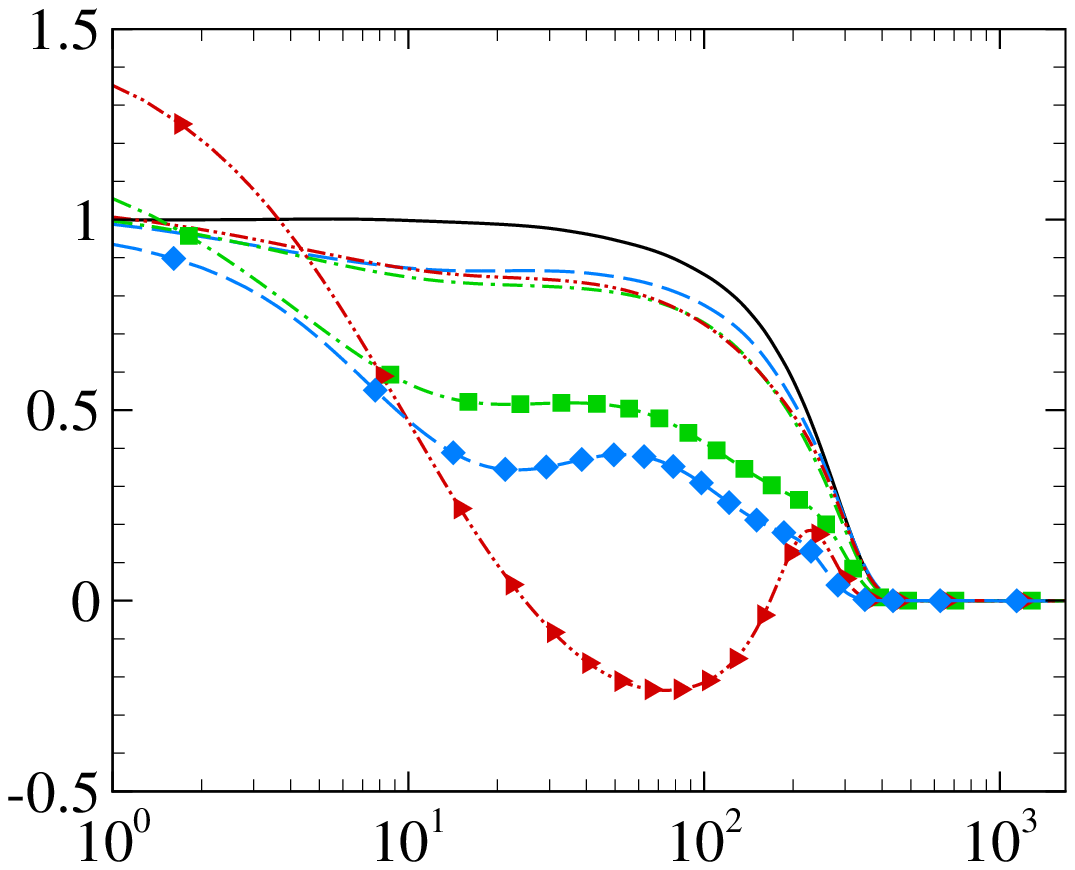}
\put(0,70){(d)}
\put(48,0){$y/\delta$}
\put(0,38){\rotatebox{90}{$\tau^+_t$}}
\end{overpic}
\caption{Wall-normal distribution of (a) viscous shear stress $\tau_v$, (b) Reynolds shear stress
$\tau_r$, (c) particle force $\tau_p$ and (d) total stress $\tau_t$.
Line legends refer to Table~\ref{tab:param}.}
\label{fig:mombal}
\end{figure}

The mean velocity distribution both depends on and determines the mean streamwise momentum transport.
Under the conditions of the temporally statistically steady, the streamwise mean momentum equation 
can be formulated as
\begin{equation}
\frac{\partial \bar \tau_{12}}{\partial y} - \frac{\partial \overline{\rho u''_1 u''_1}}{\partial y}
+ \bar F_{p,1} = \frac{\partial \bar \rho \tilde u_1 \tilde u_1}{\partial x} + 
\frac{\partial \bar \rho \tilde u_1 \tilde u_1}{\partial y} 
- \frac{\partial \bar \tau_{11}}{\partial x}
+ \frac{\partial \overline{\rho u''_1 u''_1}}{\partial x},
\label{eqn:meanmom}
\end{equation}
with the right-hand-side representing the spatial development and mean flow convection.
Integrating this formula in the wall-normal direction, the left-hand-side terms can be cast as
\begin{equation}
\tau_v = \tau_{12} =\bar \mu \left( \frac{\partial \bar u_1}{\partial y} + 
\frac{\partial \bar u_2}{\partial x} \right),~~
\tau_r = - \overline{\rho u''_1 u''_2}, ~~
\tau_p = \int^\delta_y F_{p,1} dy
\end{equation}
namely the viscous shear stress, the Reynolds shear stress and the particle shear stress.
Note that the integration is performed from the outer edge of the boundary layer to the wall
to avoid non-zero $\tau_p$ in the free-stream, which is misleading and unphysical.
Their summation is referred to as the total shear stress 
\begin{equation}
\tau_t = \tau_v + \tau_r + \tau_p.
\end{equation}
It should be equivalent to the integration of the streamwise development terms
on the right-hand-side of Equation~\eqref{eqn:meanmom} but in the opposite sign.

Figure~\ref{fig:mombal} shows the wall-normal distribution of the shear stresses given above.
Consistent with the previous observations in the mean velocity, these shear stresses are only
weakly affected in low mass loading cases P1-F02, P2-F02 and P3-F02, but are strongly
modulated in the rest of the cases, the degree of which is dependent on
both the mass loadings $\varphi_m$ and the particle Stokes number $St^+$.
As the $\varphi_m$ increases, the $\tau_v$ is decreased below $y^+ \approx 10$ but 
is increased above it. 
The $\tau_t$ is decreased monotonically with the $\varphi_m$.
In cases P1-F06 and P2-F06, the peaks in the near-wall region still exist but are significantly 
lower, suggesting that the turbulent motions are only disturbed to be weakened in these cases.
In case P3-F14, however, the near-wall peak completely disappears, 
leaving only comparatively high values near the edge of the boundary layer,
in agreement with our observations from the instantaneous flow fields that the turbulence
is only active in the outer region for case P3-F14.
This also supports our previous statement that the near-wall turbulent motions are completely 
different in cases P1-F06 and P3-F14, despite the similarity in the mean velocity profiles.
The particle shear stress $\tau_p$ constitutes less than 10\% for cases with $\varphi_m=0.17$,
but finitely by approximately -10\% in case P2-F06 and -30\% in case P1-F06 in the buffer layer
and P3-F14 in the logarithmic layer, respectively.
Moreover, in case P3-F14, this term constitutes approximately 40\% in the viscous sublayer.
The important role played by particles in case P1-F06 can be attributed to the effective response
of particles to the fluid motions, and that in case P3-F14 merely to the high mass loading.
Further increasing the mass loading for particle populations P1 and P2 will probably give the same
results, but the critical value of $\varphi_m$ resulting in the flow laminarization depends
on the particle Stokes number, as pointed out by~\citet{muramulla2020disruption}.
Correspondingly, the summation of these terms, i.e. the total shear stress $\tau_t$ is 
merely slightly reduced in cases P1-F02, P2-F02 and P3-F02, but highly affected in the other 
three cases with moderate and high mass loadings, showing non-monotonic trend of variation
along the wall-normal direction.
Considering such significant variations, it is probably necessary to reconsider the
strategy of turbulent modelling and predictions in particle-laden flows.

\subsection{Skin friction and its decomposition} \label{subsec:skin}

\begin{figure}[tb!]
\centering
\begin{overpic}[width=0.8\textwidth]{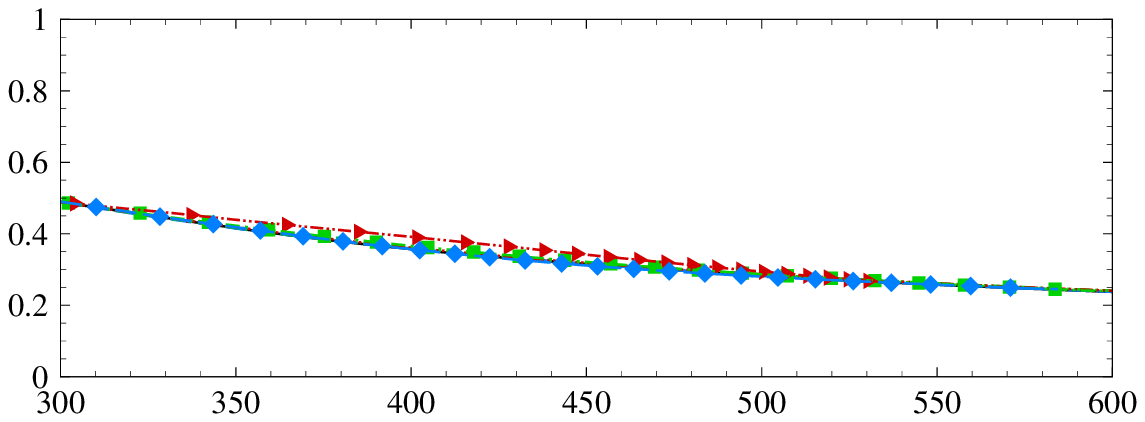}
\put(2,34){(a)}
\put(2,12){\rotatebox{90}{$C_V \times 10^3$}}
\end{overpic}\\
\begin{overpic}[width=0.8\textwidth]{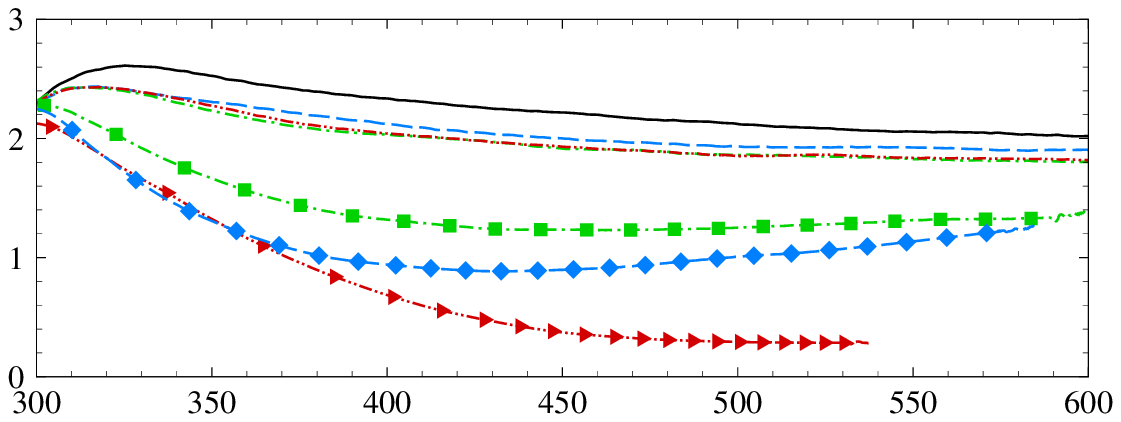}
\put(2,34){(b)}
\put(2,12){\rotatebox{90}{$C_R \times 10^3$}}
\end{overpic}\\
\begin{overpic}[width=0.8\textwidth]{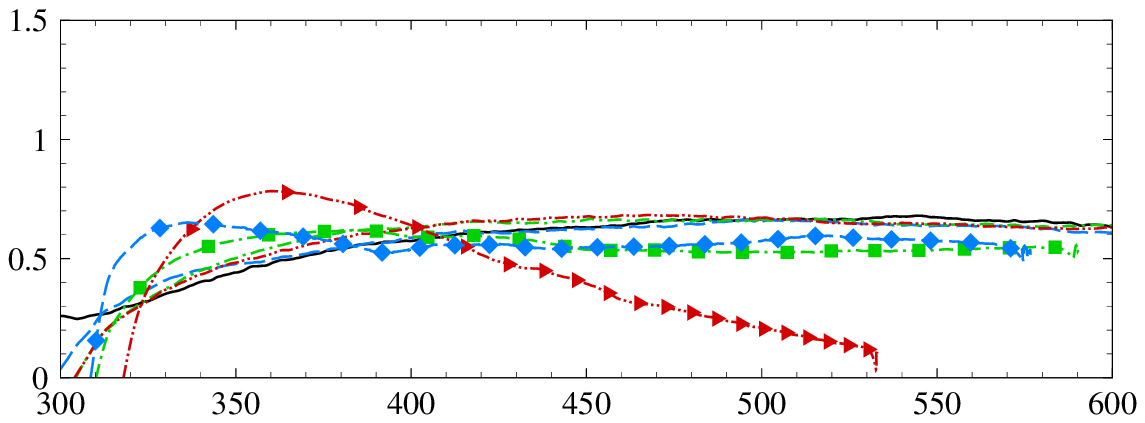}
\put(2,34){(c)}
\put(2,12){\rotatebox{90}{$C_G \times 10^3$}}
\end{overpic}\\
\begin{overpic}[width=0.8\textwidth]{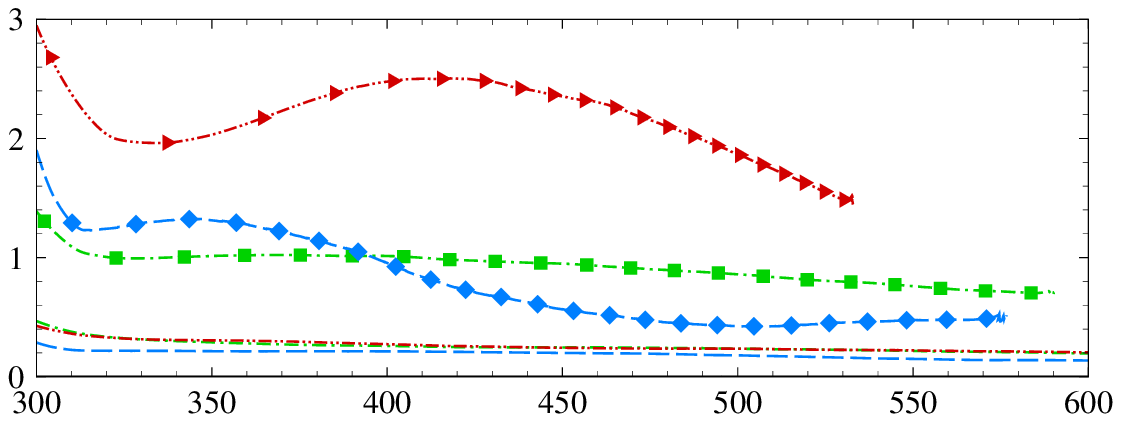}
\put(2,34){(d)}
\put(2,12){\rotatebox{90}{$C_P \times 10^3$}}
\put(50,-2){$Re_{\theta_i}$}
\end{overpic}\\
\caption{Skin friction decomposition, (a) $C_V$, (b) $C_R$, (c) $C_G$ and (d) $C_P$.
Line legends refer to Table~\ref{tab:param}.}
\label{fig:cfdecp}
\end{figure}

The alteration of the momentum balance will further result in the variation of skin friction.
In figure~\ref{fig:cfdecp} we present the distributions of the skin friction $C_f$
against the incompressible counterpart of the momentum Reynolds number $Re_{\delta_2}$, defined as
\begin{equation}
Re_{\delta_2} = Re_{\theta_i} = \frac{\rho_0 U_0 \theta}{\mu_w},~~
\theta = \int^\delta_0 \frac{\bar \rho \bar u}{\rho_0 U_0} \left( 1 -\frac{\bar u}{U_0}\right) 
{\rm d} y.
\end{equation}
It has been shown by \citet{hopkins1971evaluation} that the friction law for the incompressible
turbulent boundary layers $C_{f,i} = 0.024 Re^{-1/4}_{\theta_i}$ can be applied to 
the compressible turbulence over adiabatic walls if the following factors are incorporated
\begin{equation}
C_{fi} = F_c C_f, ~~ F_c = \frac{T_w/T_0-1}{\arcsin^2 \alpha}, ~~
\alpha = \frac{T_w/T_0 -1}{\sqrt{T_w /T_0 (T_w/T_0-1)}}.
\end{equation}
For the one-way coupling case P0-F00, the skin friction $C_f$ obeys the friction law
after a streamwise extent of flow adjustment towards the equilibrium state.
For cases P1-F02, P2-F02 and P3-F02, the skin frictions are slightly reduced compared with
the case P0-F00, and the influences of the particle Stokes numbers are weak.
In cases with low and moderate mass loadings (P1-F06, P2-F06 and P3-F14), 
the skin frictions firstly overshoot to higher values, then decrease rapidly downstream,
violating their power-law correlations with $Re_{\theta_i}$ in compressible turbulent boundary 
layers.
Consistent with the previous observations, the skin friction in case P1-F06 is the most affected, 
reduced by approximately $30\%$ compared with the one-way coupling case,
even lower than case P2-F06 with the same mass loading, suggesting that the evident particle
preferential accumulation in this case modulates the near-wall turbulence significantly.
This is reasonable, for in a low Reynolds number turbulence where no large-scale and 
very-large-scale motions can be found, near-wall turbulence is, perhaps, the most crucial,
the intrusion or disruption of which thus reduces the skin friction most effectively.

\begin{figure}[tbp!]
\centering
\begin{overpic}[width=0.8\textwidth]{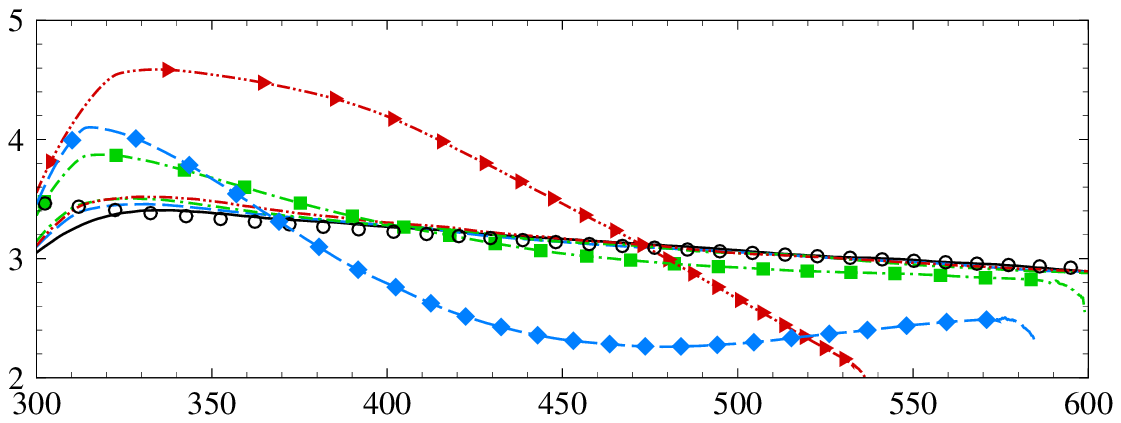}
\put(2,12){\rotatebox{90}{$C_f \times 10^3$}}
\put(50,-2){{$Re_{\theta_i}$}}
\end{overpic}
\caption{Variation of the skin friction $C_f$ with $Re_{\theta_i}$.
Line legends refer to Table~\ref{tab:param}. Symbols: turbulent friction law $C_{f,i}/F_c$.}
\label{fig:cf}
\end{figure}

To further explore the influences of the particle feedback forces and the other momentum balance 
terms on the variation of the skin friction,
we decompose the skin friction using the formula below, which is obtained by the two-fold
integration of the mean momentum equation following~\citep{wenzel2022influences},
\begin{equation}
C_f = \underbrace{\frac{2}{\rho_0 U^2_0 \delta} 
\int^\delta_0 \tau_{12} {\rm d}y}_{C_V}
- \underbrace{\frac{2}{\rho_0 U^2_0 \delta}
\int^\delta_0 \overline{\rho u''_1 u''_2} {\rm d} y}_{C_R}
+ \underbrace{\frac{2}{\rho_0 U^2_0 \delta}
\int^\delta_0 (y-\delta) \bar F_{p,1} {\rm d} y}_{C_P}
+ \underbrace{\frac{2}{\rho_0 U^2_0 \delta}
\int^\delta_0 (y-\delta) I_{x} {\rm d} y}_{C_G}
\label{eqn:fik}
\end{equation}
with $I_x$ expressed as
\begin{equation}
I_x= - \frac{\partial \bar \rho \tilde u_1 \tilde u_1}{\partial x} 
- \frac{\partial \overline{\rho u''_1 u''_1}}{\partial x}
- \frac{\partial \bar \rho \tilde u_1 \tilde u_2}{\partial y} 
- \frac{\partial \bar p}{\partial x}
+ \frac{\partial \bar \tau_{11}}{\partial x}.
\end{equation}
The terms $C_V$, $C_R$, $C_P$ and $C_G$ on the right-hand-side of equation~\eqref{eqn:fik}
originate from the integration of the viscous shear stress, Reynolds shear stress, particle
feedback force and the mean flow convection.

\begin{table}[tb!]
\centering
\caption{Contribution of $C_V$, $C_R$, $C_G$ and $C_P$ to the skin friction at $Re_{\theta_i}=450$.}
\begin{tabular*}{0.6\textwidth}{@{\extracolsep{\fill}}ccccc}
\hline
Case & $C_{V}/C_f$  &  $C_{R}/C_f$  &  $C_{G}/C_f$  &  $C_{P}/C_f$ \\ \hline
P0-F00 & 9.97\%   &  70.10\%  &  19.93\%  &  0.00\%  \\
P1-F02 & 10.05\%  &  49.73\%  &  33.87\%  &  6.35\%  \\
P1-F06 & 13.40\%  &  38.67\%  &  23.59\%  &  24.38\%  \\
P2-F02 & 10.13\%  &  47.49\%  &  34.57\%  &  7.81\%  \\
P2-F06 & 10.55\%  &  31.74\%  &  26.62\%  &  31.09\% \\
P3-F02 & 10.05\%  &  47.50\%  &  34.79\%  &  7.66\%  \\
P3-F14 & 9.87\%   &  8.87\%   &  13.26\%  &  68.00\% \\
\hline
\end{tabular*}
\label{tab:cfdecp}
\end{table}

The results are reported in figure~\ref{fig:cfdecp}, and their respective contributions
to the total skin friction at $Re_{\theta_i}=450$ are listed in Table~\ref{tab:cfdecp}.
For the one-way coupling case P0-F00, the contribution of each term to the skin friction
is consistent with that reported in previous studies
in canonical turbulent boundary layers~\citep{wenzel2022influences},
in that the viscous shear stress $C_V$ contribute by approximately $10\%$, 
the Reynolds shear stress $C_R$ by approximately $70\%$, 
and the mean flow convection $C_G$ by approximately $20\%$.
In the two-way coupling cases, the contributions of the viscous shear stress $C_V$ remain almost 
unchanged for all the cases considered.
The contribution from the Reynolds shear stress $C_R$ decreases due to the lower intensity of 
the near-wall turbulence, the trend of which is identical to those given in 
figure~\ref{fig:mombal}(b).
The mean flow convection term $C_G$ is not much affected in low and moderate mass loading cases,
manifesting the trend of converging to a constant value at $Re_{\theta_i} > 450$,
while for high mass loading case P3-F14, this term decreases monotonically with $Re_{\theta_i}$, 
suggesting that the particles are inhibiting the spatial development of the boundary layer.
The particle feedback force term increases with the mass loading.
Its contribution to case P3-F14 is the highest, but varies non-monotonically with $Re_{\theta_i}$, 
increasing to reach a peak value first then decreasing rapidly.
This term is much higher than the other components, contributing approximately 
68\% to the skin friction at $Re_{\theta_i} = 450$.
In the other cases, however, its contribution remains almost a constant value at 
$Re_{\theta_i} > 450$, being approximately 7\% in cases with $\varphi_m = 0.17$,
24\% in case P1-F06 and 31\% in case P2-F06.

\begin{figure}[tb!]
\centering
\begin{overpic}[width=0.8\textwidth]{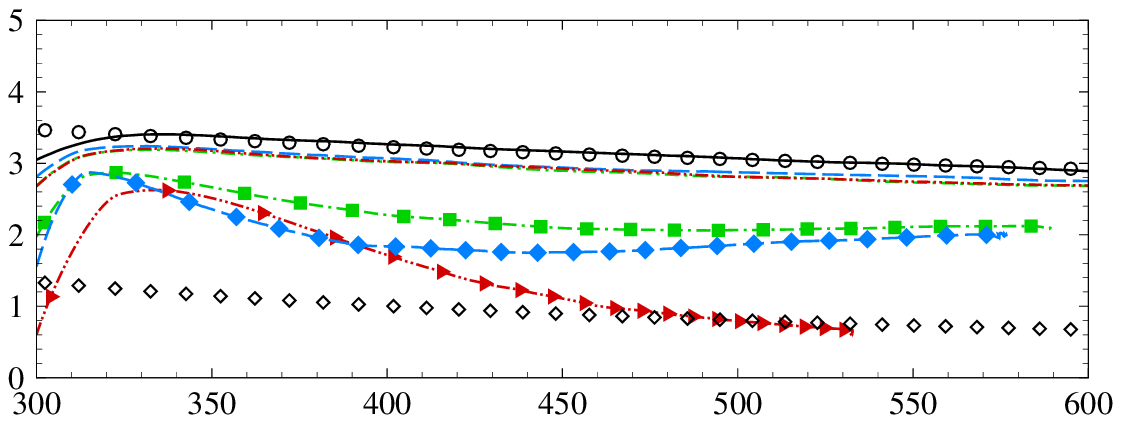}
\put(2,5){\rotatebox{90}{$(C_V+C_R+C_G) \times 10^3$}}
\put(50,0){$Re_{\theta_i}$}
\end{overpic}\\
\caption{Variation of $C_V+C_R+C_G$ with $Re_{\theta_i}$,
line legends refer to Table~\ref{tab:param},
circles: turbulent friction law, diamonds: laminar friction law.}
\label{fig:cfdecp2}
\end{figure}

Adding the $C_V$, $C_R$ and $C_G$ terms that are irrelevant to the particle feedback forces,
as shown in figure~\ref{fig:cfdecp2}, we found that the result of case P3-F14, in which
the near-wall flow is laminarized, gradually falls to the friction law of laminar
boundary layers $C_{f,i}=0.44/Re_{\theta_i}$.
This is interesting and probably useful for probing the skin friction laws for particulate flows,
especially those with high mass loadings where the near-wall turbulence is suppressed,
but more numerical simulations will needed to achieve this task.

\subsection{Kinetic energy transport}  \label{subsec:tke}

In this subsection, we discuss the mean and turbulent kinetic energy and their transport
subject to the influences of the feedback forces from the particles.

\begin{figure}[tb!]
\centering
\begin{overpic}[width=0.5\textwidth]{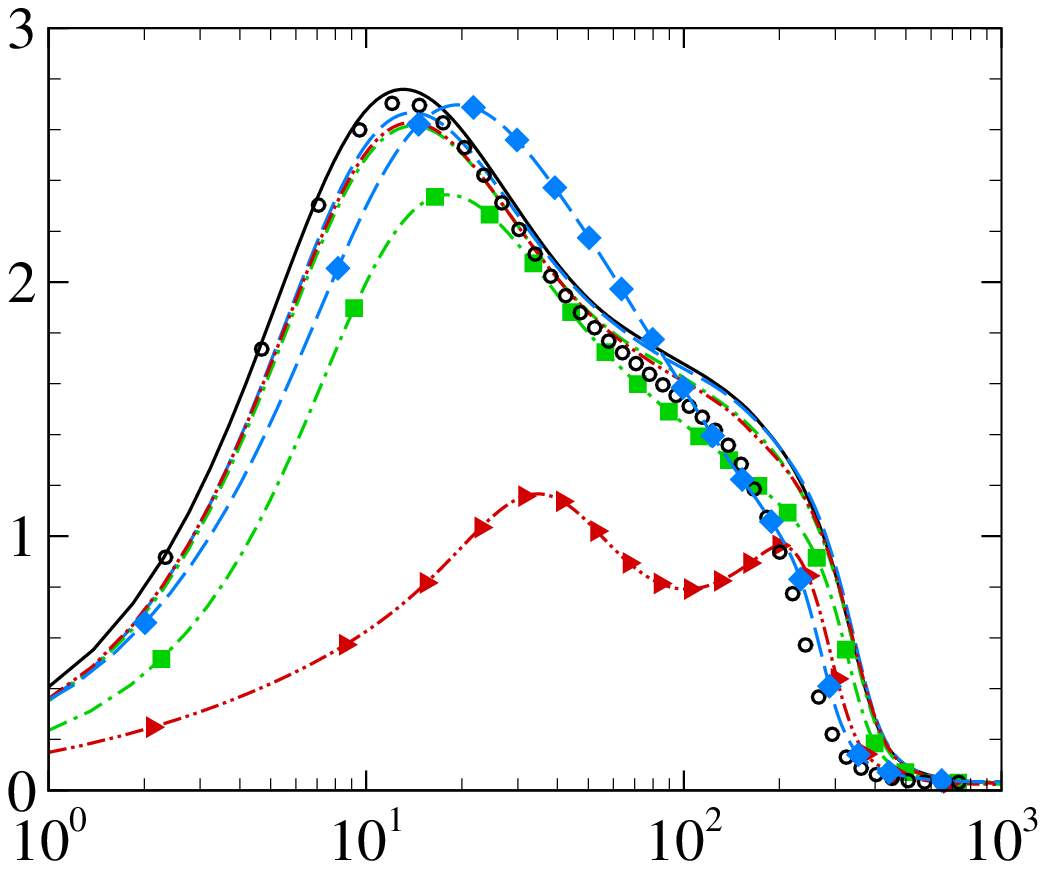}
\put(0,70){(a)}
\put(50,0){$y^+$}
\put(0,38){\rotatebox{90}{$\sqrt{R^+_{11}}$}}
\end{overpic}~
\begin{overpic}[width=0.5\textwidth]{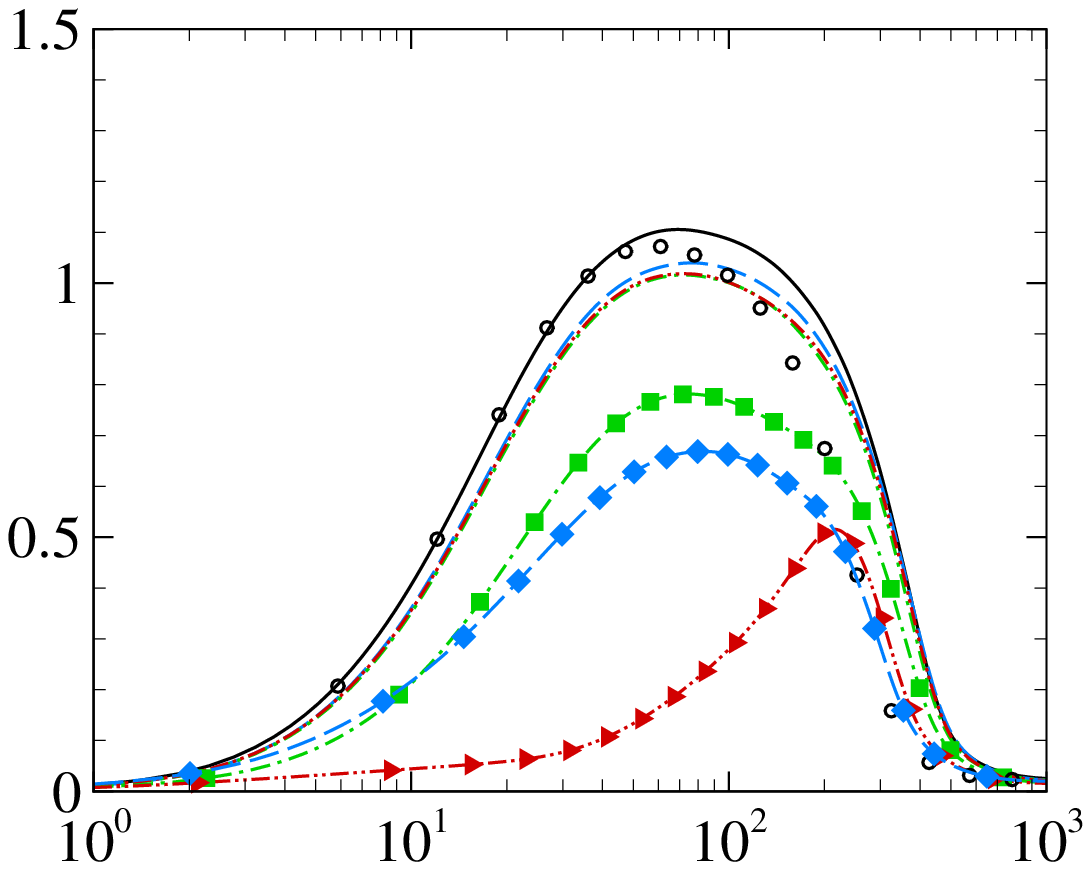}
\put(0,70){(b)}
\put(50,0){$y^+$}
\put(0,38){\rotatebox{90}{$\sqrt{R^+_{22}}$}}
\end{overpic}\\[1.0ex]
\begin{overpic}[width=0.5\textwidth]{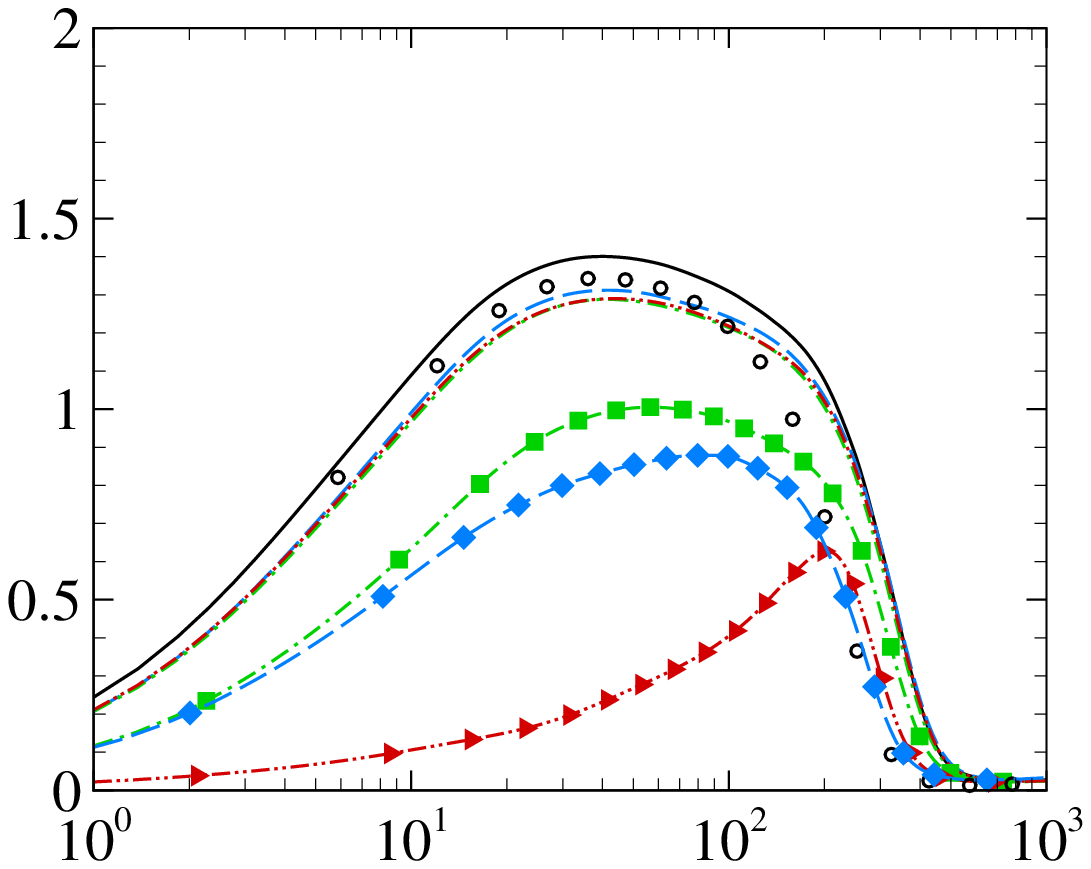}
\put(0,70){(c)}
\put(50,0){$y^+$}
\put(0,38){\rotatebox{90}{$\sqrt{R^+_{33}}$}}
\end{overpic}~
\begin{overpic}[width=0.5\textwidth]{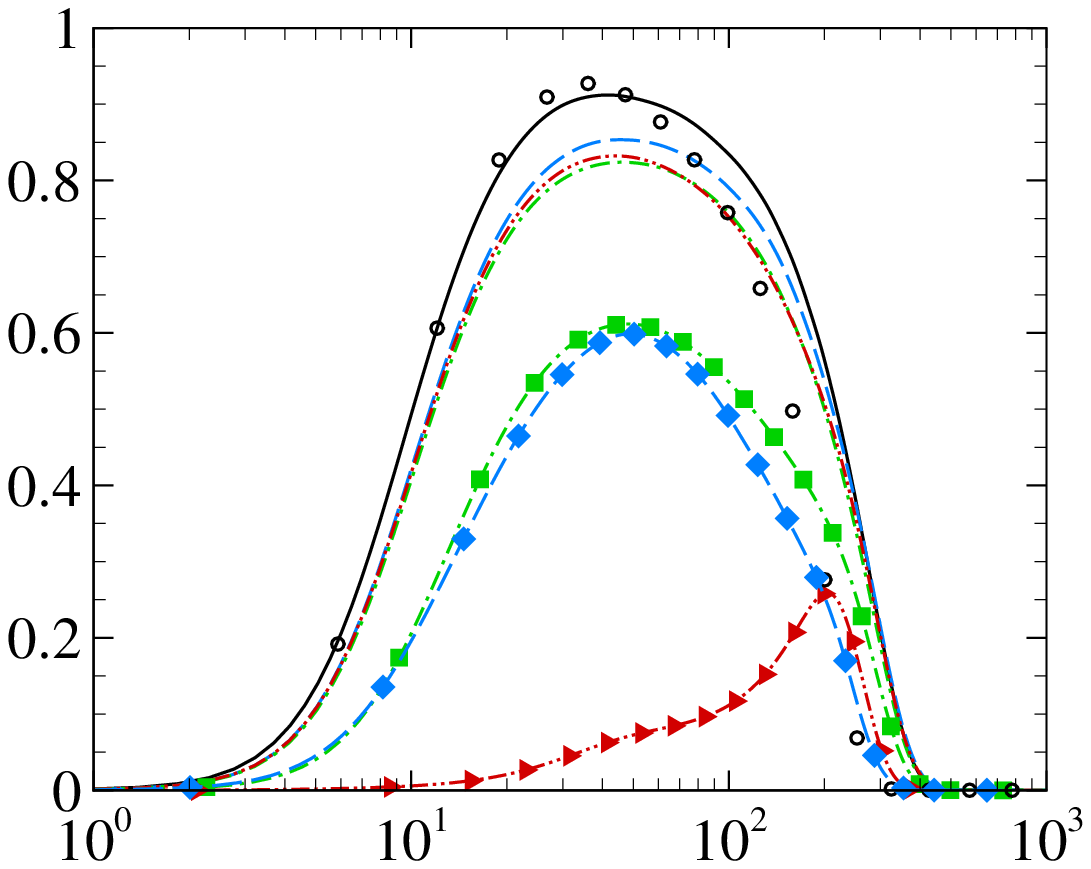}
\put(0,70){(d)}
\put(50,0){$y^+$}
\put(0,35){\rotatebox{90}{$-R^+_{12}$}}
\end{overpic}\\
\caption{Distributions of (a) $\sqrt{R^+_{11}}$, (b) $\sqrt{R^+_{22}}$, (c) $\sqrt{R^+_{33}}$,
(d) Reynolds shear stress $-R^+_{12}$. Line legends refer to Table~\ref{tab:param}. 
Black circles: data reported by \citet{pirozzoli2011turbulence} at $M_0=2$ and $Re_\tau=250$.}
\label{fig:rey}
\end{figure}

\begin{figure}[tb!]
\centering
\begin{overpic}[width=0.45\textwidth,trim={0.2cm 0.2cm 0.2cm 0.2cm},clip]{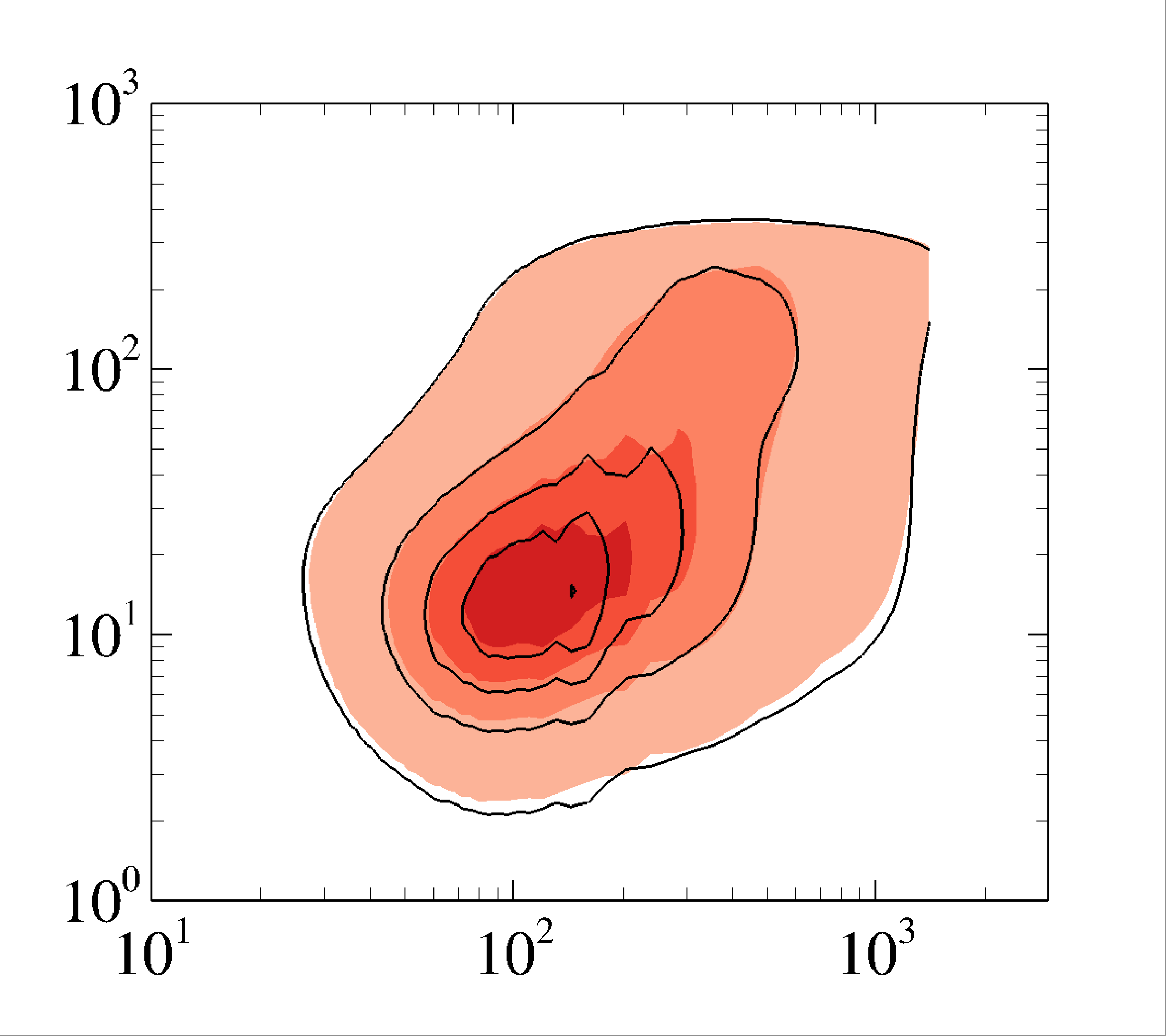}
\put(-2,80){(a)}
\put(48,0){$\lambda^+_z$}
\put(0,45){\rotatebox{90}{$y^+$}}
\end{overpic}~
\begin{overpic}[width=0.45\textwidth,trim={0.2cm 0.2cm 0.2cm 0.2cm},clip]{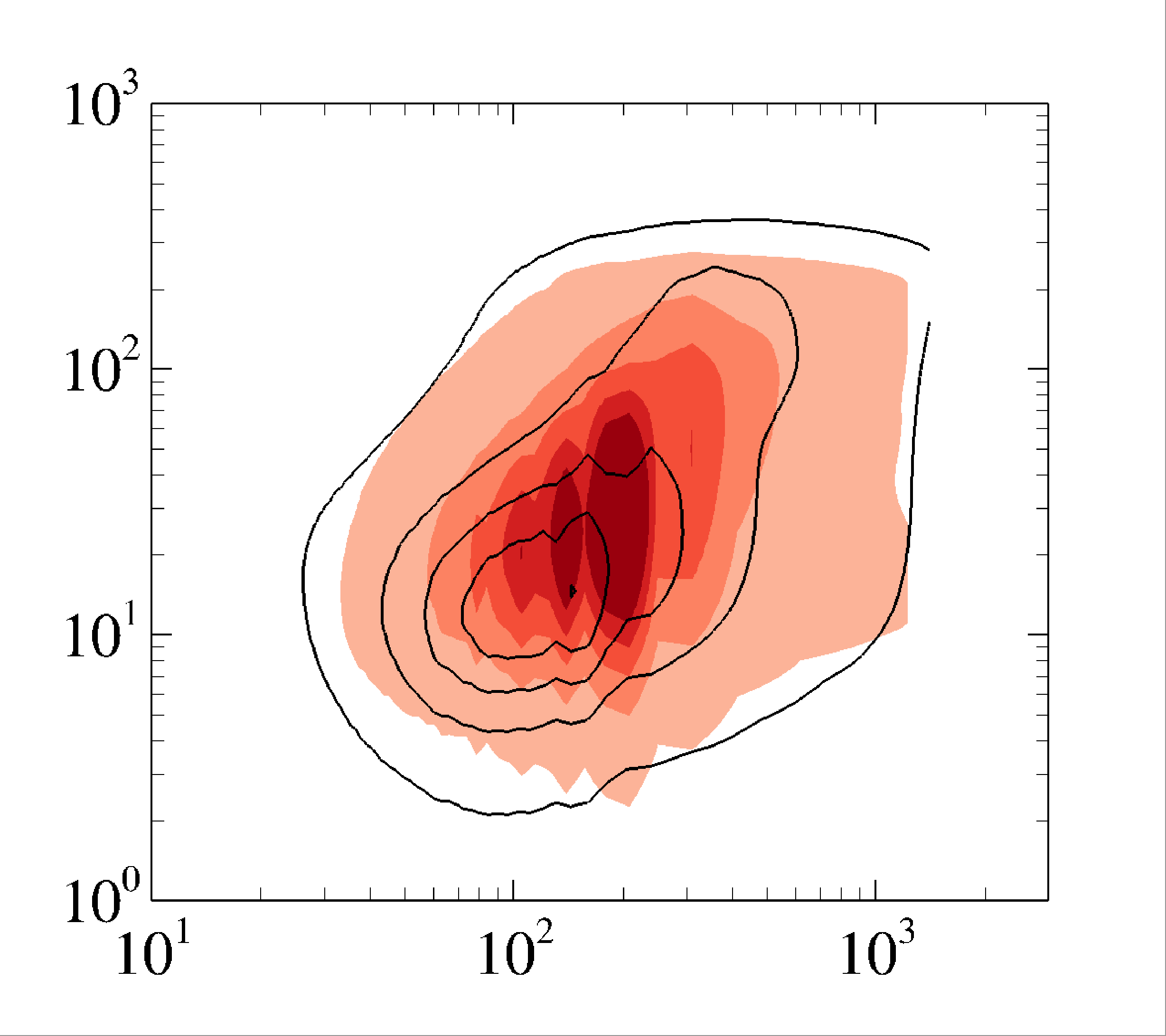}
\put(-2,80){(b)}
\put(48,0){$\lambda^+_z$}
\put(0,45){\rotatebox{90}{$y^+$}}
\end{overpic}\\[1.0ex]
\begin{overpic}[width=0.45\textwidth,trim={0.2cm 0.2cm 0.2cm 0.2cm},clip]{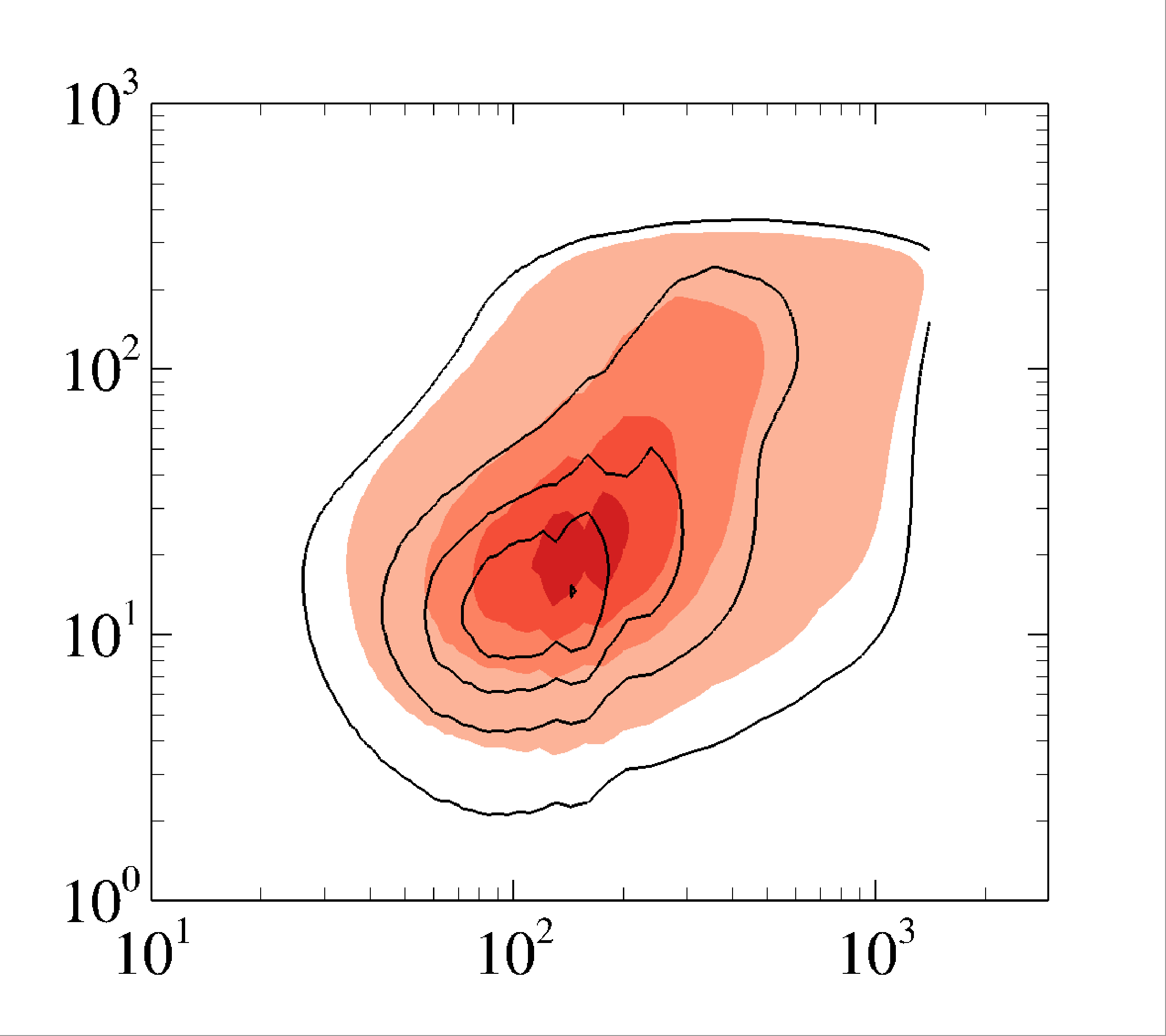}
\put(-2,80){(c)}
\put(48,0){$\lambda^+_z$}
\put(0,45){\rotatebox{90}{$y^+$}}
\end{overpic}~
\begin{overpic}[width=0.45\textwidth,trim={0.2cm 0.2cm 0.2cm 0.2cm},clip]{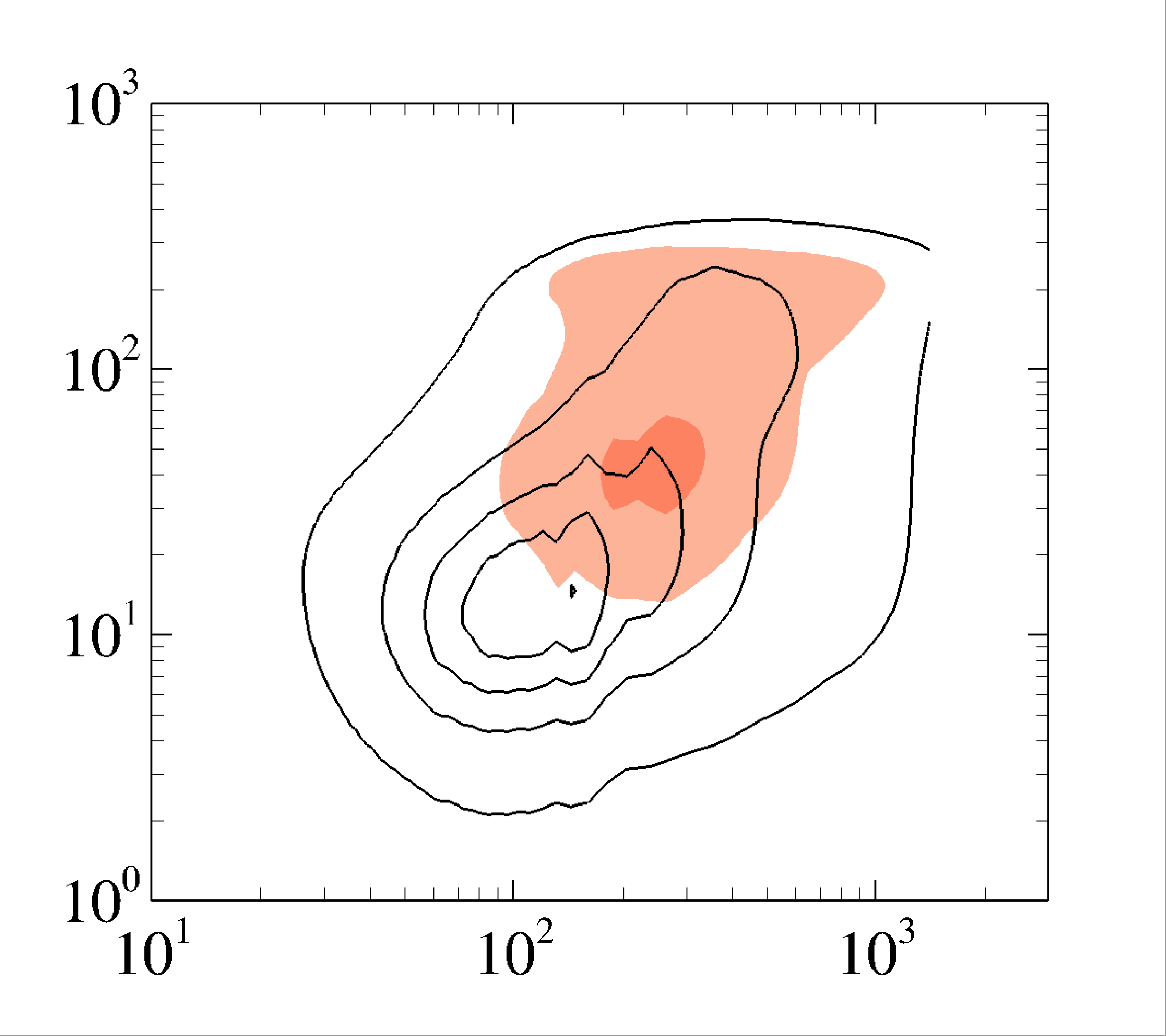}
\put(-2,80){(d)}
\put(48,0){$\lambda^+_z$}
\put(0,45){\rotatebox{90}{$y^+$}}
\end{overpic}\\[1.0ex]
\caption{Pre-multiplied spanwise spectra of turbulent kinetic energy, lines: case P0-F00,
flooded: (a) P1-F02, (b) P1-F06, (c) P2-F06, (d) P3-F14.
Contour levels: (0.2, 0.8, 1.4, 2.0, 2.6).}
\label{fig:spec}
\end{figure}

The density-weighted velocity fluctuation intensities, evaluated by the square root of 
the Reynolds normal stresses, are shown in figure~\ref{fig:rey}, along with the Reynolds shear
stress.
The pre-multiplied spanwise spectra that reflect the scale-by-scale kinetic energy are displayed
in figure~\ref{fig:spec}.
For cases P1-F02, P2-F02 and P3-F02 with low mass loadings, the near-wall high turbulent 
intensities persist, but the magnitudes are lower and
the spanwise characteristic length scales are slightly wider, 
consistent with our previous observations in the instantaneous near-wall turbulence
and the results in incompressible wall-bounded turbulence~\citep{zhou2020non,lee2015modification,
richter2013momentum}.
In cases with moderate mass loadings, the wall-normal and spanwise velocity
fluctuation intensities are further decreased, with those in case P1-F06 slightly lower than 
those in case P2-F06.
The streamwise velocity fluctuation intensities are increased in the former, 
thus resulting in stronger anisotropy, but this trend is reversed in the latter.
From the turbulent kinetic energy spectra, we found that the peaks are reached at
larger spanwise widths and higher wall-normal locations,
with the spectra intensity stronger in case P1-F06.
Both the phenomena above indicate that the particle population P2 suppresses the near-wall
turbulence at all scales, but the population P1 suppresses the turbulent motions 
at the scale of $\lambda^+_z = 100$ but enhances those at the scale of $\lambda^+_z = 200$.
Such a scale-dependent turbulence modulation will be discussed subsequently by showing
the work of the particle feedback force.
In case P3-F14, the wall-normal and spanwise velocity fluctuation intensities and 
the Reynolds shear stress are highly reduced below $y^+ \approx 200$.
The near-wall peaks completely disappear, only showing high intensities above $y^+ \approx 100$.
This conforms to the features of the turbulent fluctuations reflected by the flow structures as 
presented in figure~\ref{fig:instyz}.
The streamwise velocity fluctuation intensity in this high mass loading case manifests
a near-wall peak at $y^+ \approx 40$
with the spanwise characteristic scale of $\lambda^+_z \approx 200$.
This intense near-wall peak is different from the rest of the Reynolds stress components.
Since the sweeping and ejections and the streak meandering are absent in the inner region,
such a high intensity can only be attributed to the particle feedback force, as can be envisaged
as follows.
The feedback force is expected to be positive, generated in the process of 
the particle sweeping towards the wall and decelerating, thereby leading to the acceleration
of the fluid.
Due to their high inertia, the particles are decelerated tardily, hitting and bouncing off the wall
before they follow the trace of quiescent fluid motions and, at least a part of them,
reentry into the active turbulent regions~\citep{vreman2007turbulence}.

\begin{figure}[tb!]
\centering
\begin{overpic}[width=0.5\textwidth]{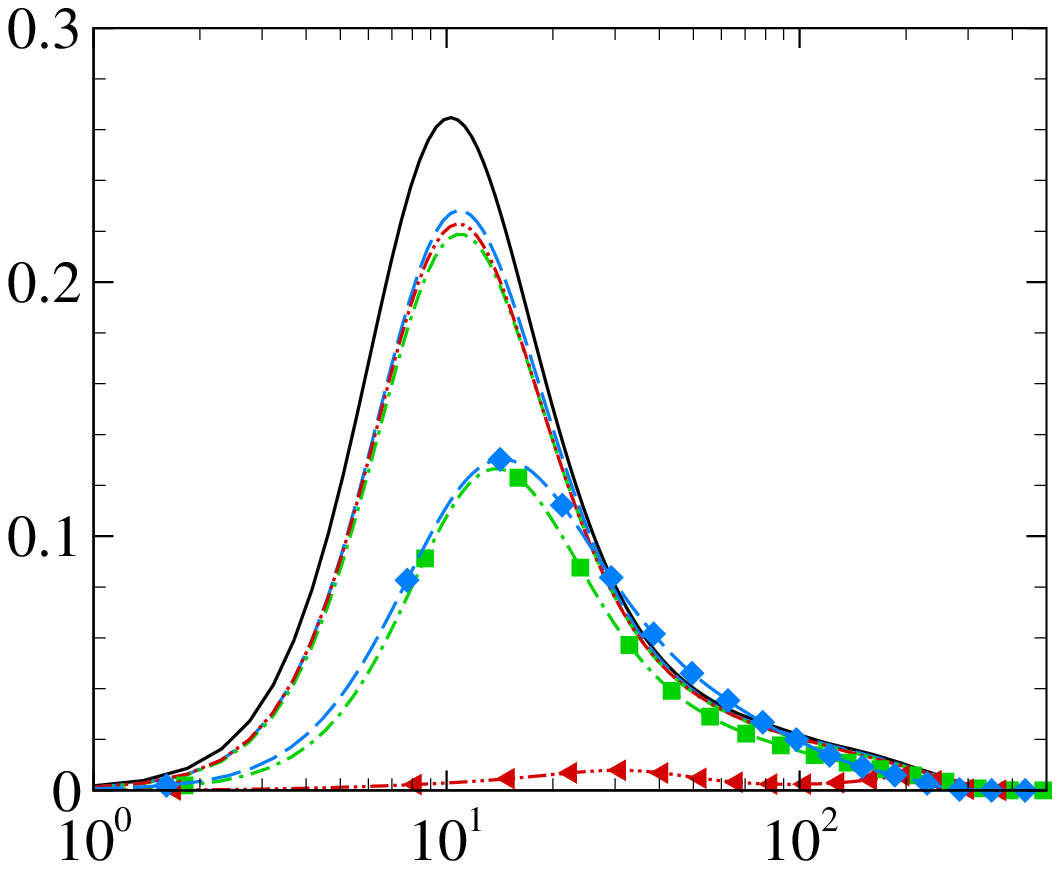}
\put(0,70){(a)}
\put(50,0){$y^+$}
\put(-2,38){\rotatebox{90}{$P^+_K$}}
\end{overpic}~
\begin{overpic}[width=0.5\textwidth]{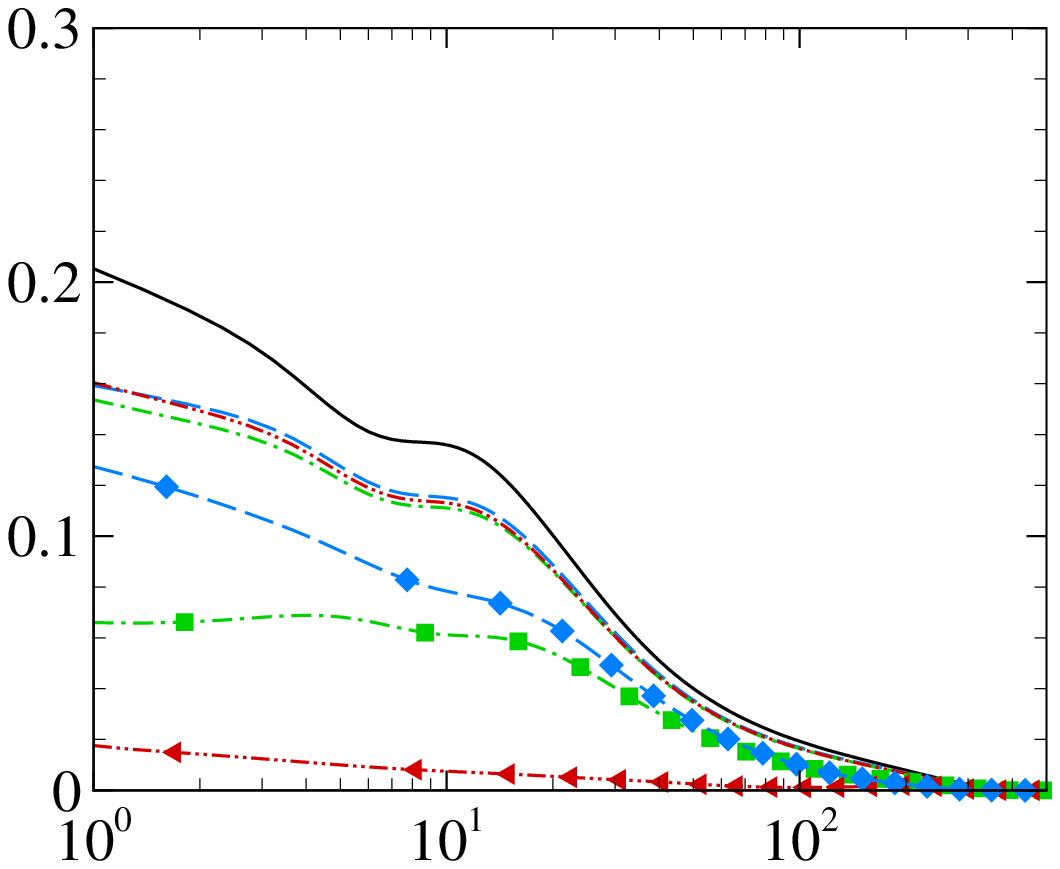}
\put(0,70){(b)}
\put(50,0){$y^+$}
\put(-2,38){\rotatebox{90}{$\varepsilon^+_K$}}
\end{overpic}\\[2.0ex]
\begin{overpic}[width=0.5\textwidth]{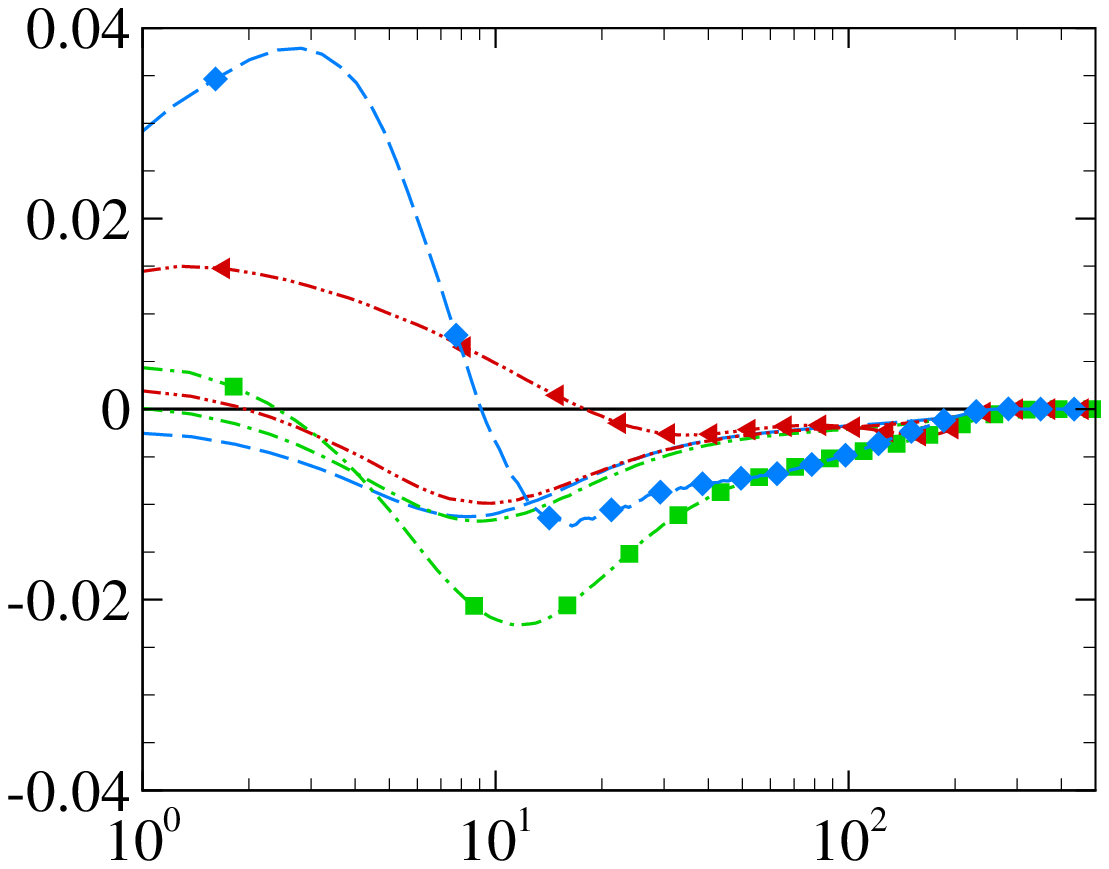}
\put(0,70){(c)}
\put(50,0){$y^+$}
\put(-2,38){\rotatebox{90}{$\overline{F'_{p,i} u'_i}^+$}}
\end{overpic}~
\begin{overpic}[width=0.5\textwidth]{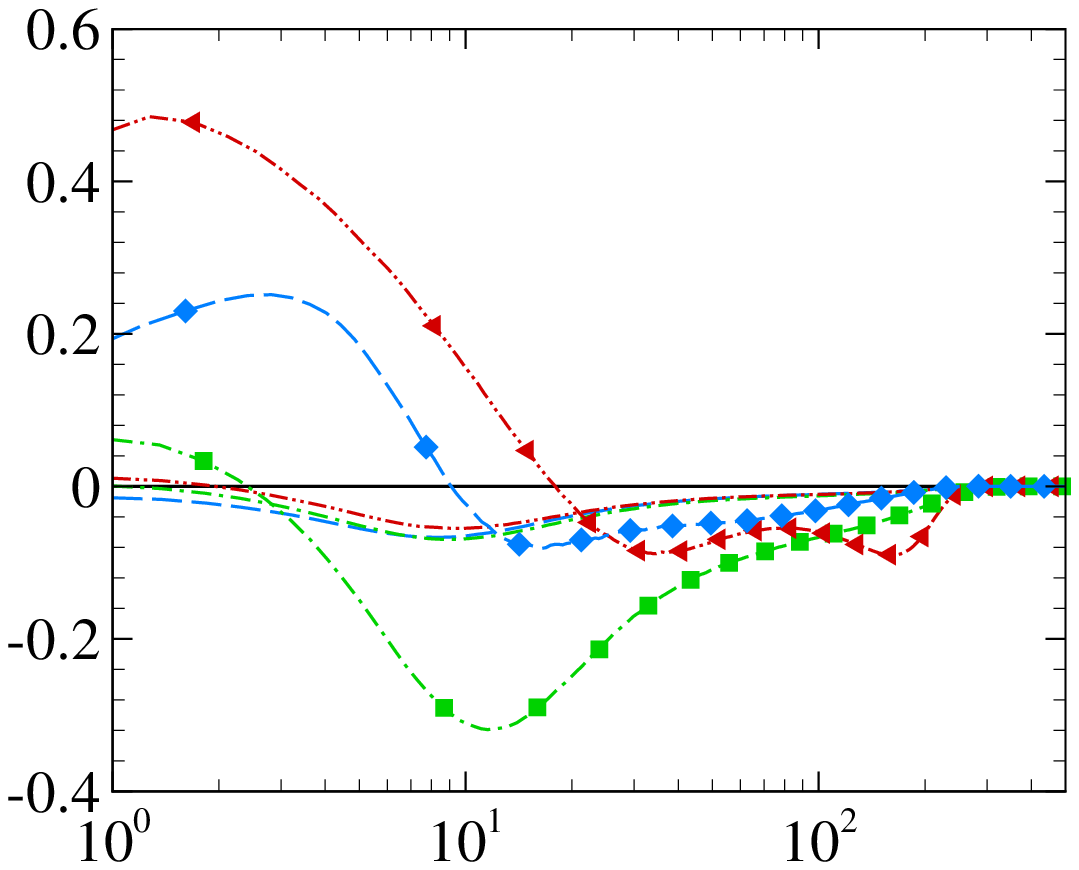}
\put(0,70){(d)}
\put(50,0){$y^+$}
\put(-2,38){\rotatebox{90}{$\overline{F'_{p,i} u'_i}/\varepsilon_{K,w}$}}
\end{overpic}\\[2.0ex]
\caption{Wall-normal distribution of (a) TKE production $P^+_K$,
(b) TKE dissipation $\varepsilon^+_t$, 
(c) TKE production by particle force $\overline{F'_{p,i} u'_i}^+$ and 
(d) $\overline{F'_{p,i} u'_i}$ normalized by TKE dissipation on the wall.
Line legends refer to Table~\ref{tab:param}.}
\label{fig:kebal}
\end{figure}

We further examine the turbulent kinetic energy transport and the roles played by
the particle forces.
The turbulent kinetic energy transport equations can be cast as
\begin{equation}
\frac{\partial K_t}{\partial t} + \frac{\partial \bar u_j K_t}{\partial x_j}= 
\underbrace{-\overline{\rho u''_i u''_j} \frac{\partial \tilde u_i}{\partial x_j}}_{P_K}
\underbrace{+\frac{\partial}{\partial x_j} \left(-\overline{\rho u''_k u''_k u''_j} 
- \overline{p' u''_j} +\overline{\tau'_{ij} u''_i} \right)}_{D_K}
\underbrace{+\overline{\tau'_{ij} \frac{\partial u'_i}{\partial x_j}}}_{\varepsilon_K}
+\overline{F'_{p,i} u''_i}
\end{equation}
where the right-hand-side corresponds to the turbulent kinetic energy production by the mean shear,
spatial diffusions, viscous dissipation and the work by fluctuating particle forces.
In the following discussions, we neglect the convective terms and the spatial diffusion terms,
for these terms do not generate energy in the integral sense.

In figure~\ref{fig:kebal}(a) we present the production term that extracts energy from
the mean flow to turbulence.
Although neither the mean flow nor the Reynolds stresses are much affected by the low mass loading,
the production term is reduced by approximately $15\%$ compared with the one-way coupling case.
In cases P1-F06 and P2-F06, the production term is further reduced, but the near-wall peaks
still exist.
In case P3-F14 with the high mass loading, the production term almost falls to zero,
especially in the near-wall region, further proving our inference that the near-wall high
turbulent intensities are not generated by turbulent motions of the fluid.
This is also consistent with the findings in the previous study~\citep{muramulla2020disruption} 
that the decrement in the production is responsible for the flow laminarization.
The trend of variation of the turbulent viscous dissipation (figure~\ref{fig:kebal}(b))
is similar to the turbulent production, except that the peaks are located at the wall.
The work of the particle forces, as shown in figure~\ref{fig:kebal}(c),
is negative in low mass loading cases and case P2-F06 except for a small region below $y^+=3$,
indicating that the particles are extracting energy from the fluid.
In cases P1-F06 and P3-F14, the region of the positive work extends to $y^+ \approx 10$
and $y^+ \approx 20$, respectively,
and as displayed in figure~\ref{fig:kebal}(d), this term counteracts approximately 20\% and 50\% of
the viscous dissipation at the wall.

\begin{figure}[tb!]
\centering
\begin{overpic}[width=0.45\textwidth,trim={0.2cm 0.2cm 0.2cm 0.2cm},clip]{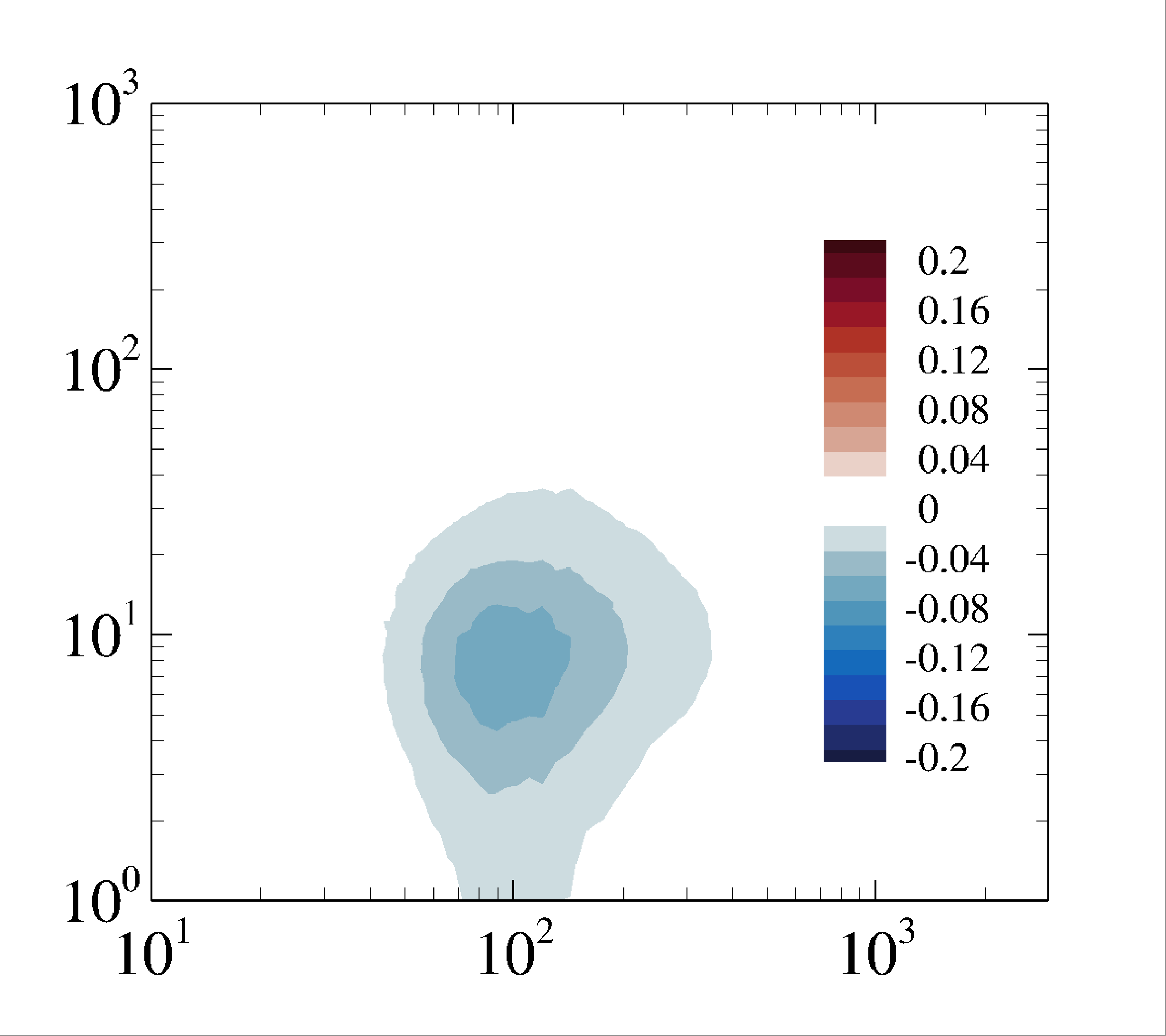}
\put(-2,80){(a)}
\put(48,0){$\lambda^+_z$}
\put(0,45){\rotatebox{90}{$y^+$}}
\end{overpic}~
\begin{overpic}[width=0.45\textwidth,trim={0.2cm 0.2cm 0.2cm 0.2cm},clip]{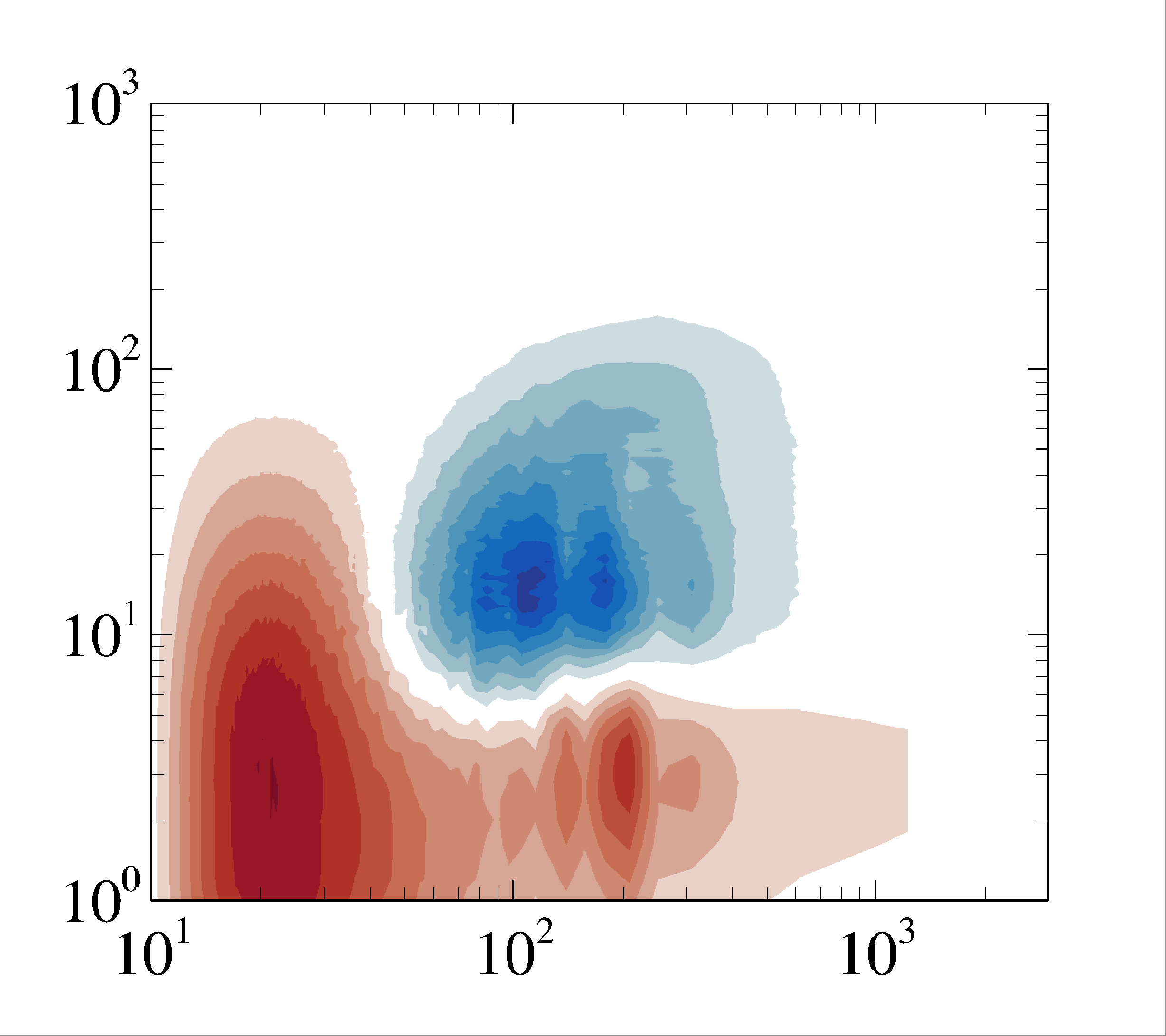}
\put(-2,80){(b)}
\put(48,0){$\lambda^+_z$}
\put(0,45){\rotatebox{90}{$y^+$}}
\end{overpic}\\[2.0ex]
\begin{overpic}[width=0.45\textwidth,trim={0.2cm 0.2cm 0.2cm 0.2cm},clip]{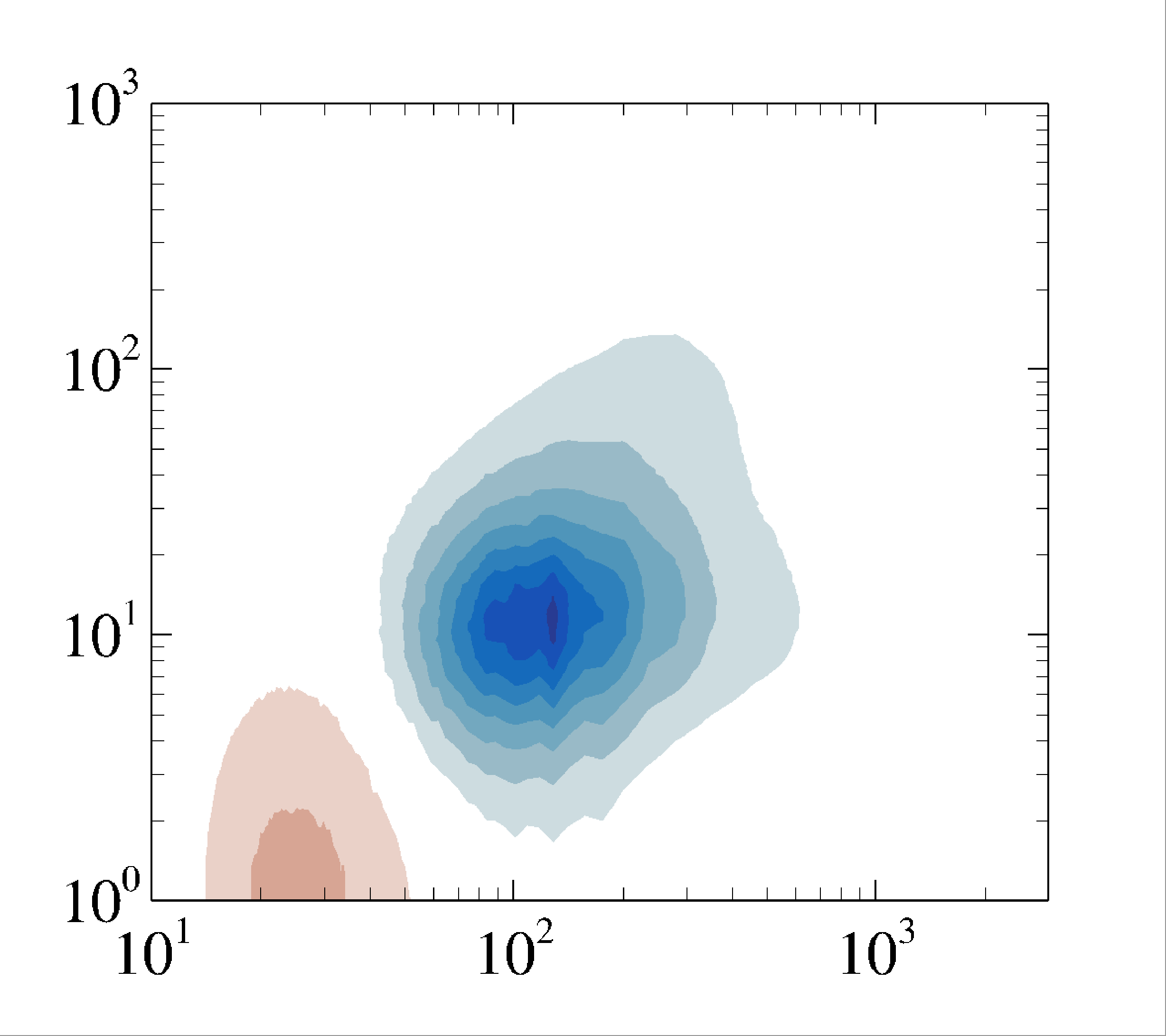}
\put(-2,80){(c)}
\put(48,0){$\lambda^+_z$}
\put(0,45){\rotatebox{90}{$y^+$}}
\end{overpic}~
\begin{overpic}[width=0.45\textwidth,trim={0.2cm 0.2cm 0.2cm 0.2cm},clip]{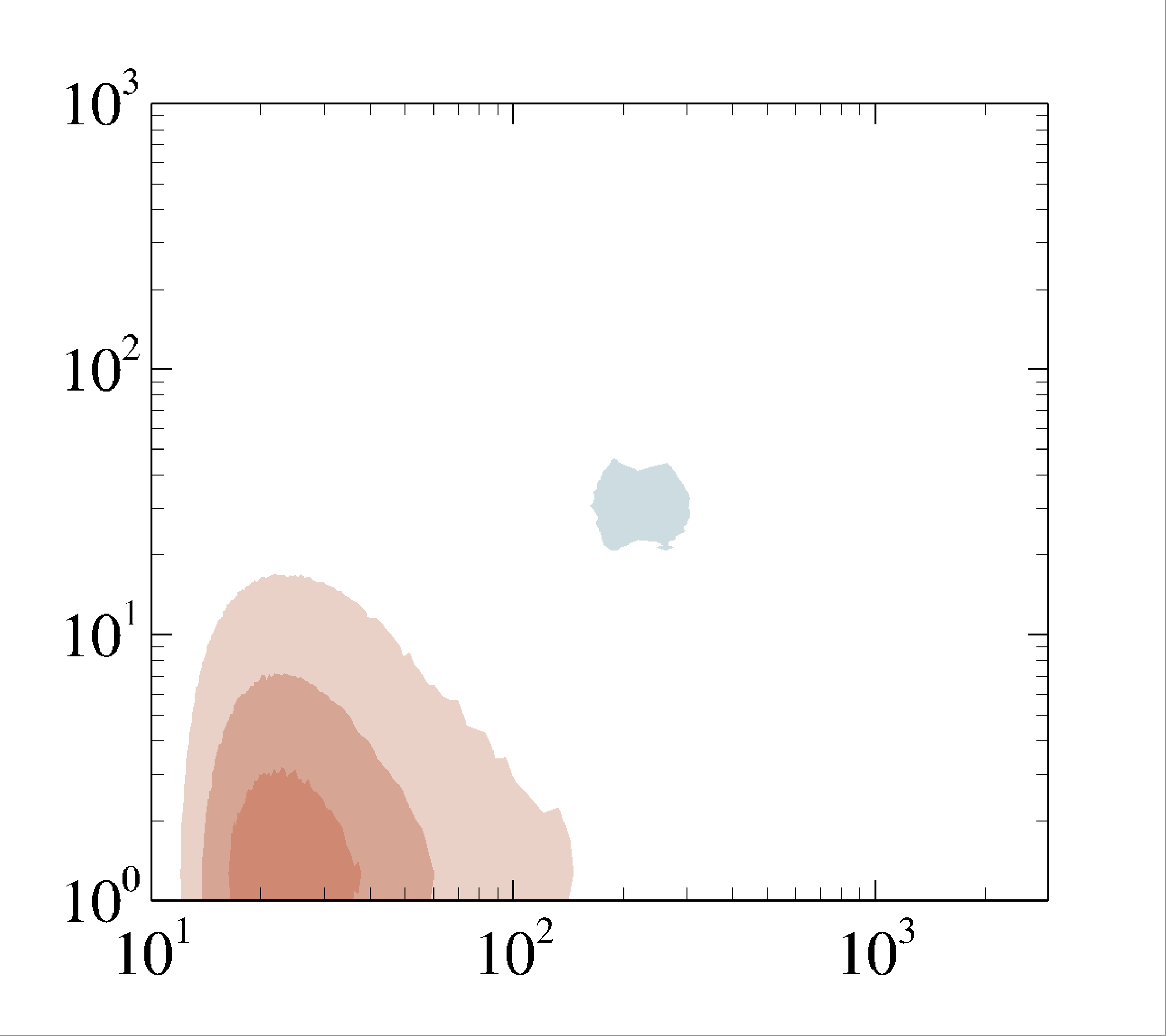}
\put(-2,80){(d)}
\put(48,0){$\lambda^+_z$}
\put(0,45){\rotatebox{90}{$y^+$}}
\end{overpic}\\
\caption{Pre-multiplied spanwise spectra of the TKE particle force work $k_z E^+_{F'_{p,i} u'_i}$,
(a) case P1-F02, (b) case P1-F06, (c) case P2-F06, (d) case P3-F14.}
\label{fig:fuspec}
\end{figure}
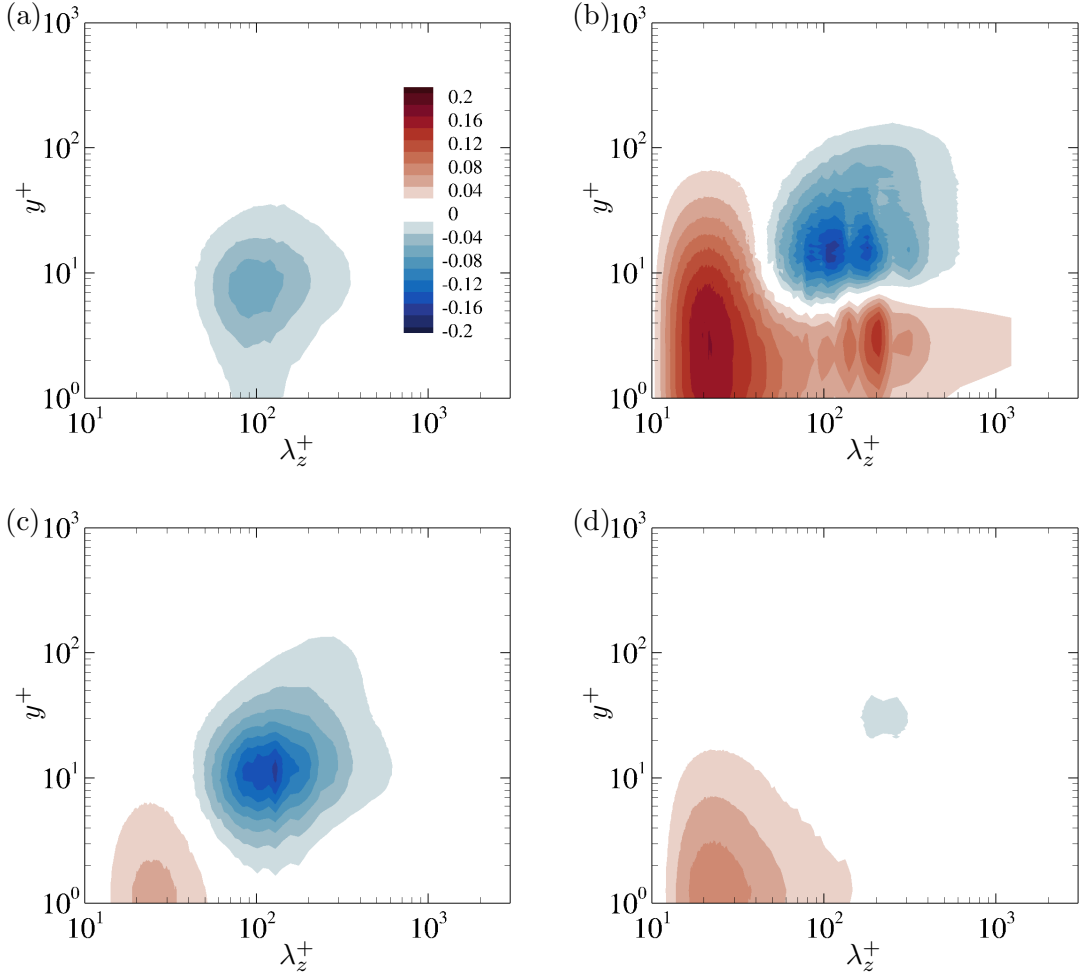

To further illustrate the work of the particle feedback forces on the turbulent kinetic energy
in the scale space, in figure~\ref{fig:fuspec} we present the pre-multiplied spanwise 
co-spectra between the particle forces and the velocity fluctuations.
For the low mass loading cases (only case P1-F02 is shown in figure~\ref{fig:fuspec}(a)), 
the values of the co-spectra are negative, with the peaks located at $y^+ \approx 5$ 
and $\lambda^+_z \approx 100$, corresponding to the locations and 
the spanwise characteristic length scales of the particle streaks.
Recalling figure~\ref{fig:instxz}(b), we infer that the particles that cluster as velocity streaks
are mostly being decelerated, and reversely, feeding back positive forces to the fluid, 
thereby leading to the negative work when acting on the negative velocity fluctuations.
In the previous study of \citet{dritselis2008numerical}, it has also been demonstrated 
via conditional average that
the particles generate the feedback forces and torques that counteract the streamwise
vortices in the cross-stream plane.
Both of these phenomena point to the elucidation that the negative work of the particles 
should be ascribed as the reason for the reduction of the turbulent intensities and 
the more coherent streaky structures in the near-wall region.
In cases P1-F06 and P2-F06 with moderate mass loading, the negative regions of the co-spectra 
are strengthened, with the peaks shifted slightly upwards and towards the larger spanwise scales.
Moreover, a small near-wall peak appears at small-scale $\lambda^+_z \approx 25$ in case P2-F06.
In case P1-F06, such positive regions are much more intense in comparison, and cover a much
wider range in the wall-normal direction at small scales and in the scale space in the near-wall
region.
Considering that the peak of the spectra is reached at $y^+ \approx 25$ and 
$\lambda^+_z \approx 200$, we conjecture that the intense energy production at small scales
and the near-wall region are transferred to such scales by the nonlinear inter-scale transfer 
and the spatial diffusion. Details are left to be explored in our future work.
In case P3-F14, the negative spectra almost disappear, while the near-wall positive spectra
intensities are slightly stronger compared with those in case P2-F06.
Since there is no obvious small-scale particle clustering in either of these cases, 
this term can be ascribed to the comparatively discrete particle force distributions,
as shown in figure~\ref{fig:uf}(c) the mixed accelerating and decelerating particles
without showing obvious spatial separation between them as in other cases.

\begin{figure}[tb!]
\centering
\begin{overpic}[width=0.5\textwidth]{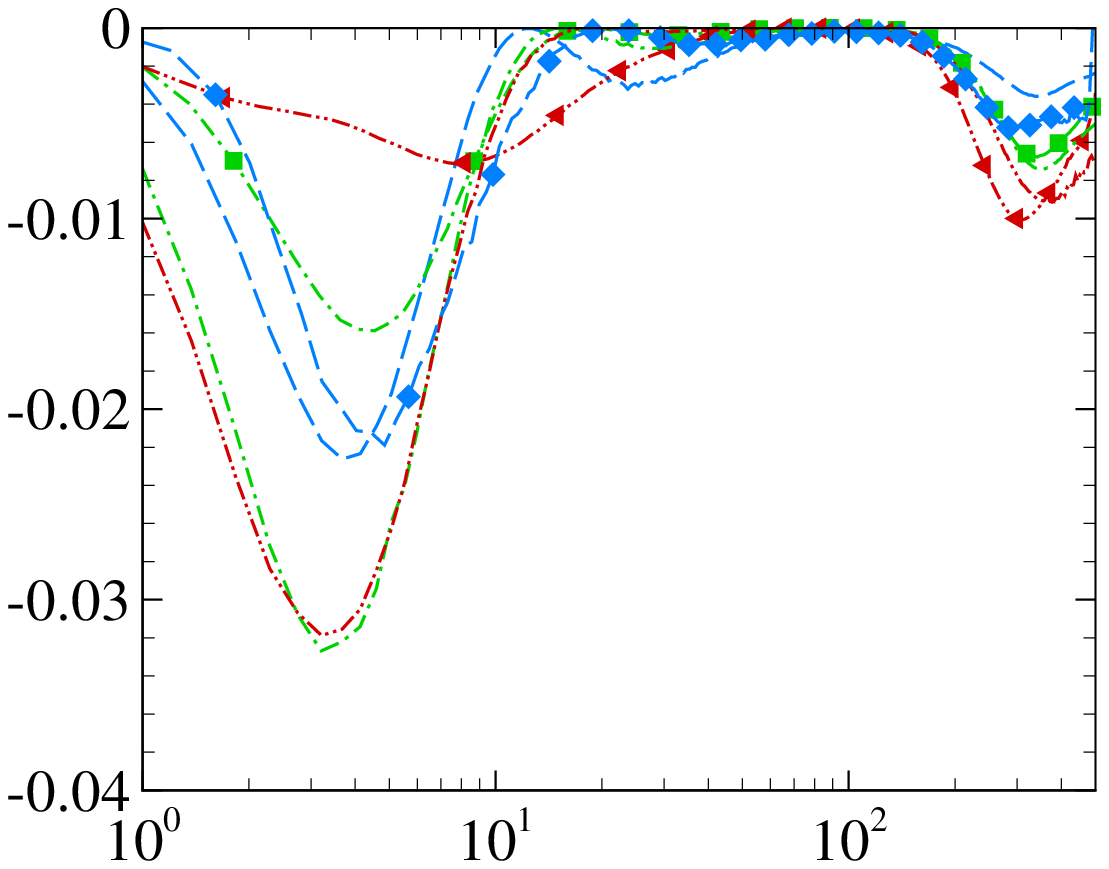}
\put(0,70){(a)}
\put(50,0){$y^+$}
\put(-2,35){\rotatebox{90}{$w^+_{p,m}$}}
\end{overpic}~
\begin{overpic}[width=0.5\textwidth]{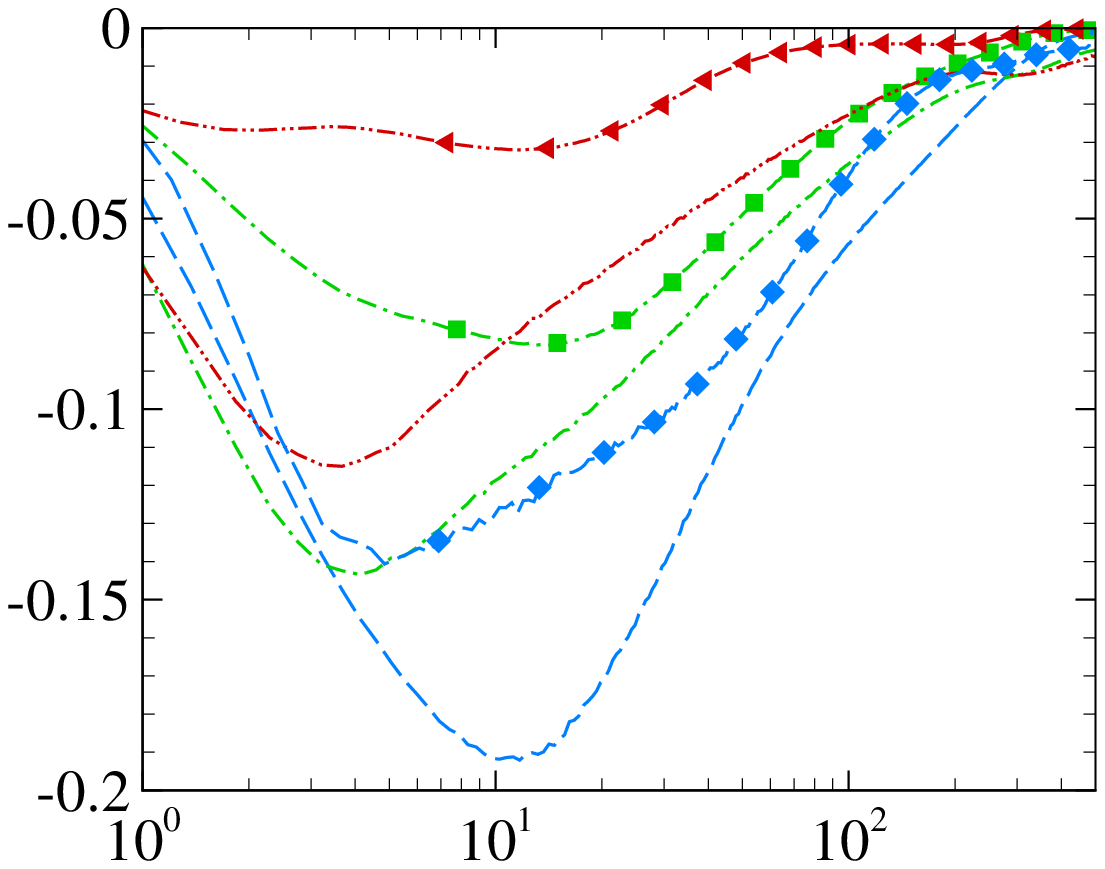}
\put(0,70){(b)}
\put(50,0){$y^+$}
\put(-2,35){\rotatebox{90}{$w^+_{p,f}$}}
\end{overpic}\\
\begin{overpic}[width=0.5\textwidth]{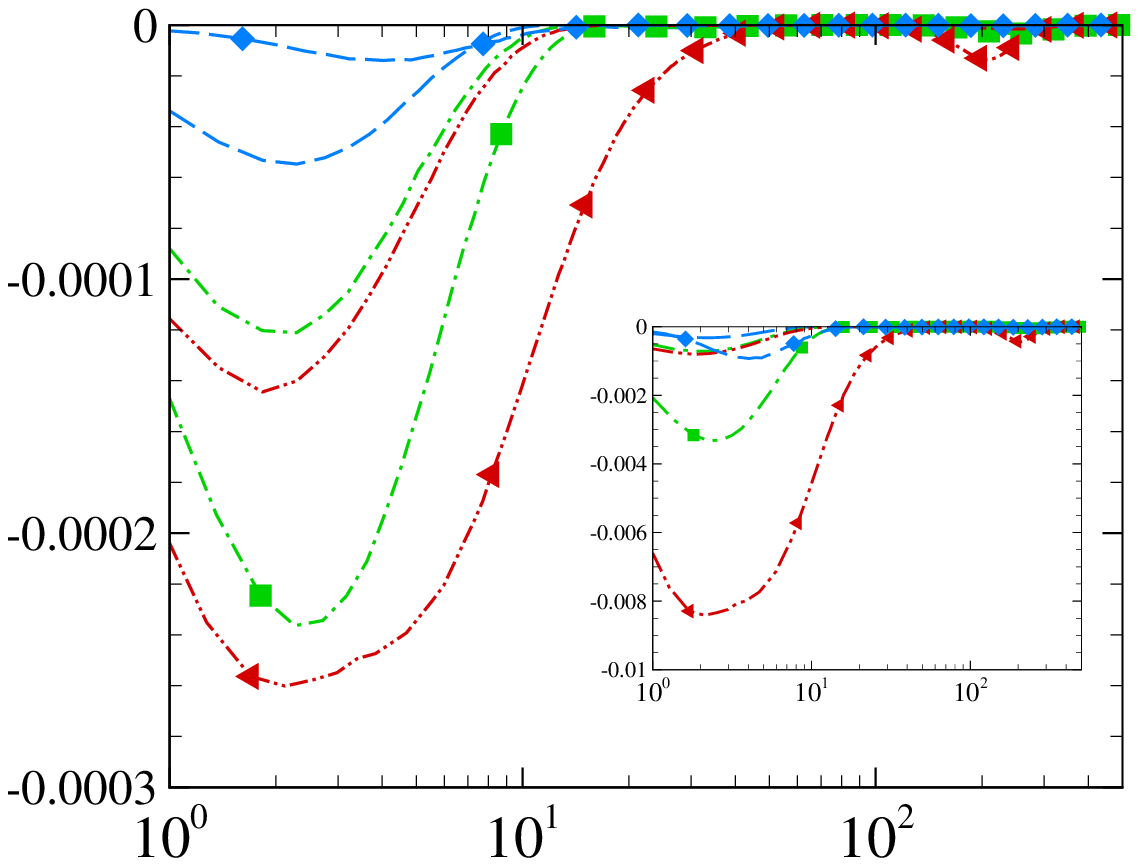}
\put(0,70){(c)}
\put(50,0){$y^+$}
\put(-7,25){\rotatebox{90}{$ \bar F^+_{p,i} (\bar v_i-\bar u_{p,i})^+$}}
\end{overpic}~
\begin{overpic}[width=0.5\textwidth]{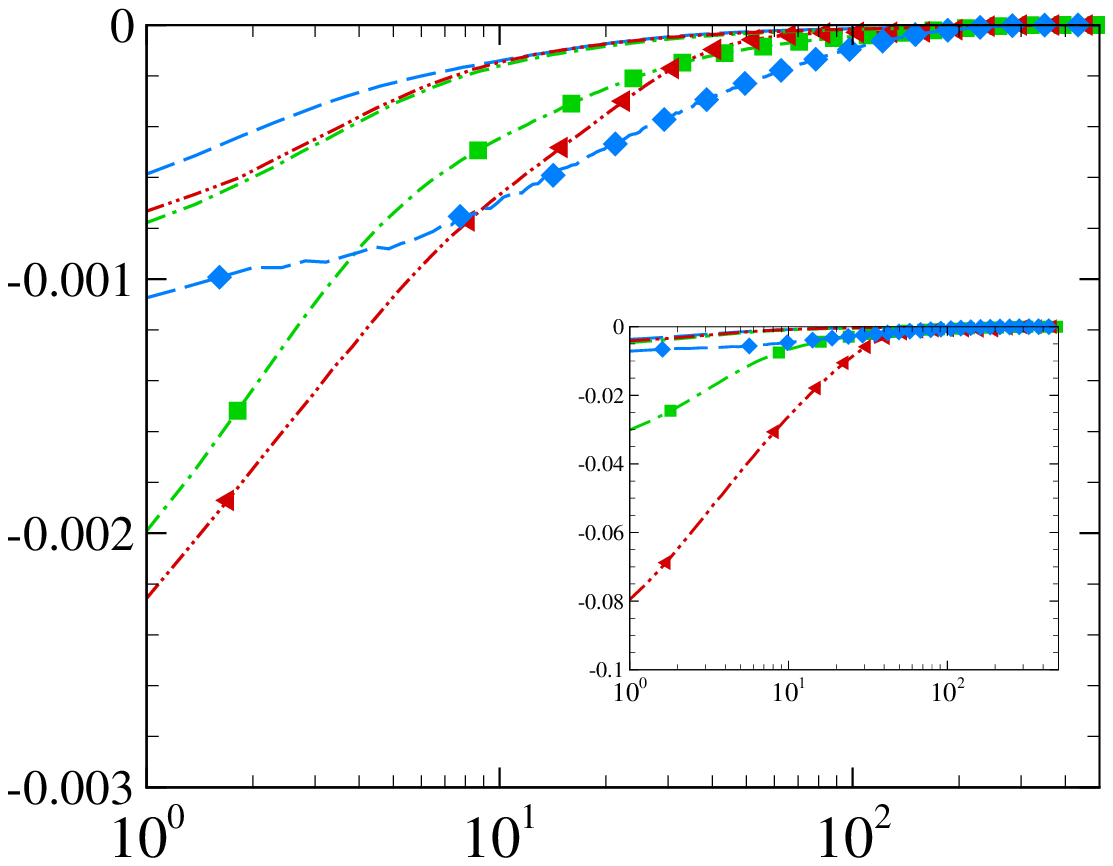}
\put(0,70){(d)}
\put(50,0){$y^+$}
\put(-7,30){\rotatebox{90}{$\overline{F'_{p,i} (v'_i - u'_{p,i})}^+$}}
\end{overpic}\\
\caption{Particle force dissipation
(a) $w^+_{p,m}$, (b) $w^+_{p,f}$, (c) $W^+_{p,m}$, (b) $W^+_{p,f}$,
insets of (c,d): normalized by wall viscous dissipation $\varepsilon_w$.
Line legends refer to Table~\ref{tab:param}.}
\label{fig:pfene}
\end{figure}

The works of the forces acting on the particles and the feedback forces acting on the fluid
do not add up to zero, as demonstrated by \citet{zhao2013interphasial}.
Specifically, the work of the fluid on a single particle per unit mass writes as
\begin{equation}
w_{fp} = F_{p,i} v_i/m_p = \frac{f_D}{\tau_p} (u_{p,i} - v_i) v_i
\end{equation}
and reversely that on the particle as
\begin{equation}
w_{pf} = -F_{p,i} u_{p,i}/m_p = -\frac{f_D}{\tau_p} (u_{p,i} - v_i) u_{p,i},
\end{equation}
and their summation
\begin{equation}
w_p = w_{fp} + w_{pf} = - \frac{f_D}{\tau_p} (u_{p,i} - v_i)^2
\end{equation}
is negative, suggesting that the forces between the fluid and particles dissipate their 
total kinetic energy, which is referred to as the particle dissipation.
In figure~\ref{fig:pfene} (a,b) we present the wall-normal distribution of the mean and fluctuating
particle dissipation $w_{p,m}$ and $w_{p,f}$.
In general, the latter is higher than the former.
The $w_{p,f}$, in particular, is decreasing with both the Stokes number $St^+$ and the mass loading
$\varphi_m$.
In the integral sense, the particle dissipation on the unit mass of fluid, 
denoted as $W_{p,m}$ and $W_{p,f}$, 
is also primarily constituted by the latter, namely the slip velocity fluctuations,
as shown in figure~\ref{fig:pfene} (c,d).
Both $W_{p,m}$ and $W_{p,f}$ are the most intense near the wall, and
under viscous scales, the orders of the magnitudes of these terms are $O(10^{-3})$ and $O(10^{-4})$,
respectively, suggesting their relatively small contribution compared with the mean kinetic energy
dissipation.
Compared with the turbulent kinetic energy viscous dissipation,
$W_{p,f}$ constitute finitely of approximately 8\% at the wall in case P3-F14 and less than 3\% 
in case P2-F06 and less than 1\% in other cases.

The particle dissipation reflects the kinetic energy loss due to the particle-fluid interaction
which, if only the drag force is considered, is produced due to the fluid viscosity.
Therefore, in the cases where the heat exchanges between the fluid and the particles are 
incorporated, namely the two-way force and heat coupling, the particle dissipation should be 
added either to the internal/total energy equation of particles or fluid, or both, 
depending on the type of particles considered, otherwise there will be total energy loss 
in such particulate systems.

\subsection{Mean temperature}  \label{subsec:temp}

\begin{figure}[tb!]
\centering
\begin{overpic}[width=0.5\textwidth]{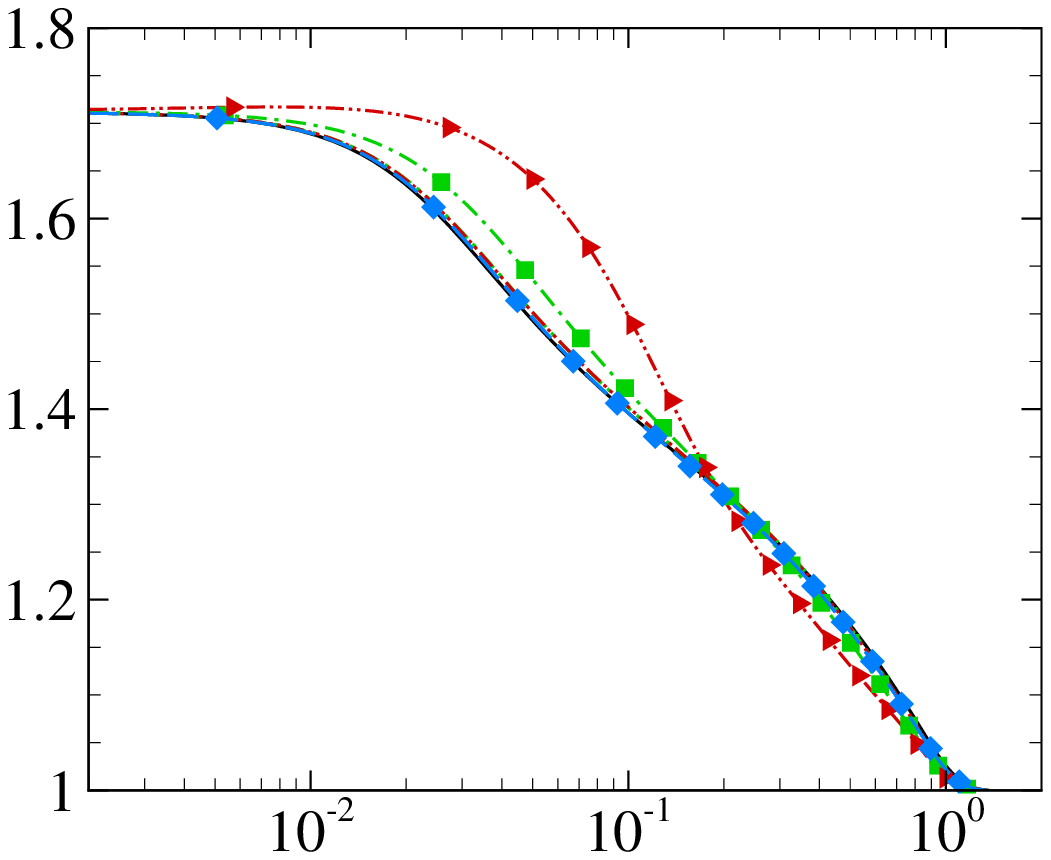}
\put(0,70){(a)}
\put(48,0){$y/\delta_0$}
\put(0,35){\rotatebox{90}{$\bar T/T_0$}}
\end{overpic}~
\begin{overpic}[width=0.5\textwidth]{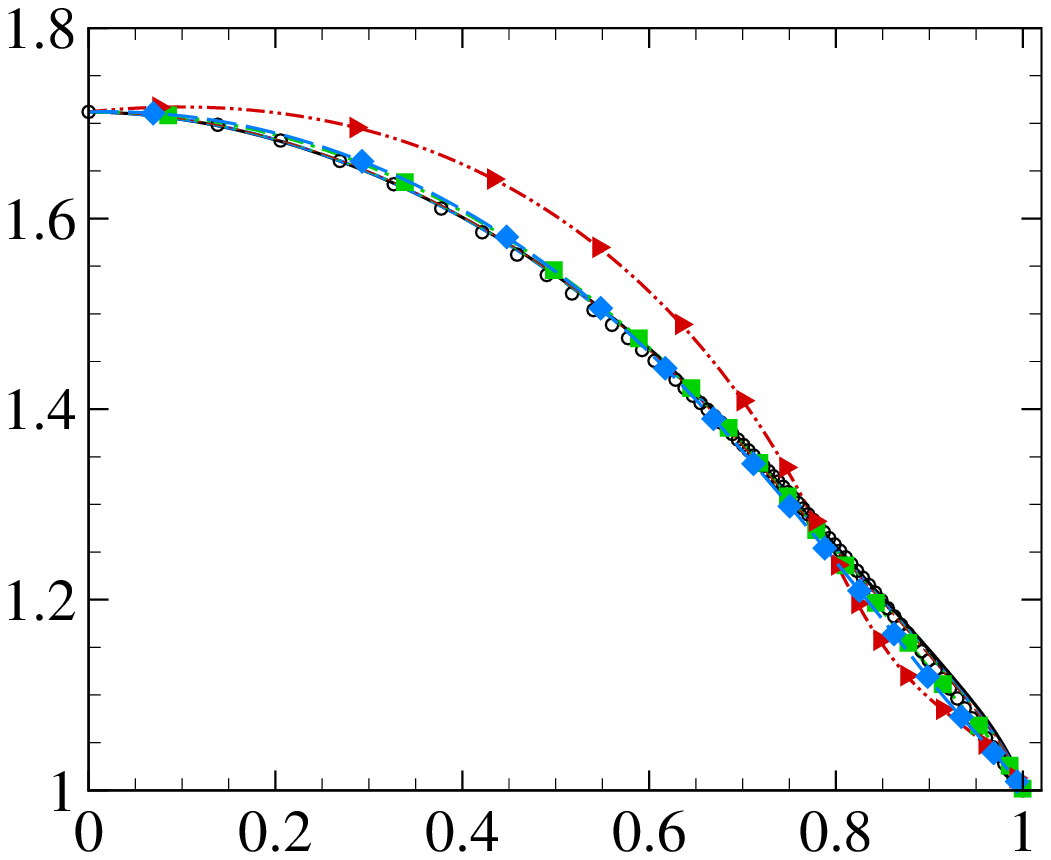}
\put(0,70){(b)}
\put(48,0){$\bar u/U_0$}
\put(0,35){\rotatebox{90}{$\bar T/T_0$}}
\end{overpic}\\[2.0ex]
\begin{overpic}[width=0.5\textwidth]{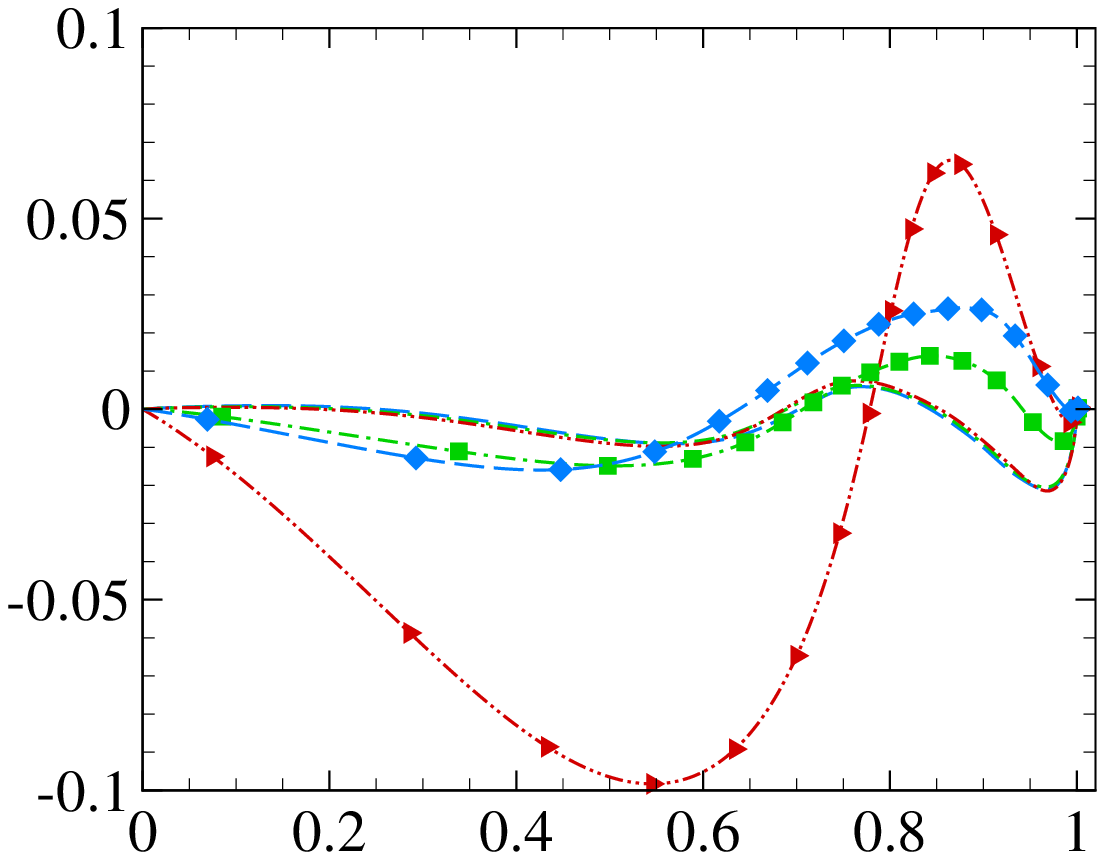}
\put(0,70){(c)}
\put(48,0){$\bar u/U_0$}
\put(0,38){\rotatebox{90}{$K_p$}}
\end{overpic}
\caption{Mean temperature distribution, (a) $\bar T/T_0$ against $y$, 
(b) $\bar T/T_0$ against $\bar u$ and (c) compensating kinetic energy $K_p$.
Line legends refer to Table~\ref{tab:param}, 
black circles in (b): GRA in equation~\eqref{eqn:gra}.}
\label{fig:meanT}
\end{figure}

Lastly, we discuss the two-way coupling effects on the mean temperature,
as shown in figure~\ref{fig:meanT}(a).
Compared with the one-way coupling case, the mean temperature is not much affected in the cases 
with low particle mass loadings and case P1-F06.
In moderate and high mass loading cases P2-F06 and P3-F14, however, 
the mean temperature is augmented near the wall below $y=0.2\delta_0$ while abated above 
that location.
This leads to the disparity from the generalized Reynolds analogy (GRA)~\citep{zhang2014generalized},
which is written as 
\begin{equation}
\frac{\bar T}{T_0} = \frac{\bar T_w}{T_0} + \frac{T_{rg}-T_w}{T_0} \frac{\bar u}{U_0}
+\frac{T_\infty - T_{rg}}{T_0} \left( \frac{\bar u}{U_0} \right)
\label{eqn:gra}
\end{equation}
with the generalized recovery temperature $T_{rg} = T_0 + r_g U^2_0/(2C_p)$ 
and $r_g$ the generalized recovery coefficient.
Obviously, in the case of canonical compressible wall-bounded turbulence, the mean temperature
is a quadratic function of the mean velocity, dependent on the wall temperature and the free-stream
Mach numbers.
As illustrated in figure~\ref{fig:meanT}(b), this is approximately the case for low and moderate
particle mass loadings, but evidently not for the high mass loading case P3-F14.

The differences from the GRA have also been found in our previous studies regarding
the compressible turbulent boundary layers subject to wall disturbances
~\citep{yu2023effects,yu2023compressibility,yu2024turbulent}, 
where the deviation of the relation between the mean temperature and mean velocity 
can be ascribed to the disparities in the boundary conditions.
For the presently considered cases, however, such deviations should be attributed to
the particle feedback forces.
We start from the resemblance between the mean momentum and total enthalpy equations.
Under the assumption of $Pr=1$, the mean momentum and total enthalpy equations can be approximately 
cast as
\begin{equation}
\bar \rho \bar u_1 \frac{\partial \bar u_1}{\partial x} + 
\bar \rho \bar u_2 \frac{\partial \bar u_1}{\partial y} \approx 
\frac{\partial}{\partial y} \left( \bar \mu \frac{\partial \bar u_1}{\partial y} 
- \overline{\rho u''_1 u''_2} \right) + \bar F_{p,1}
\end{equation}
\begin{equation}
\bar \rho \bar u_1 \frac{\partial \bar H}{\partial x} + 
\bar \rho \bar u_2 \frac{\partial \bar H}{\partial y} \approx 
\frac{\partial}{\partial y} \left( \bar \mu \frac{\partial \bar H}{\partial y} 
- \overline{\rho H' u''_2} \right) + \bar F_{p,1} \bar u_1.
\end{equation}
with the streamwise Reynolds stress and viscous stress neglected.
Multiplying the former by $U_w$ and subtracting it from the latter, we have
\begin{equation}
\left( \bar \rho \bar u_1 \frac{\partial }{\partial x} 
+ \bar \rho \bar u_2 \frac{\partial }{\partial y} 
- \frac{\partial}{\partial y} \left( \bar \mu \frac{\partial }{\partial y}\right) \right) 
(\bar H - U_w \bar u) = \bar F_{p,1} (\bar u_1-U_w).
\label{eqn:subs}
\end{equation}
In the cases without particle feedback forces, the right-hand-side is zero, 
then we have the solution
\begin{equation}
\bar H - \bar H_w = U_w \bar u.
\end{equation}
Here, $H_w$ is the total enthalpy at the wall and $U_w = -Pr q_{yw}/\tau_w$.
\citet{zhang2014generalized} further suggested that $\bar H_g = C_p T + r_g \bar u^2_1/2$ 
should be used instead of $\bar H$ to incorporate the non-unity of $Pr$.
This gives the GRA above in equation~\eqref{eqn:gra}.

If the right-hand-side term is non-negligible, the generalized total enthalpy $\bar H_g$
should be further expressed as
\begin{equation}
\bar H_g = C_p T + r_g \bar u^2_1/2 + K_p,
\end{equation}
where $K_p$ represents the kinetic energy transported by particles, hereinafter referred to as the 
compensating kinetic energy.
The $K_p$ is a solution of the following equation
\begin{equation}
\left( \bar \rho \bar u_1 \frac{\partial }{\partial x} 
+ \bar \rho \bar u_2 \frac{\partial }{\partial y} 
- \frac{\partial}{\partial y} \left( \bar \mu \frac{\partial }{\partial y}\right) \right) 
K_p = \bar F_{p,1} (\bar u_1-U_w),
\label{eqn:subs2}
\end{equation}
but it cannot be easily solved because the operator on the left-hand-side also depends on
the particle feedback force on the right-hand-side.
Statistically, we can obtain the $K_p$ by subtracting the original GRA from the DNS results,
as is shown in figure~\ref{fig:fuspec}(c).
It is negative in the inner region and positive in the outer region, corresponding to
the influences of the kinetic energy transfer between the particles and the fluid.

The compensating kinetic energy $K_p$ is by definition the kinetic energy loss caused by 
the particle forces. This can also be supported by their same trend of variation along the
wall-normal direction (comparing figures~\ref{fig:meanT}(c) and \ref{fig:pfene}(d)).
The transport of the kinetic energy suggests that the work of the particle forces is balanced
by the viscous dissipation, which transfers the kinetic energy to the internal energy,
thus leading to the increment of the local temperature.
From another perspective, in the cases where near-wall turbulence remains active, 
the vertical turbulent heat flux further redistributes the internal energy, 
rendering the mean temperature-velocity relation
similar to the canonical compressible turbulent boundary layers.
On the contrary, in high mass loading cases where the near-wall turbulence is laminarized,
the high temperature close to the wall cannot be transported towards the free-stream.
These factors could probably synergize, resulting in a higher temperature in the near-wall region.



\section{Conclusions} \label{sec:con}

In the current investigation, we conducted direct numerical simulations of particle-laden 
compressible turbulent boundary layers at a free-stream Mach number of 2.0. 
We focused on the two-way force coupling between the particles and the fluid, 
particularly examining how particles modulate momentum and kinetic energy transfer across 
various mass loadings and particle Stokes numbers.

The presence of particles leads to a suppression of turbulence and laminarization, 
characterized not only by weaker, wider, and more coherent velocity streaks but also 
by reductions in skin friction, turbulent kinetic energy, and Reynolds shear stress. 
The extent of turbulence suppression is primarily dependent on the mass loadings, 
but also on the particle Stokes number, which influences the preferential accumulation and 
clustering of particles near the wall.
In cases of the highest mass loading examined in this study, 
near-wall turbulence is entirely laminarized. 
The remnant turbulent fluctuations are induced by particles that are swept down 
from the outer region.

The analysis of mean momentum transfer and skin friction decomposition reveals that 
the presence of particles minimally affects viscous shear stress. 
However, the contribution from Reynolds stress diminishes while that from 
particle feedback force escalates with increasing mass loading. 
Through skin friction decomposition, it was observed that the aggregate of terms, 
excluding the particle feedback force, transitions from a turbulent empirical relationship 
to a laminar one as mass loading increases.

The distributions of turbulent kinetic energy and pre-multiplied spanwise spectra 
conclusively demonstrate that moderate mass loadings disrupt near-wall turbulence, 
exhibiting a broader characteristic spanwise length scale at elevated wall-normal positions. 
In the case with high mass loading, near-wall turbulence is entirely eradicated. 
Analysis of turbulent kinetic energy transport indicates significant suppression of 
both production and dissipation, with particle-induced work predominating the production of 
turbulent kinetic energy in the near-wall region and at smaller scales for moderate and 
high mass loadings. 
Particle dissipation, induced by the interaction between particles and fluid due to viscosity, 
is substantial relative to viscous dissipation and merits consideration in simulations 
involving two-way force and heat coupling. 
Additionally, it was found that the mean temperature is elevated in the near-wall region and 
reduced in the outer region in cases of high mass loading, 
attributable to particle feedback forces and the diminution of near-wall turbulent motions.


\appendix

\section{Validation of the numerical solver} \label{sec:val}

We validate our two-way coupling particle-laden compressible turbulence solver
by performing direct numerical simulations for turbulent channel flows at the bulk Mach number 
$M_b= u_b/a_w=0.5$ and friction Reynolds number $Re_\tau = \rho_w u_\tau h \approx 180$, 
with $u_b$ the bulk velocity, $a_w$ the acoustic velocity at the wall and 
$h$ the half channel height.
The turbulence is monitored to be temporally fully developed before $4 \times 10^6$ particles
with the diameters of $d_p = 0.004 h$ and the density ratio
$\rho_p = 1042 \rho_b$ ($\rho_b$ is the fluid bulk density) are randomly injected into the channel
with the size of $12h$, $2h$ and $6h$ in the streamwise, wall-normal and spanwise directions.
The particle Stokes number $St^+$ is approximately $30$ and the mass loading is 
$\varphi_m \approx 1.0$.

\begin{figure}[tb!]
\centering
\begin{overpic}[width=0.5\textwidth]{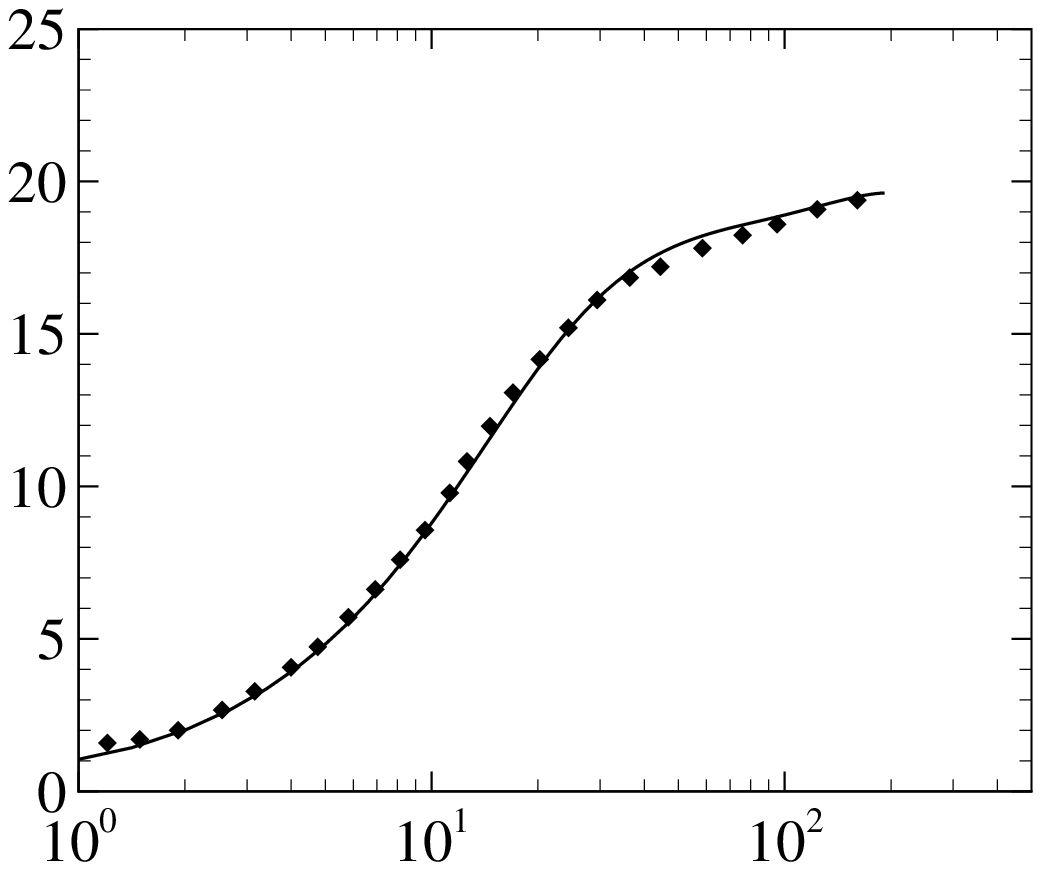}
\put(0,70){(a)}
\put(50,0){$y^+$}
\put(0,38){\rotatebox{90}{$\bar u^+_1$}}
\end{overpic}~
\begin{overpic}[width=0.5\textwidth]{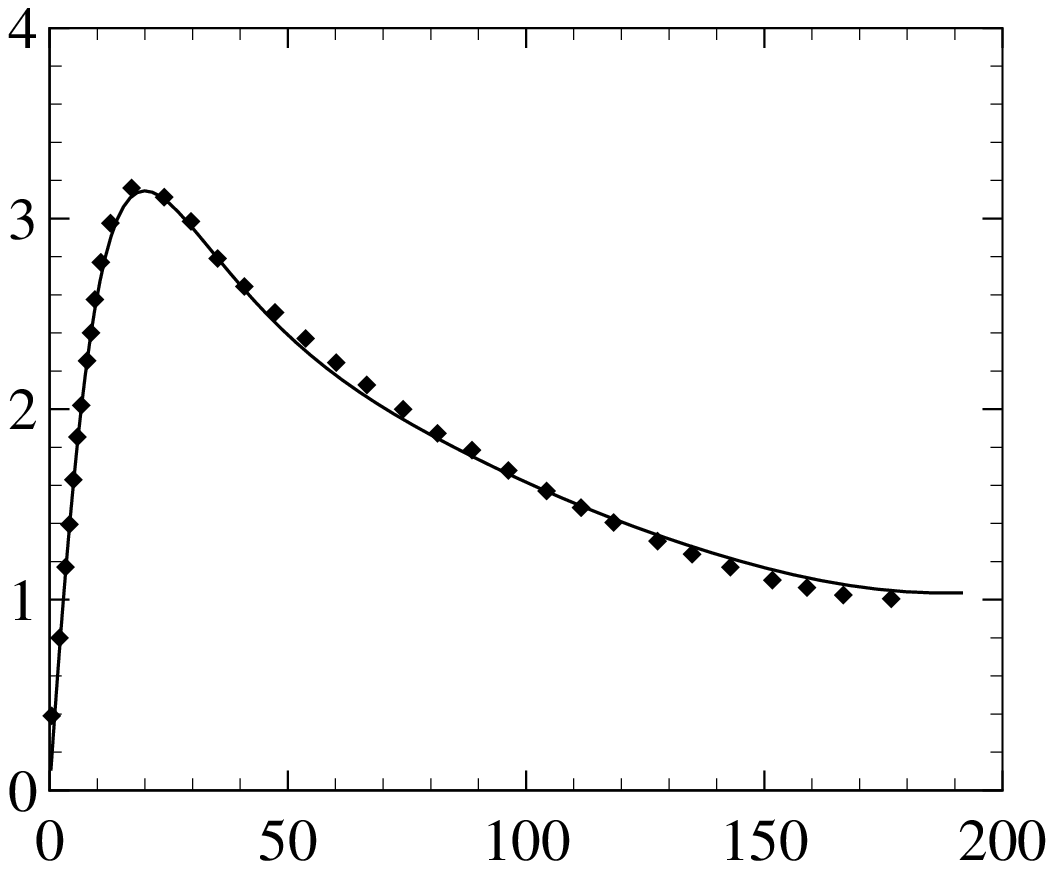}
\put(0,70){(b)}
\put(50,0){$y^+$}
\put(0,38){\rotatebox{90}{$u'^+_{1,rms}$}}
\end{overpic}\\[2.0ex]
\begin{overpic}[width=0.5\textwidth]{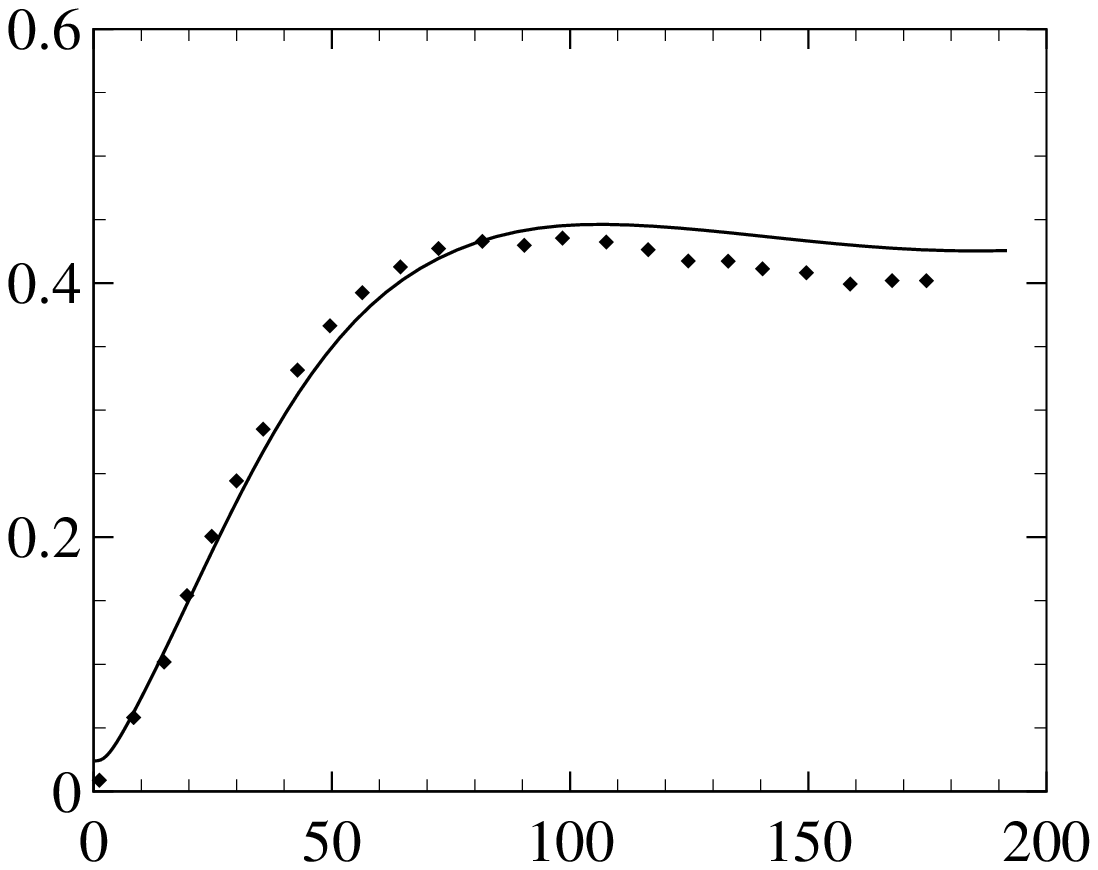}
\put(0,70){(a)}
\put(50,0){$y^+$}
\put(0,38){\rotatebox{90}{$u'^+_{2,rms}$}}
\end{overpic}~
\begin{overpic}[width=0.5\textwidth]{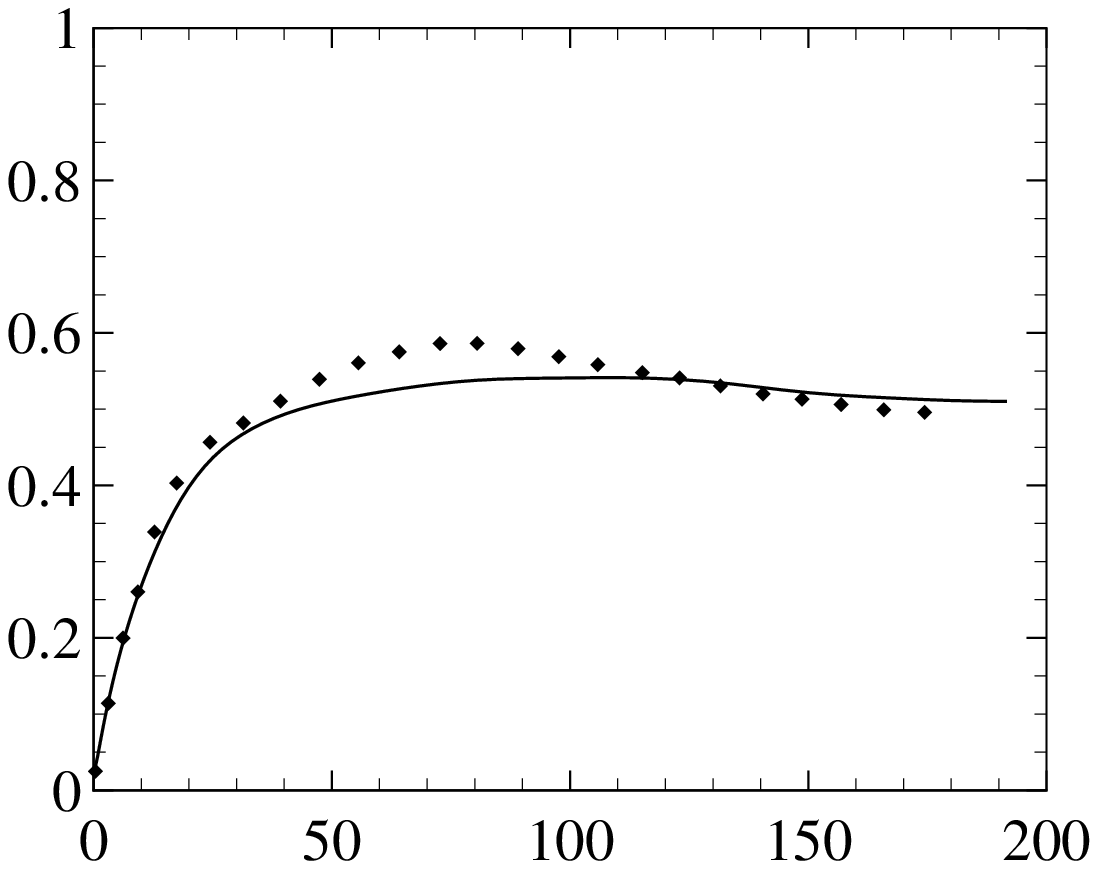}
\put(0,70){(a)}
\put(50,0){$y^+$}
\put(0,38){\rotatebox{90}{$u'^+_{3,rms}$}}
\end{overpic}\\
\caption{Wall-normal distribution of the (a) mean velocity $\bar u^+_1$ and
(b) $u'^+_{1,rms}$, (c) $u'^+_{2,rms}$, (d) $u'^+_{3,rms}$.
Lines: present simulation, symbols: reported by~\citet{zhao2010turbulence}.}
\label{fig:val}
\end{figure}

We compare our simulation results with those reported by~\citet{zhao2010turbulence},
as shown in Figure~\ref{fig:val}.
Both the mean velocity and the root-mean-square of the velocity fluctuations collapse well
with the reference data, proving our numerical solver valid.

\bibliography{bibfile}

\end{document}